\documentclass[onecolumn,secnumarabic,amssymb, notitlepage, nobibnotes, aps, prd]{revtex4-1}

\usepackage{amsmath}
\usepackage{amssymb}
\usepackage{graphicx}
\usepackage[cmyk]{xcolor}
\usepackage{float}
\usepackage{epstopdf}
\usepackage{caption}
\usepackage{subcaption}
\usepackage[section]{placeins}
\usepackage{soul}
\usepackage{color}
\usepackage{multirow}

\makeatletter
\g@addto@macro\@floatboxreset\centering
\makeatother
\usepackage{adjustbox}
\adjustboxset*{center}

\makeatletter
\AtBeginDocument{%
  \expandafter\renewcommand\expandafter\subsection\expandafter{%
    \expandafter\@fb@secFB\subsection
  }%
}
\makeatother

\begin{document}

\title{A boundary element method for the solution of finite mobility ratio immiscible displacement in a Hele-Shaw cell}

\author{S.J. Jackson}%
\author{D. Stevens}%
\author{H. Power}%
\email[henry.power@nottingham.ac.uk]{}
\author{D. Giddings}%
\affiliation{Faculty of Engineering, Division of Energy and Sustainability, University of Nottingham, UK}
\date{February 17, 2015}

This is the peer reviewed version of the following article: [Jackson, SJ, Stevens, D, Power, H, and Giddings, D (2015), A boundary element method for the solution of finite mobility ratio immiscible displacement in a Hele‐Shaw cell. Int. J. Numer. Meth. Fluids, 78, 521– 551], which has been published in final form at [https://doi.org/10.1002/fld.4028]. This article may be used for non-commercial purposes in accordance with Wiley Terms and Conditions for Use of Self-Archived Versions.\\

\begin{abstract}

In this paper, the interaction between two immiscible fluids with a finite mobility ratio is investigated numerically within a Hele-Shaw cell. Fingering instabilities initiated at the interface between a low viscosity fluid and a high viscosity fluid are analysed at varying capillary numbers and mobility ratios using a finite mobility ratio model. 

The present work is motivated by the possible development of interfacial instabilities that can occur in porous media during the process of  $CO_2$ sequestration, but does not pretend to analyse this complex problem. Instead, we present a detailed study of the analogous problem occurring in a Hele-Shaw cell, giving indications of possible plume patterns that can develop during the $CO_2$ injection.

The numerical scheme utilises a boundary element method in which the normal velocity at the interface of the two fluids is directly computed through the evaluation of a hypersingular integral. The boundary integral equation is solved using a Neumann convergent series with cubic B-Spline boundary discretisation, exhibiting 6th order spatial convergence. The convergent series allows the long term non-linear dynamics of growing viscous fingers to be explored accurately and efficiently.

Simulations in low mobility ratio regimes reveal large differences in fingering patterns compared to those predicted by previous high mobility ratio models. Most significantly, classical finger shielding between competing fingers is inhibited. Secondary fingers can possess significant velocity, allowing greater interaction with primary fingers compared to high mobility ratio flows. Eventually, this interaction can lead to base thinning and the breaking of fingers into separate bubbles.

\end{abstract}

\maketitle

\section{Introduction}

Viscous fingering occurs during the displacement of a high viscosity fluid by a low viscosity fluid, in which interfacial instabilities may arise and subsequently evolve to form complex interface topologies. Perturbations greater than a certain wavelength create instabilities along the fluid interface and promote the growth of long fingers which penetrate into the more viscous fluid. Since the work of Saffman and Taylor in 1958 \cite{saffman1958}, there has been extensive research on viscous fingering occurring in Hele-Shaw cells, where the fluid flows between two thinly separated plates. The mobility of a fluid within a Hele-Shaw cell is defined by the cell separation and the viscosity, giving rise to an intrinsic permeability, analogous to that in porous media flows. 

Although the present work is motivated by the possible development of interfacial instabilities that can occur in porous media during the displacement of resident brine due to the injection of supercritical $CO_2$ in carbon sequestration, it does not pretend to analyse this complex problem and instead presents a detailed study of the analogous problem occurring in a Hele-Shaw cell, giving some indications of possible plume patterns that can occur during the $CO_2$ injection.

\begin{figure}[H]
\centering
\adjincludegraphics[scale=0.8]{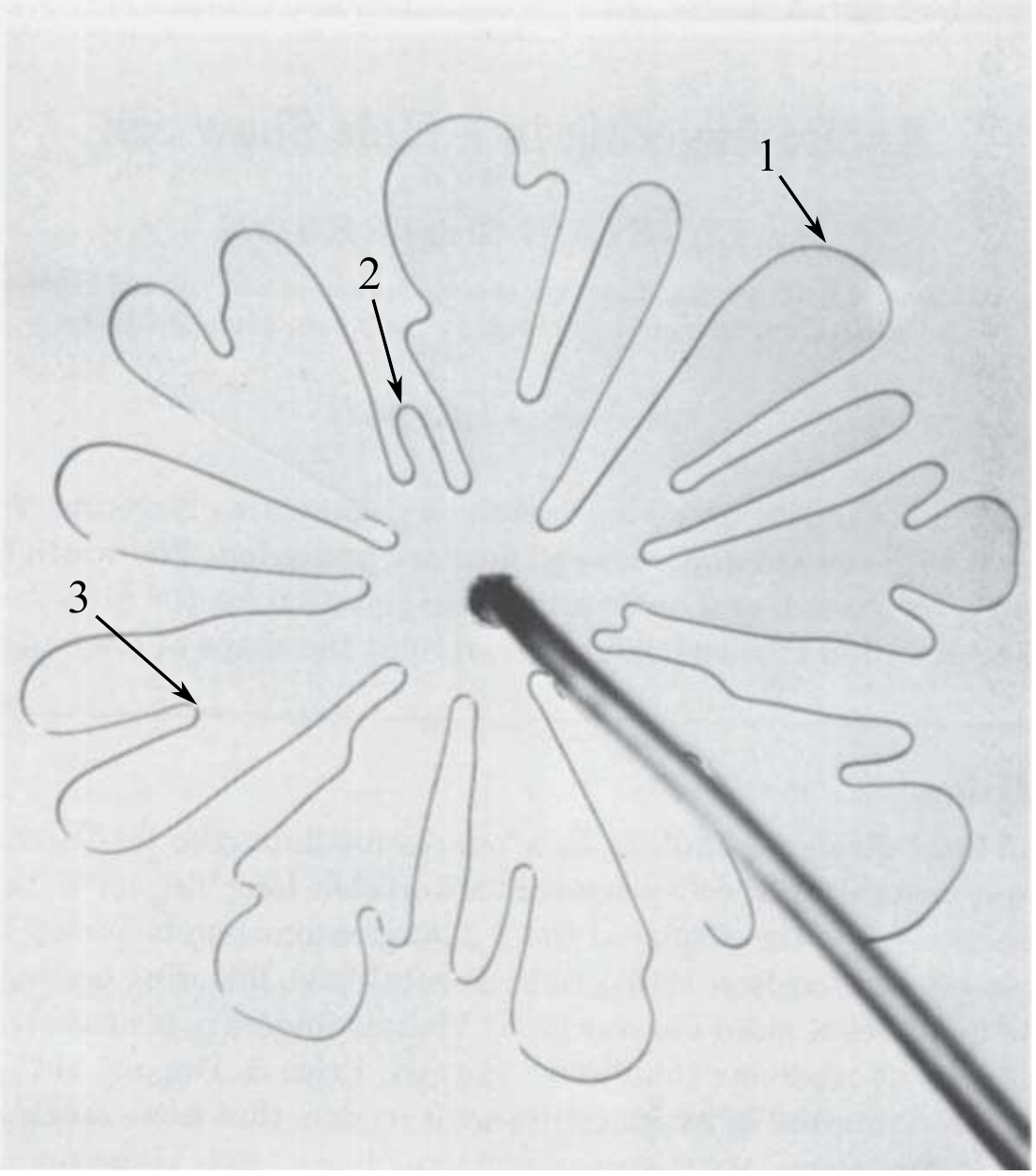}
\centering
\vspace{-10pt}
\caption{Viscous fingering occurring during the radial injection of air into a Hele-Shaw cell filled with glycerine \cite{patterson1981}. Numbers correspond to the three basic fingering mechanisms: 1 - Spreading; 2 - Shielding; 3 - Splitting. Figure reproduced with permission from the Journal of Fluid Mechanics.}
\label{spread}
\end{figure}

Viscous fingering is typically described by three main mechanisms; spreading, shielding and splitting, which occur over a wide range of length scales \cite{homsy1987}\cite{howison1986}. Figure (\ref{spread}) shows these three mechanisms occurring in a Hele-Shaw cell experiment performed by Patterson in 1981 \cite{patterson1981}. As a finger grows due to the radial injection of an inner, less viscous (more mobile) fluid, the front of the advancing finger is spread continuously, creating a fan like structure with an increasingly flat front. This growth holds in the linear regime, until the flat section at the front of the advancing finger becomes larger than the critical length scale of bifurcation. At this point, tip-splitting occurs, whereby the tip will bifurcate into smaller fingers creating a more convoluted surface. These fingers will grow and compete with each other, with larger fingers shielding the growth of smaller fingers in a non-linear regime controlled by interfacial dynamics.

The early stages of viscous finger growth occur in a linear regime, where perturbations greater than the critical wavelength form into separate fingers. The fastest growing fingers have a wavelength proportional to the square root of the capillary number of the flow, relating the viscous driving forces to surface tension forces \cite{patterson1981}\cite{tanveer2000}\cite{maxworthy1989}. Linear stability analysis gives good correspondence with both radial and channel flows during the early stages of finger growth. However, when the coupling of the different modes of perturbation becomes significant, the regime becomes weakly non-linear and processes such as spreading and splitting occur \cite{miranda1998}. Including second order terms in the Fourier decomposition of the modes of perturbation, \cite{miranda1998} shows that harmonic and sub-harmonic perturbation modes are responsible for tip splitting and finger competition respectively. Past this weakly non-linear stage, during the later stages of finger evolution, finger growth is only fully described by the full coupling of modes in the Fourier decomposition of the perturbation. To this extent, it is necessary to use numerical methods to fully explore the non-linear regime, as current stability analyses insufficiently describe the interaction between fingers and their resulting non-linear growth. 

There is substantial experimental evidence previously reported in the literature, where it is shown that the processes of shielding, spreading, and splitting are also present during viscous fingering evolution in porous media, and determine the pattern of the fluids interface. Chouke et al. observed the formation fingering patterns in immiscible displacement in porous media, which show variation of the length scales with increasing velocity and viscosity contrast, i.e. with increasing capillary number \cite{chouke1959}. Fingering takes place on many scales, including a macroscopic one, suggesting the existence of a characteristic macroscopic length scale or wavelength. 



Immiscible displacement is characterised by a sharp interface, across which the properties of the fluids (such as viscosity and density) vary discontinuously \cite{homsy1987}. One such flow, and the motivation behind the current work, is the injection of supercritical $CO_2$ into deep subsurface aquifers containing brine. Here, the injection process can be considered immiscible, with the mobility ratio between the fluids typically of order 10, involving high characteristic capillary numbers. 

The first attempt to provide a theoretical analysis of the onset of immiscible viscous fingering in porous media was by Chouke et al. in 1959 \cite{chouke1959}. They assumed that there was complete displacement of one fluid by the other, using the similarity between Hele-Shaw and porous media flows and ignoring the zone of partial saturation or volume concentration of the displacing fluid behind the front. In the case of two dimensional immiscible displacement in a porous medium, the finger characteristic width scale predicted by a Hele-Shaw approximation under predicts the experimental observations. This has led to the hypothesis of an effective surface tension, larger than the molecular surface tension and function of the wetting conditions, that varies with the large-scale curvature at the macro-scale (see \cite{weitz1987}). The use of a modified jump condition in terms of the effective surface tension is known as Chouke’s boundary condition and the resulting interface instability analysis is referred to Hele-Shaw-Chouke theory (for more details see the review article \cite{homsy1987}). 

An alternative approach to study the viscous fingering instability of the displacement of immiscible fluids in a porous media can be obtained from the classical porous media formulation of multiphase flows in terms of the saturation index, $S_w$, where an overlapping region between the fluids is considered for $0 \leq S_w \leq 1$, without definition of the fluid interface. This type of formulation is now one of the most popular approaches used in the numerical solution of immiscible flows displacement in porous media (see \cite{garcia2003} and \cite{riaz2006}).  

In the multiphase flow approach, the variation in saturation in the overlapping region results in a gradual change of the mobility of both phases. This type of analysis is closely related to the stability of graded mobility process, see \cite{gorell1983} and \cite{hickernell1986}, where depending upon the mobility function a displacement that has an unfavourable viscosity ratio may still be linearly stable, even at infinite Capillary number. 

Both types of models for immiscible displacement in porous media, i.e. the saturation index (multiphase flow) and sharp front (Hele-Shaw-Chouke), are consistent with the main hypothesis of Darcy flow, i.e. seepage average flow. In the saturation index approach, the flow field of both fluids in the region near the front is averaged in a representative elementary volume (REV) resulting in a type of fluid mixture characterised by the saturation index, $S_w$. On the other hand, in the sharp front approach, the irregular and complex interface at the porous media is averaged in a representative smooth surface. 

In this work, we use a 2D Darcy model to analyse the simplified problem of viscous fingering in a Hele-Shaw cell, using a sharp front approach. By considering that the flow between the plates in the Hele-Shaw cell follows a Poiseuille profile, the Stokes equation can be reduced to a Darcy equation by depth-averaging across the gap. The immiscible displacement of the fluids is then described by 2D potential flow in the plane of the Hele-Shaw cell.

During immiscible displacement, the advancing front is defined by kinematic and dynamic matching conditions at the interface of the two fluids. The surface tension and curvature cause a jump in the pressure which along with continuity of normal velocities at the interface must be matched by the solutions in both fluid domains. Detailed and robust analyses of immiscible displacement in Hele-Shaw cells have been the subject of many publications in the literature including the review article "Surprises in viscous fingering" by Tanverr \cite{tanveer2000}. 

Most previous work in the literature has focussed on flow regimes where the mobility ratio of the fluids is typically very large, such as gas-oil injection occurring in enhanced oil recovery. Therefore, most numerical approaches consider only the external fluid, with an injected fluid of negligible viscosity, resulting in an infinite mobility ratio model \cite{zhao1995}\cite{degregoria1986}\cite{li2007}. Immiscible displacement with finite mobility ratio has not been as extensively explored, mainly due to difficulties associated with matching the boundary conditions for both internal and external fluids at the interface. 

Boundary element methods (BEMs) are one of the most popular techniques for solving immiscible displacement in a Hele-Shaw cell, whereby the the dimensionality of the problem is reduced by one and accurate representation of the surface is provided, explicitly tracking it through time. Although only the surface of the problem has to be discretised, a fully populated collocation matrix is generated due to the integral equations being used. This can lead to very slow solution times and poor scaling. Li \cite{li2007} uses scaling techniques to rescale time and space so that the interface can evolve significantly faster without changing the interface, allowing much longer simulated times to be run. 

In addition to BEMs for use in the limit of infinite mobility ratio, BEMs have also been applied for finite mobility ratio flows, where the viscosity of both fluids is considered, resulting in a finite mobility ratio. These methods typically solve immiscible displacement between fluids with high mobility ratio, effectively reducing the model to that of an infinite mobility ratio \cite{degregoria1986}\cite{hadavinia1995}\cite{hansen1999}. Utilising a direct boundary integral approach, \cite{hadavinia1995} and \cite{hansen1999} are able to solve directly for the surface velocity to create a finite mobility ratio model applicable in both fluid domains. \cite{hadavinia1995} evaluates a set of integral equation systems expressing the internal and external fluid domains in terms of their corresponding integral equation formulations with an auxiliary external boundary enclosing the outer fluid domain. Using an auxiliary external boundary introduces additional error into the solution, which can be reduced by moving the boundary far into the external domain at the expense of increased computational cost.

The auxiliary external boundary can be evaluated analytically in the limit that the boundary tends to infinity. When the external boundary is evaluated asymptotically at infinity, it gives rise to the solvability condition for the unique solution of the infinite mobility ratio problem \cite{jawson1977}. 

An alternative approach to deal with the external boundary is presented in \cite{hansen1999}, in which the integral equation is transformed using a Green's function and periodic boundary conditions, meaning the evaluation of an auxiliary external boundary ($ {S}_{\infty}  $) can be avoided. This comes at the expense of introducing a periodic solution in the domain and the need to solve a Cauchy weakly singular integral.


Some authors propose the use of an indirect boundary integral approach, whereby a fictitious density variable is computed before the velocity is reconstructed at the interface between the two fluids \cite{li2007}\cite{power1994}. By utilising an indirect approach solely in terms of double layer potentials, the need to evaluate an auxiliary external boundary is avoided due to the double layer potential asymptotic condition at infinity.

To ease the computational cost imposed by the front tracking methods above, alternative approaches can be used whereby the interface is captured implicitly, such as the volume of fluid method \cite{guan2003}, the diffuse interface method \cite{sun2008} and the level set method \cite{hou1997}. Here, long term dynamics can be efficiently modelled as fully populated matrices are not encountered. However, as the interface is not explicitly tracked, events that occur at a length scale smaller than the volume size or transition region cannot be accurately captured. 

Experimental results from \cite{moore2003}, along with numerical results from \cite{sun2008} and \cite{guan2003} suggest that the basic fingering mechanisms such as shielding, spreading and tip-splitting that occur in low mobility ratio flows are vastly different to those in infinite (or very high) mobility ratio flows. Due to the small fingers in the domain (typically those that have branched from the side of a primary finger) possessing significant velocity compared to those in infinite mobility ratio models, finger interaction becomes much more prominent and the resulting competition can lead to coalescence and breaking \cite{guan2003}. 

To study the interaction processes and the long term evolution of low mobility ratio flows, a non-dimensional finite mobility ratio model is developed based on a direct boundary element approach first presented in \cite{power1995}. The finite mobility ratio formulation proposed by Power \cite{power1995} has not been previously implemented in the literature. Previous work has focused on indirect methods with constant boundary elements, or infinite mobility ratio approaches \cite{power1994}\cite{power1995}. In the proposed method, the hypersingular integral arising from the single integral equation is evaluated explicitly, resulting in a second kind Fredholm equation, which can be solved through the use of an analytical Neumann series. Numerically, the Neumann series is truncated using a finite number of terms, giving rise to a convergent series solution with good agreement to the analytical Neumann series. 

The need to evaluate an auxiliary external boundary is present in all direct formulations. However, in the proposed method, by evaluating the auxiliary external boundary asymptotically at infinity, the resulting integral equation avoids direct surface integration of the auxiliary external boundary, whilst maintaining solvability of the internal and external domains. Using explicit interface tracking, the velocity of the interface can be accurately computed allowing high capillary number flows to be explored. The computational cost of the convergent series scales with the square of the number of boundary elements (quadratic scaling), meaning the long term effects of finger interaction can be examined more efficiently than previous direct numerical approaches that exhibit cubic scaling \cite{hadavinia1995}. The resulting numerical method allows the effective modelling of a moving interface in a Hele-Shaw cell, using a physically realistic mobility ratio. 

In this paper, the mathematical model is first presented, followed by the boundary element numerical method. Numerical performance and validation studies of the numerical scheme are then performed. After validation of the numerical method, results for varying mobility ratio and capillary number are shown, concluding in simulations focusing on longer term interface evolution to showcase the new finite mobility ratio approach.


\section{Mathematical Formulation}

\begin{figure}[H]
\centering
\adjincludegraphics[scale=0.5]{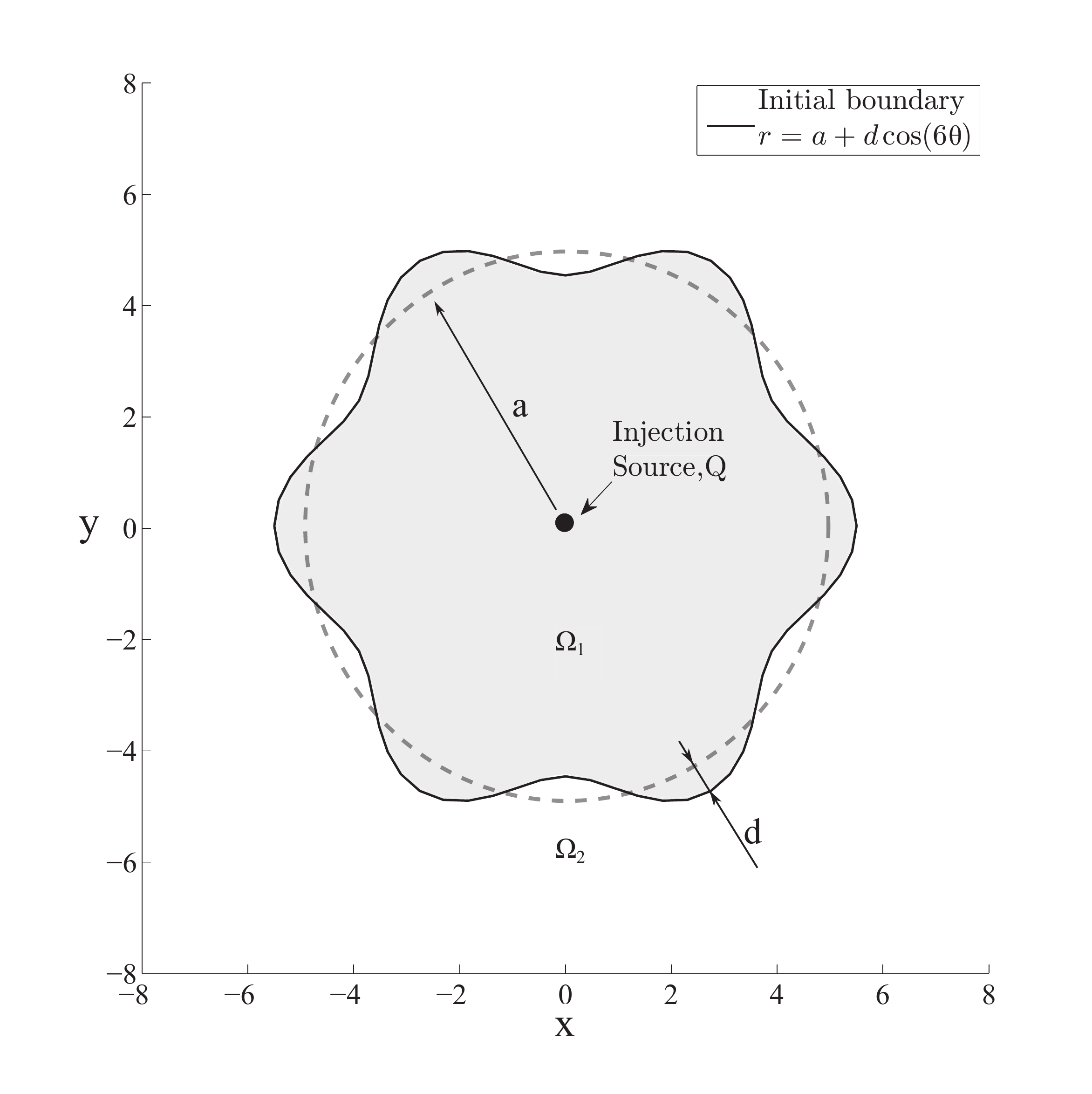}
\centering
\vspace{-20pt}
\caption{Graphical representation of the initial bubble configuration and flow domains.}
\label{perturbed}
\end{figure}

We consider a circular Hele-Shaw cell of infinite radius, in which high viscosity fluid is displaced by the injection of a less viscous fluid. The low viscosity invading fluid (such as $CO_2$) occupies region $\Omega_1$ whilst a high viscosity fluid (such as brine) occupies the external region, $\Omega_2$, shown in figure (\ref{perturbed}).

A perturbation term, given by $d \: cos(6\theta)$, is added to the initial unperturbed radius, $a$, in figure (\ref{perturbed}) to initiate instability at the interface. A symmetric perturbation of amplitude $d$ is used, so that fingering effects can be seen at several locations on the bubble. An asymmetric perturbation is used later in section \ref{long_time}, to mimic naturally occurring noise and disturbance within the system. 

To formulate the mathematical model, we introduce several non-dimensional variables. Utilising the characteristic length, time, velocity and pressure of the problem, the field variables can be represented in non-dimensional form: 

\begin{align}
 \label{non_dim1} \left( x,y,r \right)& = a \left(x',y',r' \right) \\
 \label{non_dim2}  t &= \frac{a^2}{Q} t'  \\
 \label{non_dim3} u_i &= \frac{Q}{a} u'_i    \;\;\; i = 1,2 \\
 \label{non_dim4} \left( P,\phi \right)_i &= \frac{Q}{M_2} \left(P',\phi' \right)_i   \;\;\; i = 1,2
\end{align}

In equalities (\ref{non_dim1}) - (\ref{non_dim4}), apostrophes identify non-dimensional variables. $t, u, P$ and $\phi$ represent time, two-dimensional velocity, depth averaged pressure and perturbation pressure respectively. The parameters $a, Q$ and $M_2$ are the unperturbed bubble radius, the radial injection flux and the mobility of the displaced fluid respectively. The fluid mobility in region $i$ is related to the Hele-Shaw plate separation, b, and the fluid viscosity, $\mu_i$ by: 
\begin{align}
 \label{eqn3} M_i = \frac{b^2}{12\mu_i} 
\end{align}

The $b^2/12$ term in the mobility ratio refers to the intrinsic permeability of the Hele-Shaw cell, defined by the plate separation. For the flow between two thinly separated plates in a Hele-Shaw cell, the depth averaged pressure and two dimensional velocity in each fluid region can be expressed through Darcy's law:
\begin{align}
 \label{eqn1} u_1' &= -\beta \nabla P_1'\\ 
 \label{eqn2} u_2' &= - \nabla P_2'\\ 
 \label{eqn3} &\nabla \cdot {u'} = 0 
\end{align}
In equation (\ref{eqn1}), $\beta$ is the ratio of mobilities between the two fluids:

\begin{align}
 \label{beta_eqn} \beta = \frac{M_1}{M_2}
\end{align}

For $\beta$ values greater than 1, the inner fluid is less viscous than the external fluid. In infinite mobility ratio models $\beta = \infty$. With a constant viscosity in each of the two fluid regions, equations (\ref{eqn1}) - (\ref{eqn3}) can be reduced to Laplace's equation. From this point on, the apostrophe of all dimensionless variables will be dropped for clarity, and every variable will be assumed to be in its non-dimensional form, unless otherwise stated. 

\begin{align}
 \label{eqn4} \nabla ^2 P_i (x) = 0 \;\;\; for\; every\; x \in \Omega_i , \;\;\; i = 1,2
\end{align}

To form a boundary integral equation, the pressure field can be represented as a sum of the pressures due to an injection potential source, $Q$, and a perturbation term, $\phi_i$:

\begin{align}
 \label{eqn5.0} P_1(x) = \phi_1 - \frac{1}{2 \pi \beta} \ln (r) \\
 \label{eqn5.1} P_2(x) = \phi_2 - \frac{1}{2 \pi} \ln (r)
\end{align}
In equations (\ref{eqn5.0}) - (\ref{eqn5.1}), $r$ is the non-dimensional radial distance from the collocation point, $x$, to the source point located inside the injected bubble. At a boundary point, $\xi$ on the fluid interface, S, between $\Omega_1$ and $\Omega_2$, there are two matching conditions that must be met by the advancing interface. Firstly, continuity of normal perturbation fluxes:

\begin{align}
 \label{eqn6} q = \beta \frac{\partial \phi_1}{\partial n} =  \frac{\partial \phi_2}{\partial n}
\end{align}
Secondly, the pressure jump across the interface due to the surface tension, $\gamma$:

\begin{align}
 \label{eqn7} \phi_1 - \phi_2 = \frac{1}{Ca} \left(\frac{2a}{b} + k(\xi) \right) - \left(\frac{\beta - 1}{2 \pi \beta}\right) \ln (r)  = \left(1+ \beta \right) f
\end{align}
Where:
\begin{align}
 \label{fe} f = \frac{1}{Ca(1 + \beta)} \left(\frac{2a}{b} + k(\xi) \right) - \frac{1}{1 + \beta} \left(\frac{\beta - 1}{2 \pi \beta}\right) \ln (r)
\end{align}
In equations (\ref{eqn7}) and (\ref{fe}), we have introduced the capillary number, $Ca$, which describes the ratio of viscous driving forces to surface tension forces. Classically, a modified capillary number can be used to completely describe infinite mobility ratio rectilinear Hele-Shaw flow \cite{homsy1987}. Due to the the radial setup of the Hele-Shaw injection, this modified capillary number must be adapted to adequately describe the flow regime. The capillary number produced from the dimensional analysis of the radial Hele-Shaw flow above and that presented by \cite{homsy1987} are shown below, with the classical rectilinear version shown with an apostrophe. 

\begin{align}
 \label{ca} Ca =  \frac{ 12 \mu_2 Q}{\gamma a} \left(\frac{a}{b}\right)^2   = \frac{ aQ}{\gamma M_2}   \:\: \: \: \: \: \: \: \:   Ca' = \frac{12  \mu_2 V}{\gamma} \left(\frac{L}{b}\right)^2    
\end{align}
In the modified capillary number for rectilinear flow, the half Hele-Shaw cell width, L, is used as the macroscopic length scale, with the cell plate separation, b, used as the microscopic length scale. For radial flow, we have chosen to use the initial unperturbed bubble radius, a as the macroscopic length scale as there is no characteristic cell width in the fully circular domain. The initial source injection velocity, $Q/a$ is chosen as the characteristic velocity of the problem. The capillary number presented here for radial Hele-Shaw flow, and that presented by \cite{homsy1987} are equivalent, with a difference only in the macroscopic length scale of the problem and the characteristic velocity. 

The $a/b$ scaling term in equation (\ref{ca}) relates the initial unperturbed bubble radius to the Hele-Shaw plate separation, which modifies the physical capillary number to include the effective permeability of the cell. The capillary number together with the mobility ratio uniquely describe radial Hele-Shaw flow, and as such are the main parameters used to analyse and describe different flow regimes. 

In equation (\ref{eqn7}), the contact angle of the meniscus has been assumed to be zero. The signed curvature, $k$ is considered a continuous function on the interface surface, $\eta = \eta (x,y)$ given by equation (\ref{eqn8}) below.
\begin{align}
 \label{eqn8} k = a \frac{\left( {\eta_x \eta_{yy} - \eta_y \eta_{xx}}\right)} {\left[ \left(\eta_x \right)^2 + \left(\eta_y \right)^2 \right]^\frac{3}{2}}
\end{align}

This mathematical model forms the basis for the B-Spline numerical discretisation and hypersingular integral treatment, presented in the following sections.  

\section{Numerical Method}

\subsection{B-Spline Representation}

Following from the pressure field representation in Equation (\ref{eqn5.0}) and (\ref{eqn5.1}), the perturbed pressures, $\phi_1$ and $\phi_2$ can be expressed in terms of their corresponding Green's formulae at the fluid interface \cite{jawson1977}, using the two dimensional fundamental solution, $\phi^*$.

\begin{align}
\begin{split}
 \label{eqn9} &\int_{S\infty} \phi^* (\xi,y)\frac{\partial \phi_2 (y)}{\partial n_y} dS_{y} - \int_{S\infty} \phi_2 (y)\frac{\partial \phi^* (\xi,y)}{\partial n_y} dS_{y}  \\
 + &\int_s \phi_2 (y) \frac{\partial \phi^* (\xi,y)}{\partial n_y} dS_y
  - \int_s \phi^*(\xi,y) \frac{\partial \phi_2(y)}{\partial n_y}  dS_y = \frac{1}{2} \phi_2 (\xi)\\
  \end{split}
\end{align}

\begin{align}
\label{eqn10} \int_s \phi_1 (y) \frac{\partial \phi^* (\xi,y)}{\partial n_y} dS_y -  \int_s \frac{\partial \phi_1(y)}{\partial n_y} \phi^*(\xi,y) dS_y = - \frac{1}{2} \phi_1 (\xi) 
\end{align}

The difference in sign between the two equations is due to the direction of the outward facing normal, $\vec{n}$. The continuity and discontinuity properties of the single-layer and double-layer potential are used to evaluate the integrals across the curve, S. For the external problem in equation (\ref{eqn9}), both the internal boundary at the interface and the auxiliary external boundary at infinity must be considered. The external boundary can be evaluated at a fixed location in the far field and treated as a regular surface integral, introducing extra computation and constraining the interior fluid to the region inside the external boundary \cite{hadavinia1995}. The evaluation of the external surface can be removed by utilising a Green's function with periodic boundary conditions \cite{hansen1999}.

The external boundary at infinity can also be evaluated asymptotically, considering the perturbation flux to approach zero as the radial distance from the source approaches infinity. This allows the fluid domain to extend to infinity, so the evolution of the inner fluid interface can continue unrestricted, without having to re-scale an exterior bounding surface. Batchelor \cite{batchelor1967} has shown that asymptotic evaluation introduces a constant into the equation, replacing the external boundary surface integral:
\begin{align}
 \label{new} \frac{k_{as}}{2 \pi} + \int_s \phi_2 (y) \frac{\partial \phi^* (\xi,y)}{\partial n_y} dS_y - \int_s \frac{\partial \phi_2(\xi,y)}{\partial n_y} \phi^*(\xi,y) dS_y = \frac{1}{2} \phi_2 (\xi)
\end{align}

The constant $k_{as}$ becomes an unknown variable to be found, which along with the zero-flux condition of the perturbation pressure across the interface ensures the solvability of the exterior problem:
\begin{align}
 \label{inte66} \int_s q(\xi) dS_{\xi} = 0
\end{align}

Equations (\ref{new}) and (\ref{inte66}) represent an infinite mobility ratio model for the exterior problem, considering solely the displaced fluid in region two. This model has been implemented by \cite{power1994} and more recently by \cite{power2013} including a dissolution velocity at the interface. To combine the interior and exterior boundary integral equations to produce a finite mobility ratio model applicable in both domains, the limiting value of the normal derivatives of equations (\ref{eqn10}) and (\ref{new}) must be taken. 

\begin{align}
 \label{new1} \int_s \phi_2 (y) \frac{\partial^2 \phi^* (\xi,y)}{\partial n_{\xi} \partial n_y} dS_y - \int_s \frac{\partial \phi_2(y)}{\partial n_y} \frac{\partial \phi^*(\xi,y)}{\partial n_{\xi}} dS_y = \frac{1}{2} \frac{\partial \phi_2 (\xi)}{\partial n_{\xi}}
\end{align}

\begin{align}
 \label{new2} \int_s \phi_1 (y) \frac{\partial^2 \phi^* (\xi,y)}{\partial n_{\xi} \partial n_y} dS_y - \int_s \frac{\partial \phi_1(y)}{\partial n_y} \frac{\partial \phi^*(\xi,y)}{\partial n_{\xi}} dS_y = - \frac{1}{2} \frac{\partial \phi_1 (\xi)}{\partial n_{\xi}}
\end{align}
Subtracting the above two equations and using the matching conditions (\ref{eqn6}) and (\ref{eqn7}), the following second kind Fredholm integral equation can be formed \cite{power1995}.

\begin{align}
 \label{eqn11} -\frac{1}{2}q(\xi) + \left(\frac{1 - \beta }{\beta + 1}\right) \int_s K(y,\xi)q(y)dS_y = g(\xi)
\end{align}

The regular kernel, $K(y,\xi)$ in equation (\ref{eqn11}) is given by:

\begin{align}
 \label{eqn12} K(y,\xi) = \frac{1}{2 \pi} \frac{\partial}{\partial n_{\xi}} \left(\ln \frac{1}{R(\xi,y)}\right) = \frac{1}{2 \pi} \frac{y_j - \xi_j}{R^2} n_j(\xi)
\end{align}
Noting from \cite{power1995} that,
\begin{align}
 \label{eqn13} \lim_{y \to \xi} K(y,\xi) = - \frac{ k(\xi)}{2}
\end{align}
The $R$ term in the regular Kernal is the non-dimensional absolute distance from the collocation point ($\xi$) to the point of integration on the surface ($y$). The non-homogeneous boundary term, $g(\xi)$, is given by the following hypersingular integral.

\begin{align}
 \label{eqn14} g(\xi) = \frac{\beta}{2 \pi} \int_s f(y) \frac{\partial^2}{\partial n_{\xi} \partial n_y} \left( \ln \frac{1}{R(\xi,y)}\right) dS_y
\end{align}
The second kind Fredholm equation in (\ref{eqn11}) permits an analytical Neumann series solution, owing to the fact that it is the adjoint of the corresponding indirect equation, which has been proven to have an analytical Neumann series solution (for more details see \cite{power1995}). Before the solution technique for equation (\ref{eqn11}) is given, it is worth noting that by combining the two integral equations for the different fluid domains, (\ref{new1}) and (\ref{new2}), we do not need to evaluate the constant obtained from the asymptotic evaluation of the surface integral at infinity in equation (\ref{new}). To show that the no-flux condition of the perturbation pressure across the interface is still met by equation (\ref{eqn11}), without the need to explicitly include it in the equation, we first integrate over the interface surface:

\begin{align}
 \label{inte} \int_s -\frac{1}{2}q(\xi) dS_{\xi} + \left(\frac{1 - \beta }{\beta + 1}\right) \int_s q (y) \int_s K(y,\xi)dS_{\xi} dS_{y} = \int_s g(\xi) dS_{\xi}
\end{align}
The integral of the kernel $K(y,\xi)$, over the surface has a value of $1/2$ meaning equation (\ref{inte}) can be simplified to:
\begin{align}
 \label{inte2} \frac{\beta}{\beta+1} \int_s q(\xi) dS_{\xi} = \int_s g(\xi) dS_{\xi}
\end{align}
The right hand side of equation (\ref{inte2}), can be written as:
\begin{align}
\begin{split}
 \label{inte3} \int_s g(\xi) dS_{\xi} &= \frac{\beta}{2 \pi} \int_s f(y) \int_s \frac{\partial^2}{\partial n_{\xi} \partial n_y} \left( \ln \frac{1}{R(\xi,y)}\right) dS_{\xi} dS_{y} \\
  &= \frac{\beta}{2 \pi} \int_s f(y) \int_s \frac{\partial}{\partial n_{\xi}} K(\xi,y) dS_{\xi} dS_{y} = 0
  \end{split}
\end{align}
Since, 

\begin{align}
 \label{inte5} \int_s \frac{\partial}{\partial n_{\xi}} K(\xi,y) dS_{\xi} = 0
\end{align}
Using the above expressions and equation (\ref{inte2}), it follows that the no-flux condition of the perturbed pressure at the interface has been met:
\begin{align}
 \label{inte6} \int_s q(\xi) dS_{\xi} = 0
\end{align}
The above analysis shows that the no-flux condition at the interface is met by equation (\ref{eqn11}). By combining the integral equations for each fluid domain into one single equation, the need to explicitly evaluate the no-flux condition at the interface has been avoided, at the expense of introducing a hypersingular integral. 

Equation (\ref{eqn11}) can be solved using a convergent series for $q$, as long as $0 \leq \beta < \infty$ \cite{power1995}. Using infinitely many terms results in an analytical Neumann convergent series solution. The series can be simplified by taking $ \lambda = \frac{\left(1 - \beta \right)}{\left(\beta + 1 \right)}$ and using a discrete number of terms, $m$, to truncate the solution of equation (\ref{eqn11}).

\begin{align}
 \label{eqn15} q(\xi) = q_0(\xi) + \lambda q_1(\xi) + \dots + \lambda^m q_m (\xi)
\end{align}
The terms in equation (\ref{eqn15}) can be calculated recursively, via the following formulae:
\begin{align}
 \label{neu_first} q_0(\xi) &= -2g(\xi) \\
 \label{neu_m} q_m(\xi) &= 2 \int_s K(\xi,y) q_{m-1}(y) dS_y \:\:\: for \:\:\: m \neq 0
\end{align}
The movement of the fluid-fluid interface is then calculated via a forward Euler time stepping approach, where $\Delta L_n(\xi)$ represents the dimensionless distance moved by a boundary point in a single time-step:

\begin{align}
 \label{eqn16} \frac{\Delta L_n(\xi)}{\Delta t} = q(\xi) + \frac{x_i(\xi) n_i(\xi)}{2 \pi r^2} 
\end{align}
As the surface grows according to equation (\ref{eqn16}), the number of nodal points on the bubble boundary are adaptively increased to maintain a target element size. The surface integrals in the boundary integral equations are solved by discretising the boundary using uniform cubic B-Spline boundary elements, first shown in \cite{cabral1990}. Every variable that requires interpolation along the boundary of the bubble can be represented by a B-Spline, using the following equations \cite{cabral1990}.

\begin{align}
 \label{eqn17} D_i(t) = E_0 (t) C_{i-1} + E_1 (t) C_{i} +  E_2 (t) C_{i+1} + E_3 (t) C_{i+2}
\end{align}
In equation (\ref{eqn17}), $D_i$ represents a continuous scalar field. $C_i$ represents the control points over the boundary for that particular variable. The blending functions, $E$ over each element of length $ 0 \le t \le 1 $ are given by:

\begin{equation}\label{eqn18}
  \begin{split}
    E_0 (t) &= \frac{-t^3}{6} + \frac{t^2}{2} - \frac{t}{2} + \frac{1}{6} \\
    E_1 (t) &= \frac{t^3}{2} - t^2 + \frac{2}{3} \\
    E_2 (t) &= \frac{-t^3}{2} +  \frac{t^2}{2} +\frac{t}{2} + \frac{1}{6} \\
    E_3 (t) &= \frac{t^3}{6} \\
  \end{split}
\end{equation}
Equation (\ref{eqn17}) can be formed for each nodal location along the boundary, producing a system of equations that may be solved to find the control points for each element. To close the curve of the control points, the following conditions must be met:

\begin{equation}\label{eqn19}
  \begin{split}
   C_0 &= C_N \\
    C_{N+1} &= C_1 \\
  \end{split}
\end{equation}
The system of equations can then be solved to find the control points for the variable in question. This has to be done at each time step to find the new $x$ and $y$ control points, given the new interface that has been calculated. The resulting cyclic tri-diagonal system is efficiently solved utilising the Sherman-Morrison formula \cite{press1996}.

Along with the B-Spline representation of the curve, a 4th order Lagrangian polynomial is used to compute the local curvature, $k(\xi)$ at a nodal point. A Lagrangian polynomial accurately represents second derivatives at nodal points, which are second order accurate with a B-Spline representation. The Lagrangian polynomial is adaptively fitted to the B-spline curve, reconstructing locally using the surrounding nodal points. A non-uniform polynomial was tested, but since the locations must be reconstructed using the B-Spline rather than the raw nodal positions, the accuracy was only as good as the B-spline itself.

Table \ref{curvature_approx} shows the 4th order Lagrangian polynomial and cubic B-Spline schemes used to calculate the curvature of a test function, $y = \frac{1}{2} + \frac{3}{10} sin \left(2 \pi x\right)$, which was also presented in \cite{guan2003} and \cite{chorin1985} with a 20x20 grid. There are N elements used to approximate the function, with N+1 nodal points. The Lagrangian polynomial has a much better approximation to the curvature than the cubic B-Spline scheme, which when centred on the nodal points is second order accurate. The $L_1$ error norm between the 4th order Lagrangian polynomial and the analytical solution using 20 elements is 0.003\%, with the numerical points being indistinguishable from the analytical function when viewed graphically. This is a much better approximation than the schemes used in \cite{guan2003} and \cite{chorin1985}, in which the numerical approximation shows noticeable discrepancy from the analytical function graphically.

\begin{table}[h]
\begin{tabular}{|l|l|l|l|l|l|}
\hline
\multicolumn{2}{|l|}{Number of elements, N}                                                                               & 20       & 40       & 60       & 80       \\ \hline
\multirow{2}{*}{$L_1$ Error Norm}                                                        & B-Spline             & 1.18E-03 & 3.00E-04 & 1.34E-04 & 7.58E-05 \\ \cline{2-6} 
                                                                               & 4th Order Lagrangian & 3.03E-05 & 1.80E-06 & 3.61E-07 & 1.15E-07 \\ \hline
\multirow{2}{*}{$L_2$ Error Norm}                                                        & B-Spline             & 1.76E-03 & 4.42E-04 & 1.97E-04 & 1.11E-04 \\ \cline{2-6} 
                                                                               & 4th Order Lagrangian & 3.58E-05 & 2.18E-06 & 4.32E-07 & 1.37E-07 \\ \hline
\multirow{2}{*}{\begin{tabular}[c]{@{}l@{}}$L_2$ Error Norm\\ Convergence rate\end{tabular}} & B-Spline             & --        & 1.98     & 1.98     & 1.99     \\ \cline{2-6} 
                                                                               & 4th Order Lagrangian & --        & 4.07     & 3.97     & 3.97     \\ \hline
\end{tabular}
\vspace{10pt}
\caption{$L_1$ and $L_2$ error norms of the 4th order Lagrangian polynomial and B-Spline approximations to the curvature of $y = \frac{1}{2} + \frac{3}{10} sin \left(2 \pi x\right)$.}
\label{curvature_approx}
\end{table}

%
%
After discretising the boundary into B-Splines, and effectively evaluating the curvature, the hypersingular integral in equation (\ref{eqn14}) must be handled numerically. In \cite{power1995}, the surface integral could be simplified by using constant elements. This allowed the hypersingular integral to be equated to the integral over the remaining elements making up the surface. However, in the present scheme using non-linear B-Spline elements, the hypersingular integration must be handled explicitly. B-spline boundary elements are favoured over constant, linear or quadratic elements, due to their accuracy when approximating highly curved surfaces, which are typical of the ramified patterns seen in viscous fingering phenomena. 

\subsection{Hypersingular Integral Treatment}

\par The hypersingular integral in (\ref{eqn14}) is treated using a semi-analytical approach implemented in \cite{gui1998}, first proposed by Mikhlin in 1957 {\cite{mikhlin1957}. The formulation and limiting process will not be examined here; only the final hypersingular integral formula will be presented. The hypersingular integral in equation (\ref{eqn14}) becomes an issue when the field points of integration, $y$ lie close to a collocation point, $\xi$. This is most significant on elements that coincide with the collocation point, shown in figure (\ref{hyp_point}). The hypersingular integral must be evaluated in the sense of Hadamard finite parts in order to guarantee its existence over the two elements where the hypersingular point coincides \cite{hadamard1952}. 
\begin{figure}
\begin{center}
\includegraphics[width=0.6\textwidth]{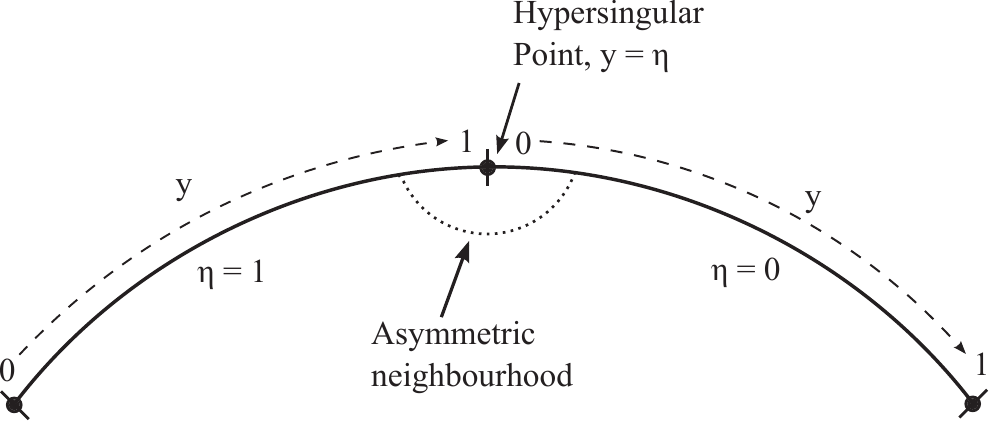}
\end{center}
\caption{Collocation point ($\xi $) on the same elements as the quadrature points (y) creating a hypersingular point ($\eta$) and corresponding hypersingular integral over these elements.}
\label{hyp_point}
\end{figure}
To effectively evaluate the hypersingular integration, firstly, equation (\ref{eqn14}) may be re-written in more convenient notation:
\begin{align}
 \label{eqn20} g(\xi) = \frac{\beta}{2 \pi} \int_s V_i(\xi,y) N^a(\xi) J_m d\xi 
\end{align}
Where,

\begin{align}
 \label{eqn21} V_i(\xi,y) = \frac{\partial^2}{\partial n_{\xi} \partial n_y} \left( \ln \frac{1}{R(\xi,y)}\right) = \frac{1}{2 \pi R^2} \left(-2 \frac{\partial R}{\partial \xi_i} \frac{\partial R}{\partial n} + n_i(y)\right)
\end{align}
The hypersingular kernel can be expanded in terms of a Laurent power series about a hypersingular point, $\eta$ \cite{gui1998}.

\begin{align}
 \label{eqn22} V_i(\xi,y)N^a(\xi)J_m = F_i(\eta,y) = \frac{F_{-2} (\eta)}{\left(y- \eta \right)^2} + \frac{F_{-1} (\eta)}{y-\eta} + O(1)
\end{align}
The $F_{-2}$ and $F_{-1}$ terms depend only on the derivatives of the B-spline shape functions, $N^a$. By introducing the above power series into the 
hypersingular boundary integral equation (\ref{eqn14}), the limits may be evaluated analytically in order to remove unbounded terms. This results in a regular integral and analytical expression.

\begin{equation}\label{eqn23}
  \begin{split}
  g(\xi) &= \frac{\beta}{2 \pi}  \sum \limits_{m=1}^2 \Biggl( \int_0^1 \left[F^m(\eta,y) - \left(\frac{F^m_{-2}(\eta)}{\left(y-\eta   \right)^2} + \frac{F^m_{-1}(\eta)}{y-\eta}  \right)\right]dy \\ 
  &+ F^m_{-1}(\eta) \ln \left| \frac{1}{\beta_m(\eta)} \right| sgn(y-\eta) - F^m_{-2}(\eta)  \left[sgn(y-\eta) \frac{\gamma_m(\eta)}{\beta^2_m (\eta)} + 1 \right]  \Biggr)
  \end{split}
\end{equation}
In equation (\ref{eqn23}), the $ \beta_m$ and $\gamma_m$ terms account for any possible distortion from an asymmetric neighbourhood around the hypersingular point \cite{gui1998}. In figure (\ref{hyp_graph}) the hypersingular function and the subtraction terms have been evaluated over two B-Spline elements. The hypersingular function, $F^m$ and the power series terms tend towards infinity when nearing the hypersingular point at $ y- \eta = 0$. However, the regular function produced by subtracting the power series terms from the hypersingular function is finite and regular at all points in the domain, meaning it can be integrated using standard Gaussian quadrature techniques.
\begin{figure}
\centering
\adjincludegraphics[scale=0.55]{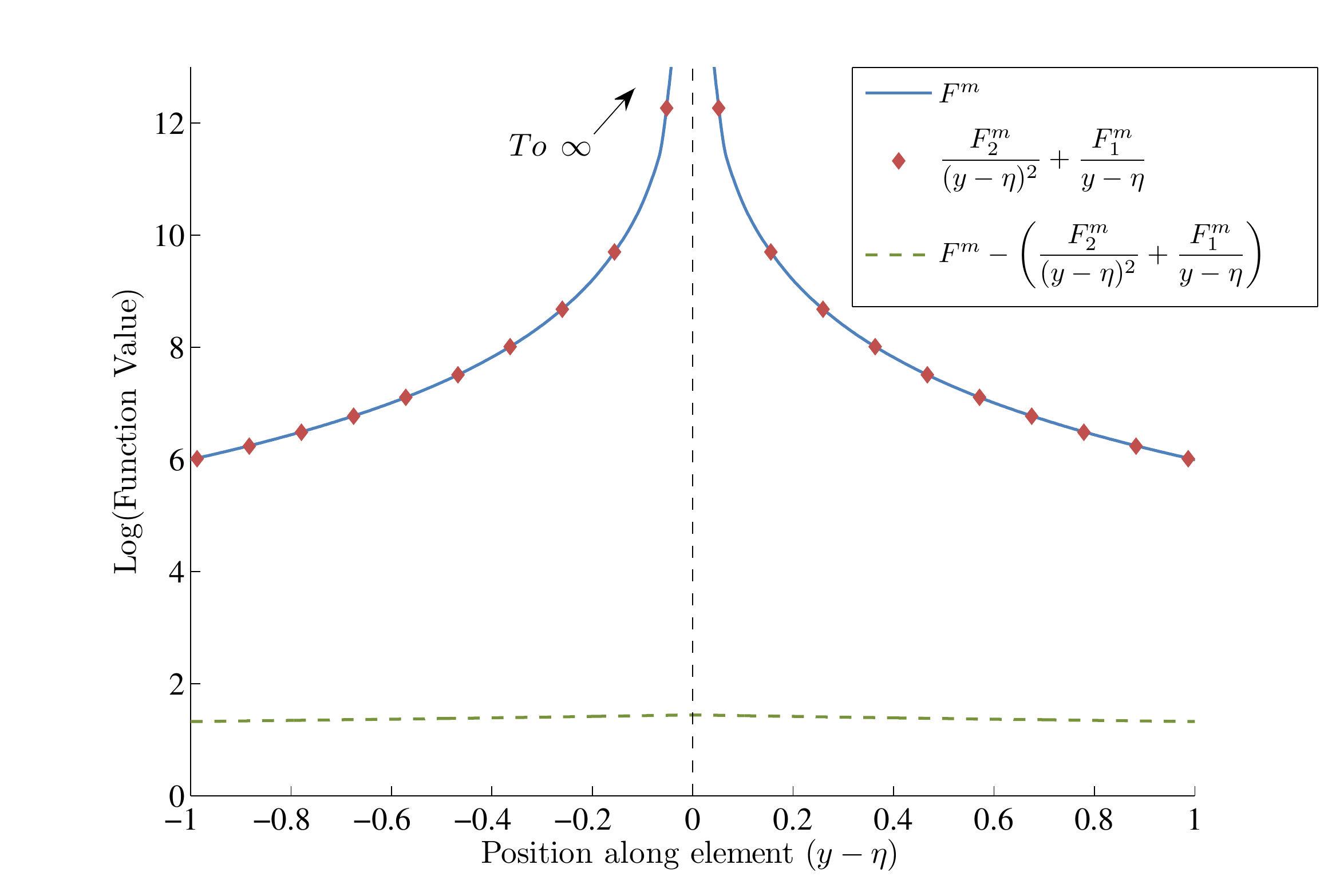}
\vspace{-20pt}
\caption{Hypersingular integrand, Power series and resulting regular function over two elements coinciding with the collocation point.}
\label{hyp_graph}
\end{figure}

Due to the lack of finite mobility ratio boundary element schemes using a low mobility ratio, a classical viscous fingering example is presented to allow comparison with the new finite mobility ratio model. In the classical fingering case presented in \cite{zhao1995}, \cite{power1995} and \cite{power2013}, air displaces oil in a fully circular Hele-Shaw cell, with a capillary number of 2000. 


To compare the single-phase solutions in \cite{zhao1995}, \cite{power1995} and \cite{power2013}, with the finite mobility ratio method developed here, the mobility ratio of the two fluids is varied between 1 and 1000. By varying the mobility ratio, the viscosity of the injected fluid is changed whilst keeping the resident oil viscosity the same. The different mobility ratio cases are compared with an infinite mobility ratio solution (created using equations (\ref{new}) and (\ref{inte66})) in figure (\ref{power_comparison}). 

\begin{figure}[h]
\centering
\adjincludegraphics[scale=0.55]{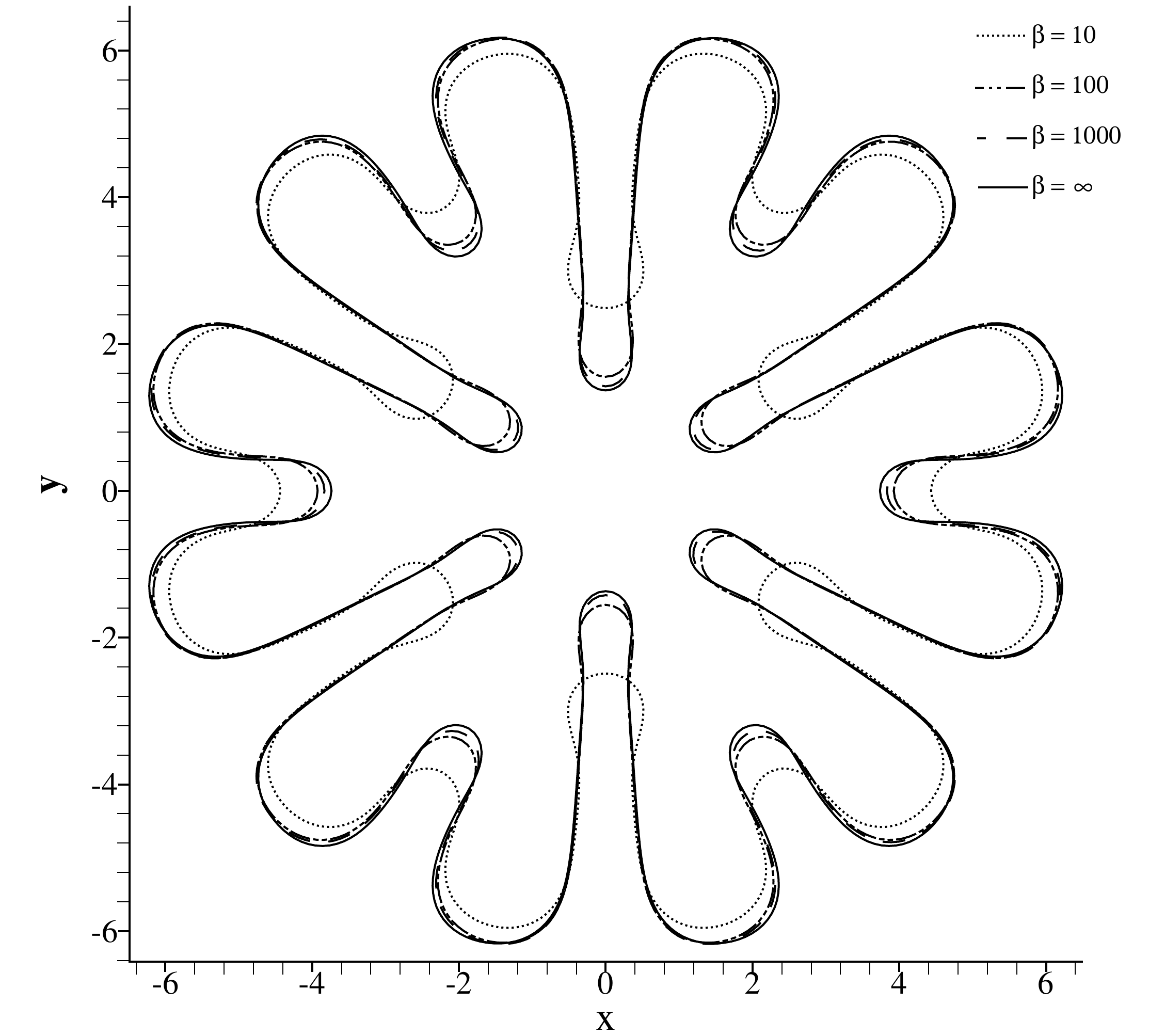}
\centering
\vspace{-20pt}
\caption{Interface of an infinite mobility ratio gas bubble at t= 80, showing the classical fingering problem presented in \cite{zhao1995}, \cite{power1995} and \cite{power2013}, and the interfaces of finite mobility bubbles utilising the new finite mobility ratio model.}
\label{power_comparison}
\end{figure}

In figure (\ref{power_comparison}) the finite mobility ratio solutions tend to the infinite mobility ratio solution when the mobility ratio is increased. The base and front of the fingers for the $\beta = 1000$ case agree very well with the $\beta = \infty$ case, with an $L_1$ error of 0.9\%. The case of $\beta = 10$ has a somewhat different shape to the infinite case, with the finger base extending much further into the liquid domain. This is due to the significant velocity possessed by the inner fluid, working to push the bases out. This process is explored more in section \ref{mob_effects}, where mobility ratio effects are examined.

\subsection{Numerical Performance}
Here we examine the numerical performance of the finite mobility ratio model using various small scale simulations. From equations (\ref{eqn15}) - (\ref{neu_m}), it can be seen that the computational cost scales with $pN^2$, where $p$ is the number of terms in the convergent series, and $N$ is the number of boundary elements. Therefore, for a fixed $p$ the scheme will exhibit close to second order scaling. This is much better than direct solvers for the corresponding matrix system (typically LU decomposition), which exhibit cubic scaling.  

During the early stages of interfacial evolution in which the number of boundary elements is low (typically $<$1000), a direct LU solver can outperform the convergent series. However, as the size of the dataset grows, the convergent series will eventually run faster than a direct LU solver due to the second order scaling. 


The number of terms used in the convergent series plays an important role in the accuracy and speed of solution. Figure (\ref{picard}) shows that as the mobility ratio of the two fluids becomes larger, the number of terms required by the convergent series to reach a desired error increases. This is because the value of $\lambda$ approaches -1, and successive terms in the convergent series do not decay as rapidly. When $\lambda = -1$ there is no unique solution to equation (\ref{eqn11}), due to a singular value in the corresponding spectrum of the integral operator. For most simulations, a residual error of $10^{-6}$ provides an acceptable convergence level. This typically gives an L$_1$ error between the interface positions obtained from the series solver and direct matrix solver of less than 0.05\%.

When the mobility ratio increases beyond 100, the number of terms required in the convergent series to get a low residual error becomes much larger ($>500$) than the number of starting elements used ($\sim100$). This means that for high mobility ratios, the convergent series solution can be significantly slower than a corresponding direct matrix solver, showing that this method is much more applicable for the solution of low mobility ratio flows. 

\begin{figure}
\centering
\adjincludegraphics[scale=0.65]{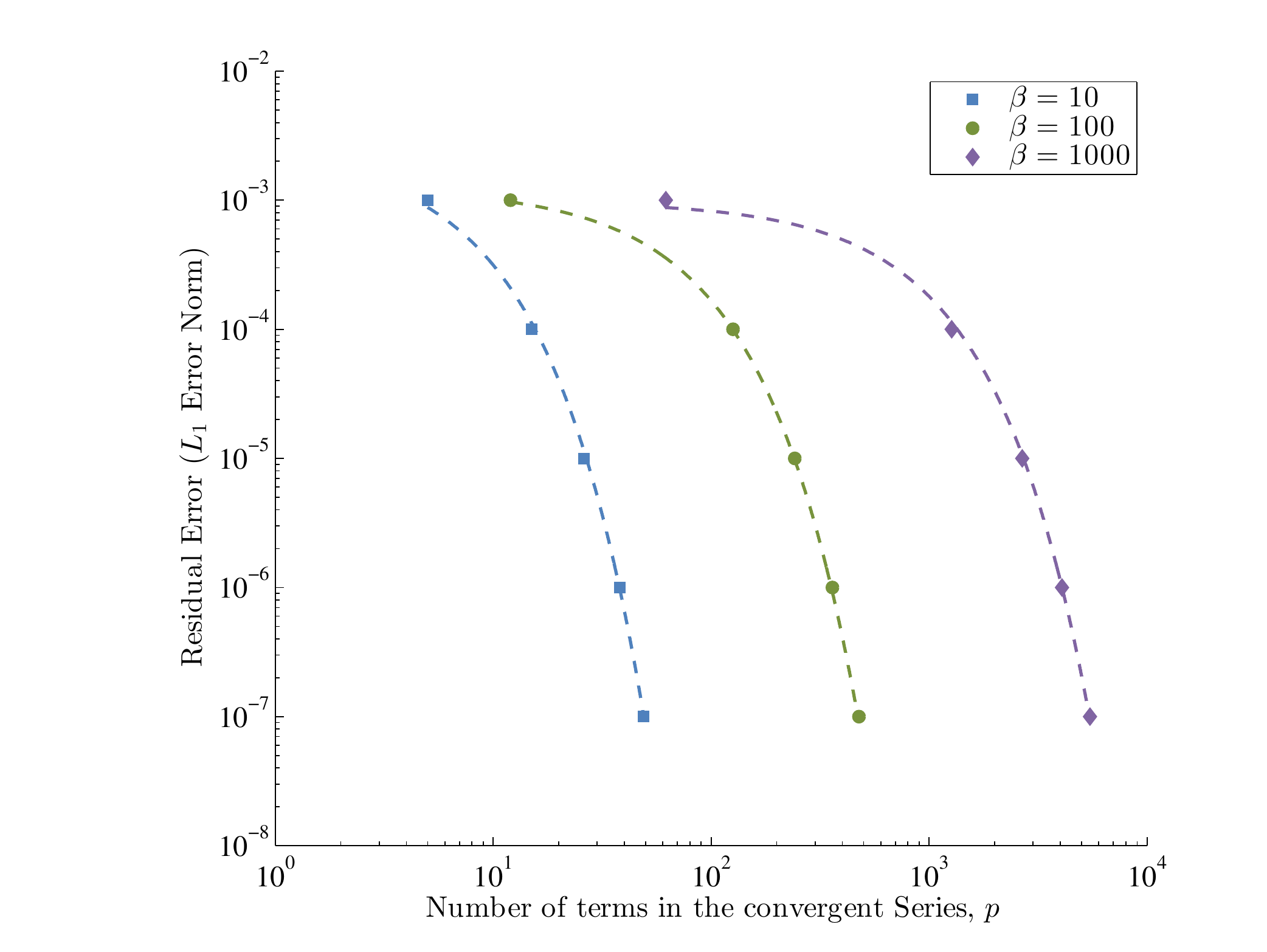}
\vspace{-20pt}
\caption{Residual error convergence with number of terms in the convergent series and varying mobility ratio.}
\label{picard}
\end{figure}

The convergence of the series solution is determined by the residual error between successive terms and is therefore largely independent from the number of boundary elements. Figure (\ref{pic_var}) shows the number of series terms ($p$) needed to achieve a L$_1$ residual error of 1x10$^{-6}$ as the number of boundary elements, N, increases during the evolution of the interface in the $\beta = $ 10 case in figure (\ref{power_comparison}). Multiple points at the same value of N correspond to outputs from different time steps in the simulation that used the same number of boundary elements in the interface profile. As the interface grows the number of elements are adaptively increased to maintain a target size. However, as the growth rate is so small per time step, the same number of elements can be used for several steps whilst still maintaining an element size under the target maximum size, resulting in several points with the same value of N. It can be seen that the number of terms needed in the convergent series varies between 30 and 50 as the number of elements is increased from 100-2000. The number of terms in the series is generally much lower than the number of boundary elements required to accurately compute the solution. 


%
\begin{figure}
\centering
\adjincludegraphics[scale=0.6]{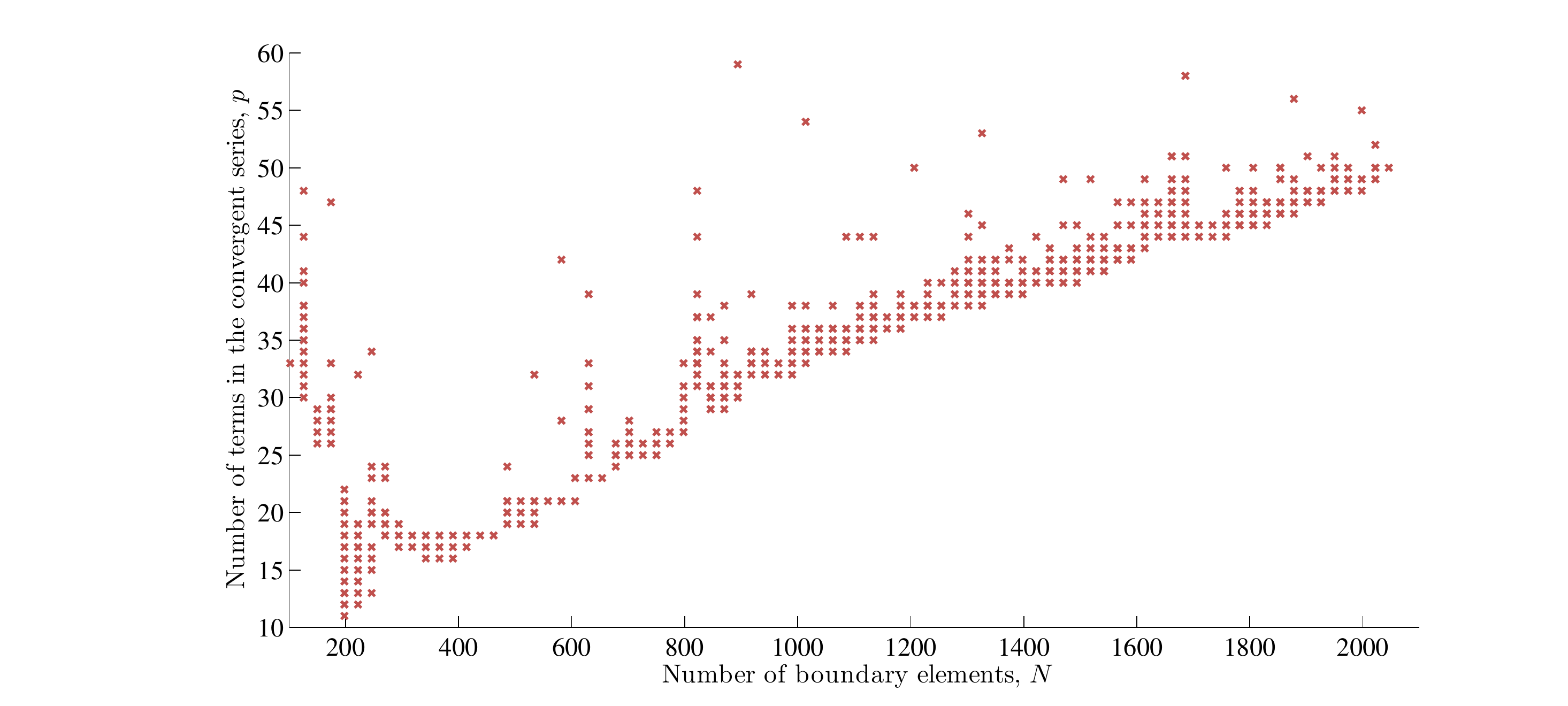}
\vspace{-20pt}
\caption{Number of terms ($p$) needed in the convergent series to produce an L$_1$ residual error of 1x10$^{-6}$, with varying numbers of boundary elements.}
\label{pic_var}
\end{figure}

Due to the small number of terms required in low mobility cases, this method is particularly well-suited to solving low mobility ratio regimes, compared to previous direct matrix solutions. As the number of terms in the series solution is typically several orders of magnitude less than the number of boundary elements, long time evolutions can be analysed quickly in comparison to direct matrix solver approaches and as such are one of the main focuses of this work.

\subsection{Numerical Stability and Convergence Analysis}
\label{numerical_Stab_section}

Along with the number of terms in the convergent series, the number of boundary elements, and time step size also affect the accuracy of the resulting solution. To investigate mesh and time independence, the model was tested under various capillary number regimes.  


Figure (\ref{num_stab}) shows the relationship between the capillary number and the solution discretisation. At certain values of mesh spacing and time step size, the solution becomes numerically unstable, with the solution quickly blowing up after only a few time steps.  
\begin{figure}
\centering
\adjincludegraphics[scale=0.7]{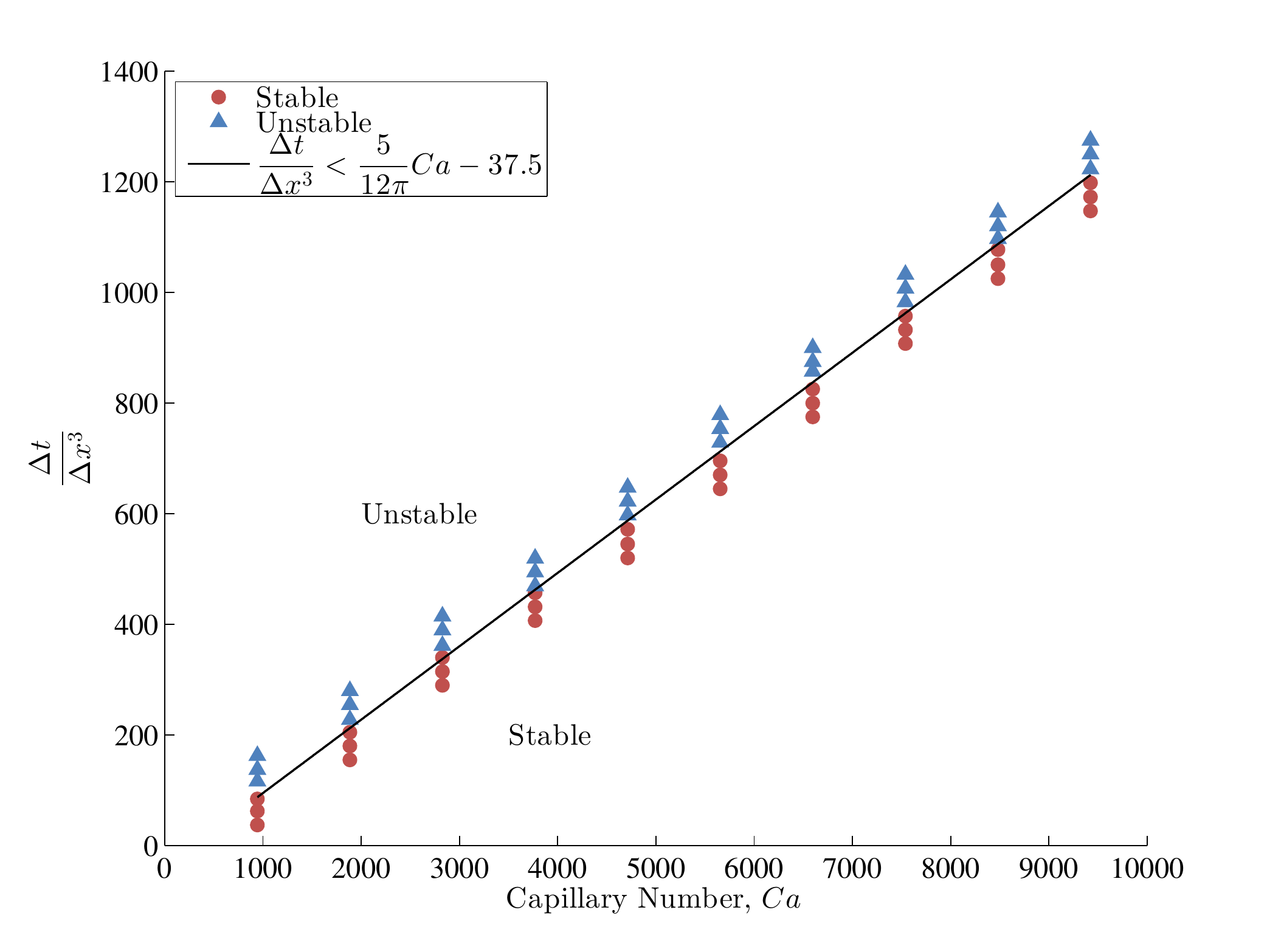}
\centering
\vspace{-20pt}
\caption{Numerical stability of the finite mobility ratio solution with least squares best fit curve giving the limit of stability.}
\label{num_stab}
\end{figure}
The proposed numerical approach is conditionally stable, as expected by the use of the explicit forward Euler time integration scheme (equation (\ref{eqn16})). When the capillary number is decreased, the solution becomes more numerically unstable and a lower $\frac{\Delta t}{\Delta x^3}$ must be used so that the solution does not blow up, as can be seen in figure (\ref{num_stab}). 


The expression represented by the line in figure (\ref{num_stab}) relates the capillary number to the mesh spacing and time step size. Equation (\ref{instability}) for the instability limit shows similarities with a Courant - Friedrichs - Lewy condition in finite difference approximation. The cubic mesh dependence comes from the cubic B-Spline discretisation that is used, with the Euler time stepping technique producing first order temporal dependence.  

\begin{align}
 \label{instability} \frac{\Delta t}{\Delta x^3} < \frac{5}{12 \pi} Ca - 37.5
\end{align}

Numerical experiments were performed to produce figure (\ref{num_stab}), whereby many simulations with varying $\frac{\Delta t}{\Delta x^3}$ were run, until the observed stability criterion became apparent. Although only six points per capillary number are shown in figure (\ref{num_stab}), many trial cases were used around the limit of stability to explicitly define the limiting value.  

The instability expression can be used for all simulations as a check to ensure that the solution is stable under the $\frac{\Delta t}{\Delta x^3}$ value being used. Fortunately, as the capillary number increases, the restriction of $\frac{\Delta t}{\Delta x^3}$ for a stable solution slackens and a more refined data set can be used without the solution becoming numerical unstable. As most of the flows and mechanisms under investigation in this paper occur at higher capillary numbers, the instability limit is generally not encountered frequently, but is a defining feature of the numerical method. 

\begin{figure}
\centering
\adjincludegraphics[scale=0.5]{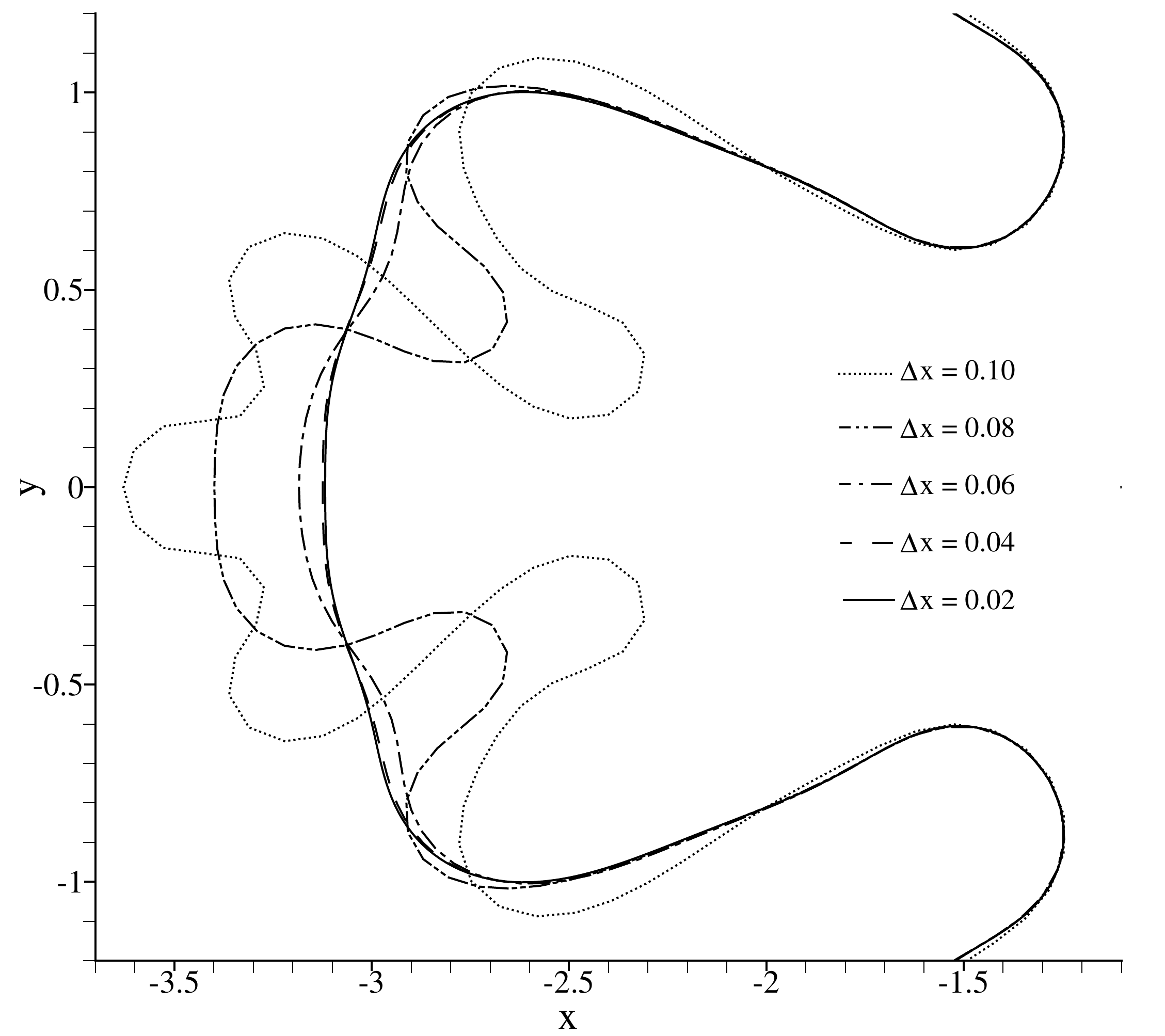}
\centering
\vspace{-20pt}
\caption{Unstable bubble interface shown at t = 20 with differing mesh resolutions. Ca = 10000.}
\label{mesh_unstab}
\end{figure}

To illustrate the physical instability of the problem caused by a high capillary number, a test case was run with identical conditions to that as the $\beta$ = 10 case in figure (\ref{power_comparison}), but with the volume flux, Q increased by a factor of 5, producing a capillary number of 10000. When using an element size of $\Delta x = 0.1$, it can be seen in figure (\ref{mesh_unstab}) that the boundary has become highly convoluted, compared to the stable shape of the solution with $\Delta x = 0.02$. The problem has become much more sensitive with the increase in capillary number, and as such large number of elements are required to accurately capture the interface. By increasing the number of elements, the solution becomes more stable and flattens out. With a factor of 2.5 decrease in the element size, the solution has effectively converged to a mesh independent solution, with no change to the boundary position. Care has to be taken when solving high capillary number systems as the highly unstable nature of the problem can permit a very different solution if the element density is not sufficiently high.

In figure (\ref{mesh_inde}), the spatial convergence of the solutions can be seen. The $L_1$ error norm quoted is the error between the position of the boundaries of the numerical solution and a mesh independent, pseudo analytical case. This pseudo analytical case was obtained using a very small time step ($\Delta t < 0.005$) and element size ($\Delta x < 0.02$), meaning that using any lower time step or element size did not change the position of the interface. To compare interfacial positions, the nodal positions (in the azimuthal direction around the solution interface) were reconstructed at the same location on the pseudo analytical case, and then the radial extents of the nodes compared.



Spatially, the solution seen in figure (\ref{mesh_inde}) converges very quickly; for large $\Delta x$ roughly 6th order is observed. This accounts for the mesh independence in figure (\ref{mesh_unstab}) whereby doubling the number of elements has produced a mesh independent solution. This excellent spatial convergence means that relatively few elements can be used to begin the simulation, with very high accuracies being achieved. 

Temporally the solutions converge linearly, as expected from the forward Euler time stepping scheme. Higher order time integration schemes were considered, such as the midpoint and Runge-Kutta methods, however, the spatial resolution was found to be much more of a limiting factor in the overall solution accuracy than the temporal resolution. A relatively low temporal resolution could be used without affecting the solution quality, and hence much greater attention was paid to the spatial resolution; in which a small change in the number of elements could create a large difference in the interface position. 


The small gain in accuracy from the higher order integration schemes was deemed unnecessary when considering the extra computation required to sub-divide the time steps and calculate the weighted average of the resulting interfacial velocities to calculate the subsequent interface position. 

\begin{figure}
\centering
\adjincludegraphics[scale=0.6]{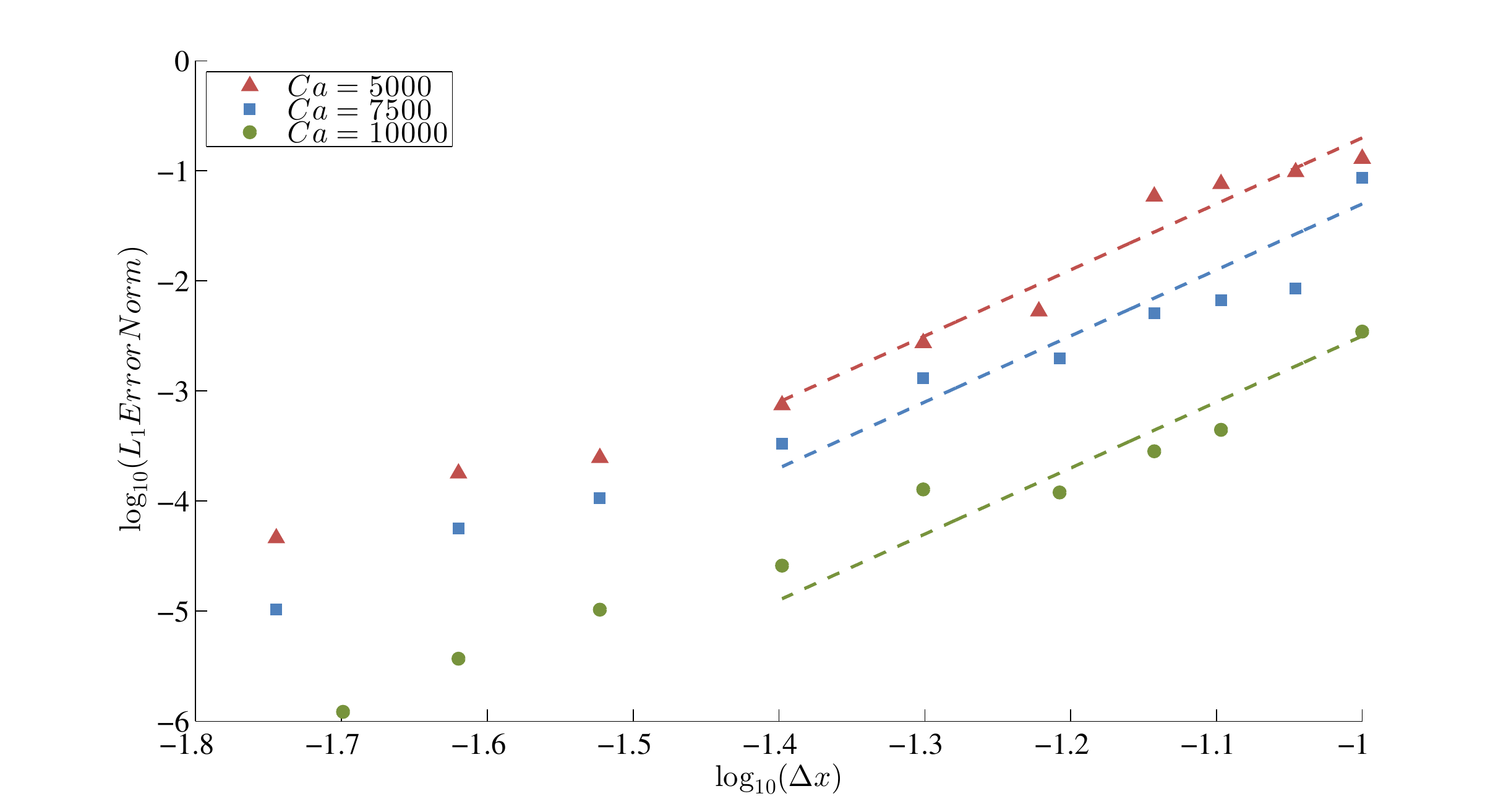}
\centering
\caption{Mesh independence study, showing the effect of mesh resolution on the $L_1$ error norm between the numerical solution and a mesh independent (pseudo analytical) case.}
\label{mesh_inde}
\end{figure}
\section{Mobility Ratio Effects}
\label{mob_effects}

This work is motivated by $CO_2$ injection and storage during carbon sequestration, in which the injection is performed supercritically, with the mobility ratio between supercritical $CO_2$ and brine of the order 10-30 \cite{ooyang2011}. 


The finite mobility ratio model developed in this work allows the efficient solution of low mobility ratio flows, and can be used to characterise the transition through a range of mobility ratios. Several finite mobility ratios are presented in figure (\ref{mobility}), in which the the capillary number is 4000. With the mobility ratio, $\beta = 1$, the initially perturbed solution stabilises after around 50s and expands to form a stable circle as there is no critical length scale for which bifurcation will occur due to there being no difference in viscosities. At all mobility ratios above 1, the interface eventually evolves to form a complex viscous fingering pattern, with higher mobility ratios promoting the onset of viscous fingering.

With a relatively low mobility ratio, such as the $\beta = 10$ case, the bases of the fingers continue to advance with time. However, once the mobility is taken higher, the finger base evolution slows as the base position approaches a near constant value. During the $\beta = 1000$ case, the finger bases effectively become stagnation points, where the interface velocity at the base drops to near zero. This characteristic is a well known feature of high mobility ratio displacement, causing highly convoluted surfaces and a much lower swept volume of the higher viscosity fluid. A consequence of the slowed base evolution and quickly growing primary fingers is that competing fingers' growth is hindered by the larger primary fingers and shielding occurs \cite{homsy1987}.

However, in the low mobility ratio regime, the growing finger bases allow secondary growing fingers to be fed by fluid, meaning that they can possess significant velocity. Shielding between competing fingers is inhibited as the fluid flow is not forced from the secondary finger into the primary finger, meaning much greater interaction and non-linear dynamics are seen between growing fingers.

\begin{figure}
        \centering
               \begin{subfigure}{0.5\textwidth}
                \includegraphics[width=\textwidth]{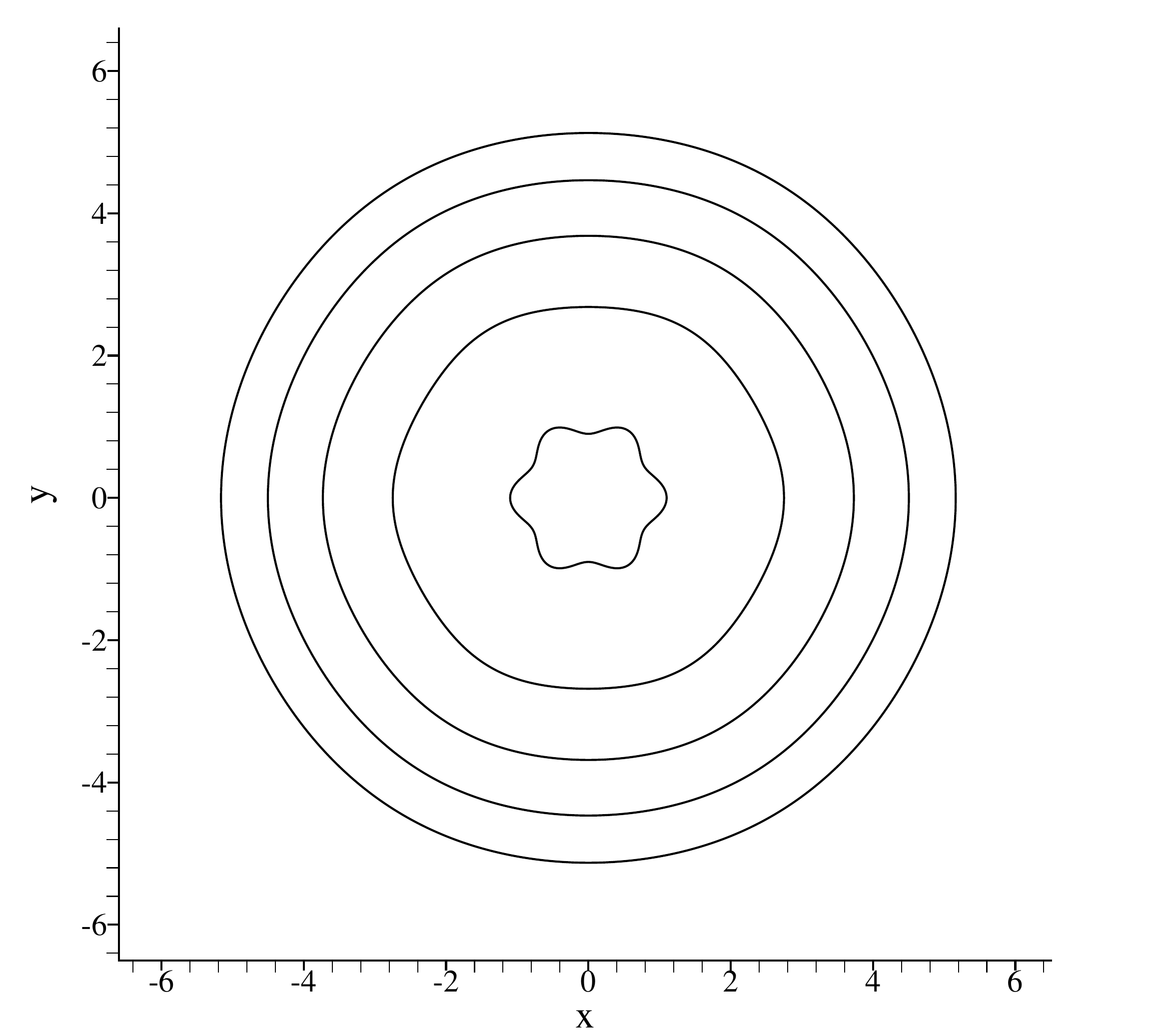}
                \caption{$\beta = 1$}
                \label{beta=1}
        \end{subfigure}%
        ~ 
        \begin{subfigure}{0.5\textwidth}
                \includegraphics[width=\textwidth]{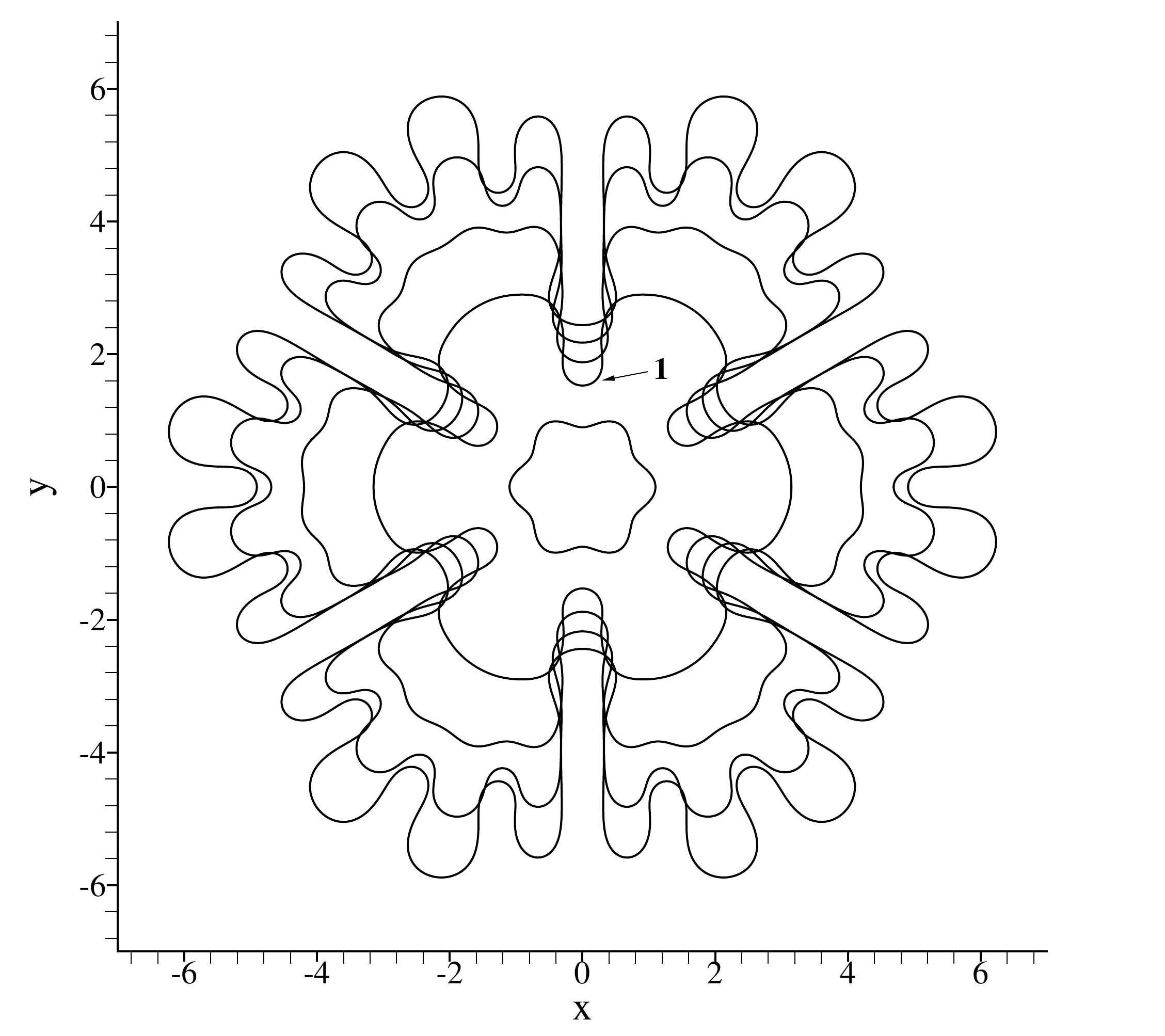}
                \caption{$\beta = 10$}
                \label{beta=10}
        \end{subfigure}%
        
        \begin{subfigure}{0.5\textwidth}
                \includegraphics[width=\textwidth]{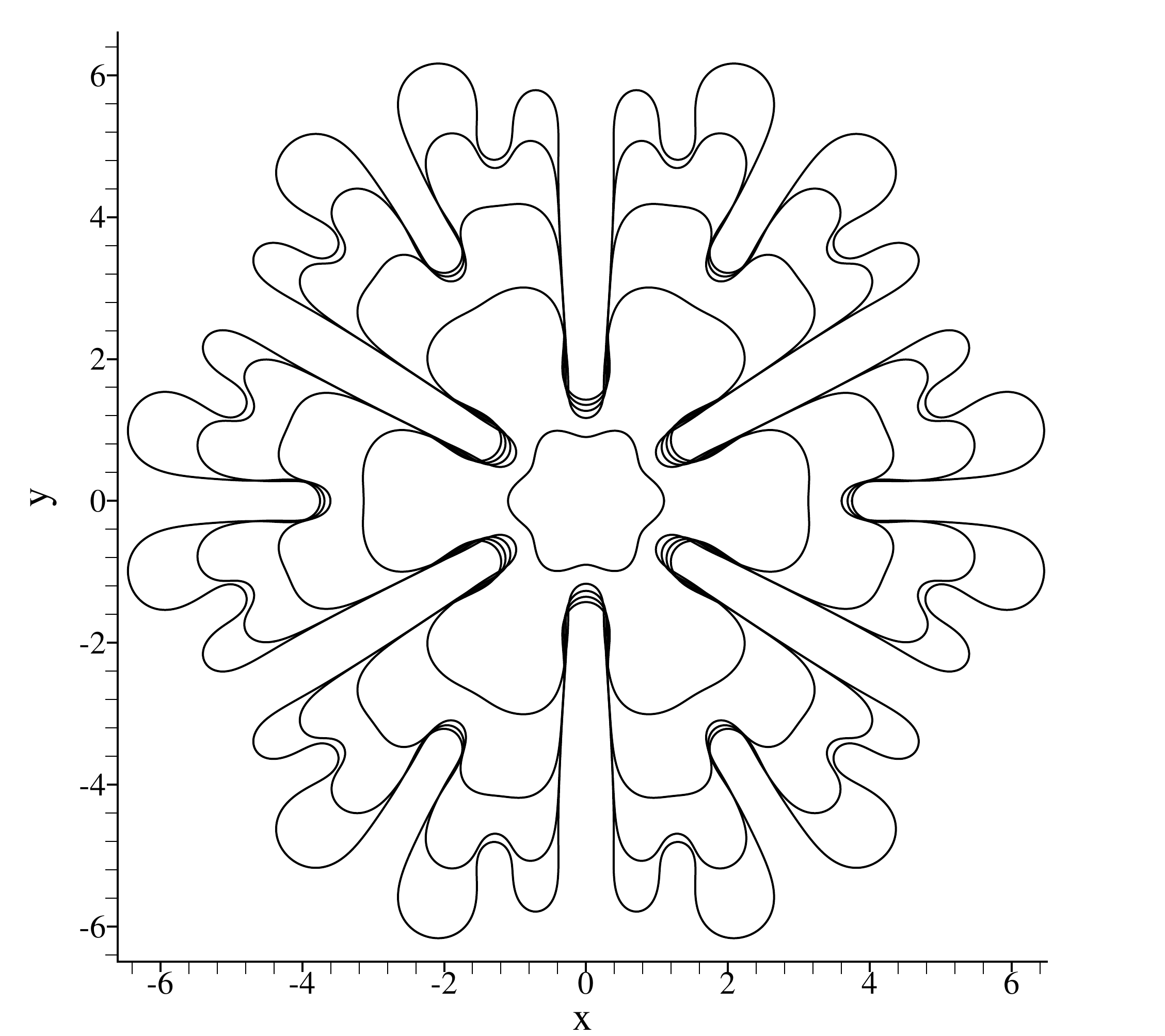}
                \caption{$\beta = 100$}
                \label{beta=100}
        \end{subfigure}%
        ~ 
                \begin{subfigure}{0.5\textwidth}
                \includegraphics[width=\textwidth]{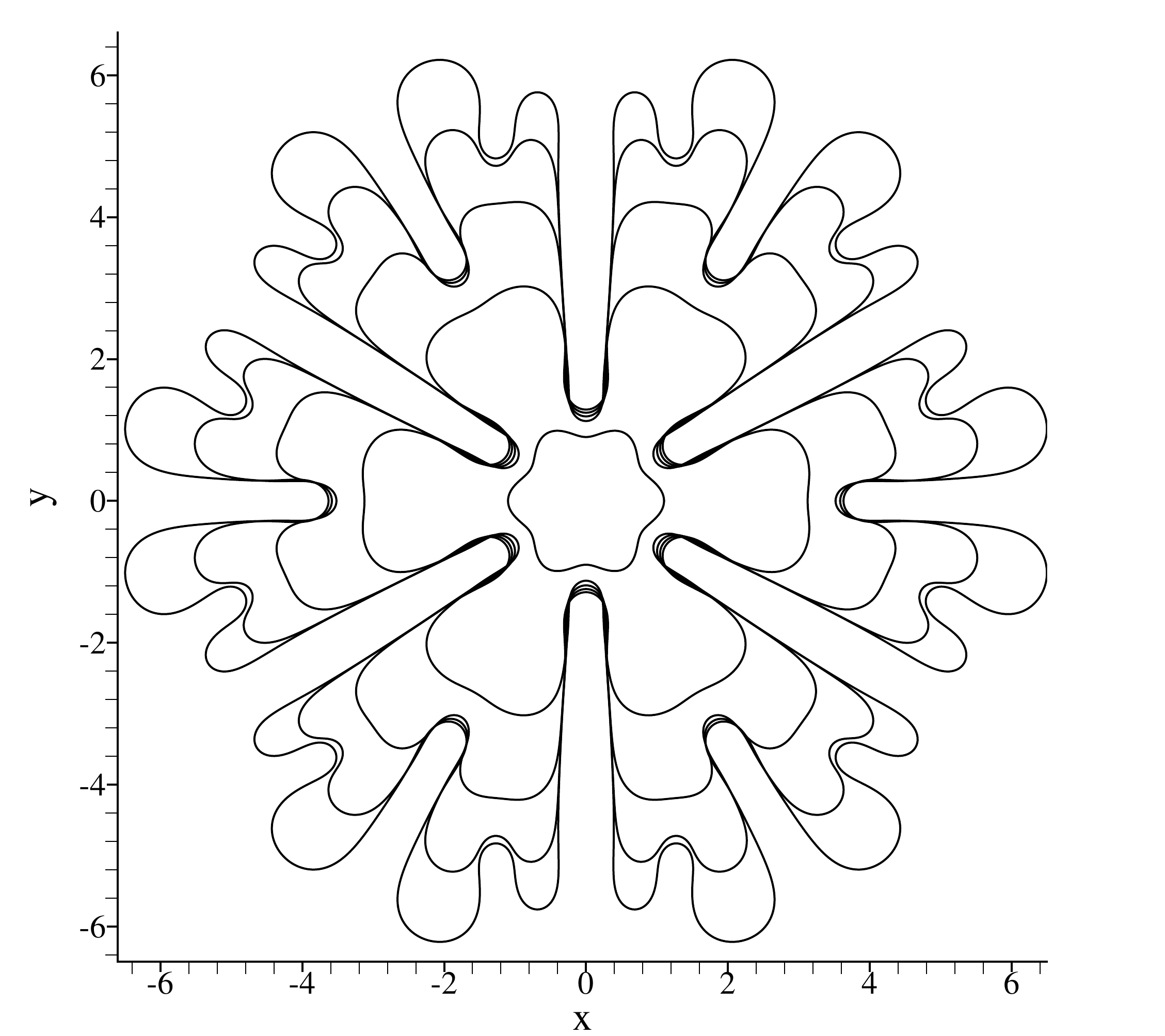}
                \caption{$\beta = 1000$}
                \label{beta=1000}
        \end{subfigure}%
        \caption{Bubble evolution plots showing the effect of varying the mobility ratio, $\beta$. Each sub-plot shows the interface of the bubble every $\Delta t=20$ from $t = 0-80$. Location 1 shows the finger base used for the velocity field analysis in figure (\ref{velocity_vectors}).}
        \label{mobility}
\end{figure}

By explicitly tracking the base position of the fingers for varying mobility ratio runs, the radial extent of the finger bases can be seen to reach an almost constant value for high mobility cases in figure (\ref{mo_stag}). The lower mobility ratio cases show a much greater evolution of the base position once the profile of the bubble has developed. This evolution continues until the non-linear regime, where the fingers interact significantly with each other affecting the base position. 

\begin{figure}
\centering
\adjincludegraphics[scale=0.5]{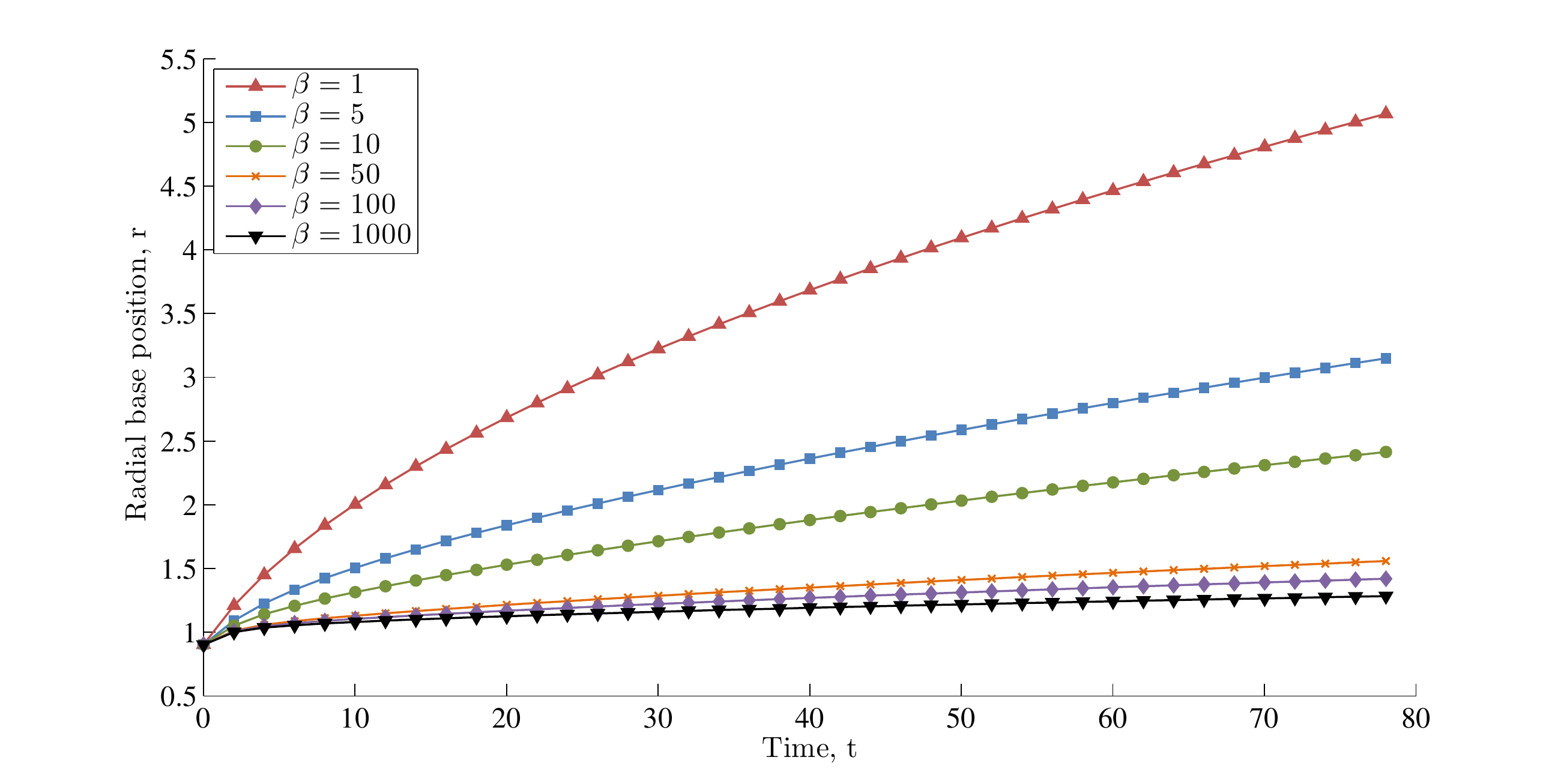}
\vspace{-20pt}
\caption{Evolution of the finger bases with time and varying mobility ratio.}
\label{mo_stag}
\end{figure}

To further emphasise the difference between the base profiles and the velocity of the fluid at the base regions, a velocity field can be generated at grid points throughout the domain. Utilising Green's formulae for the perturbed pressure in the fluid domain (as opposed to at the fluid interface, equation (\ref{eqn9})), the perturbed pressure can be reconstructed using the known interface values. Using equations (\ref{eqn5.0}),(\ref{eqn5.1}), (\ref{eqn2})  and (\ref{eqn1}) the velocity at any point in the interior or exterior domains can then be reconstructed, once the perturbed pressure and perturbed flux is known at the interface. 

Figure (\ref{velocity_vectors}) shows the velocity field generated around the finger base at location 1 in figure (\ref{mobility}), for the $\beta = 10$ and $\beta = 1000$ mobility ratio cases. In both plots, there is significant velocity in the interior domain, with the fluid flowing preferentially around the finger base, due to the high surface tension and curvature at the bottom of the base. However, in the $\beta = 1000$ case, the fluid velocity drops much more significantly in the exterior fluid close to the finger base, than in the $\beta = 10$ case. The finger base in the higher mobility example has a higher curvature which makes it harder for the fluid to displace the surface, and as such the velocity drops to near zero in the exterior fluid. This explains the near stagnation of the finger bases found in the high mobility ratio examples in figures (\ref{mobility}) and (\ref{mo_stag}). 

In figure (\ref{velocity_vectors}), the velocity of the inner fluid just inside the finger base has been labelled, along with the $x,y$ position of the vector. The velocity is approximately 7 times less in the $\beta = 1000$ case than the $\beta = 10$ case. It can also be seen that the velocity vectors immediately adjacent to the base velocity are much larger in the $\beta = 1000$, and in a tangential direction to the base profile, showing the preferential movement of the fluid around the base in higher mobility ratio flows. The velocity vectors close to the finger base in the low mobility regime have a large component in the normal direction, which gives rise to the movement of the base.

The significant velocity possessed by the finger base and the exterior fluid close to it in figure (\ref{beta10}) causes the finger base to displace and not stagnate near the starting profile. This velocity explains the greater finger interaction between competing fingers in the low mobility ratio regime, and finger breaking/coalescing mechanisms, that will be discussed in section  \ref{long_time} of this paper.

\begin{figure}
        \centering
                \begin{subfigure}{0.5\textwidth}
                \includegraphics[width=\textwidth]{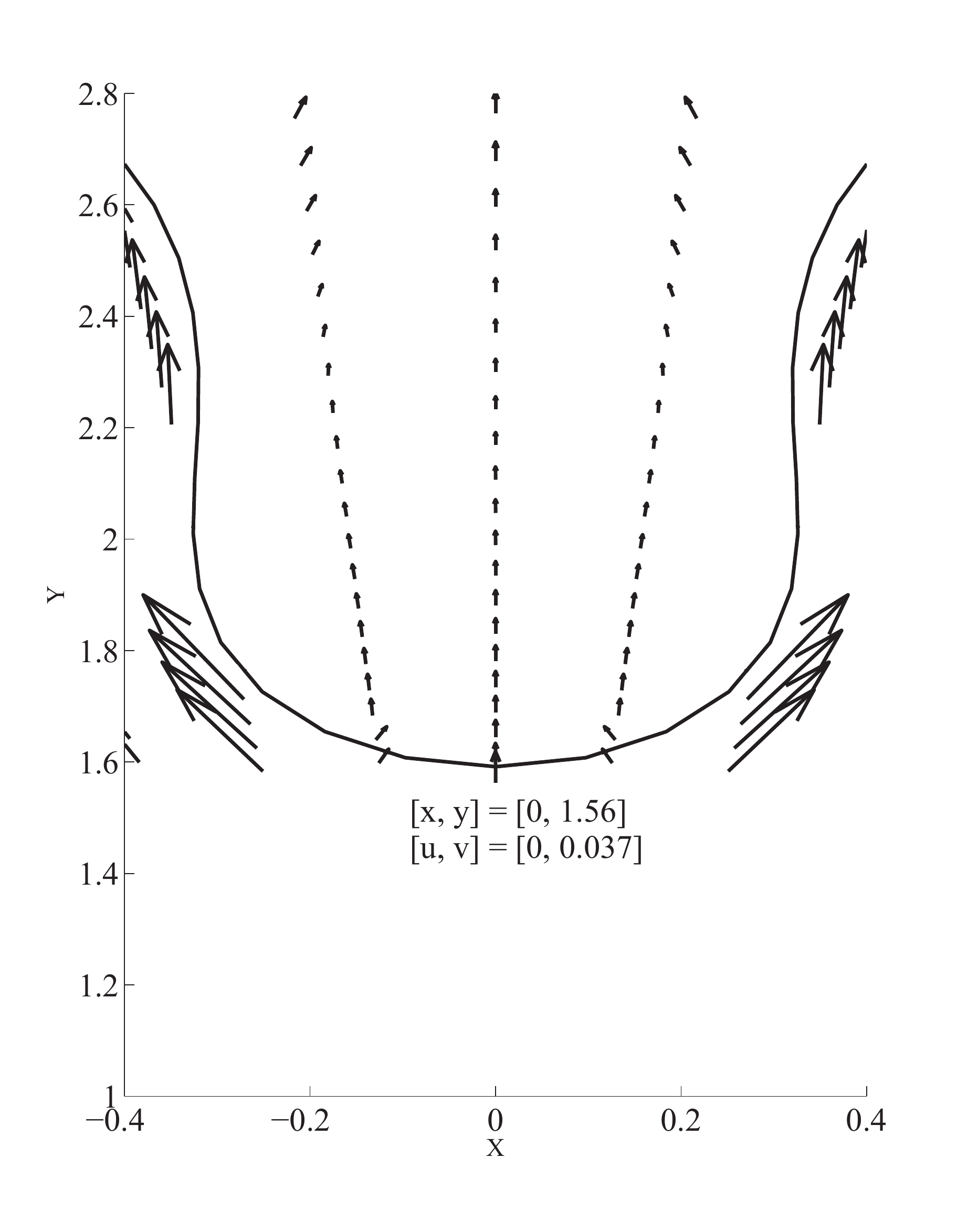}
                \caption{$\beta = 10$}
                \label{beta10}
        \end{subfigure}%
        ~ 
        \begin{subfigure}{0.5\textwidth}
                \includegraphics[width=\textwidth]{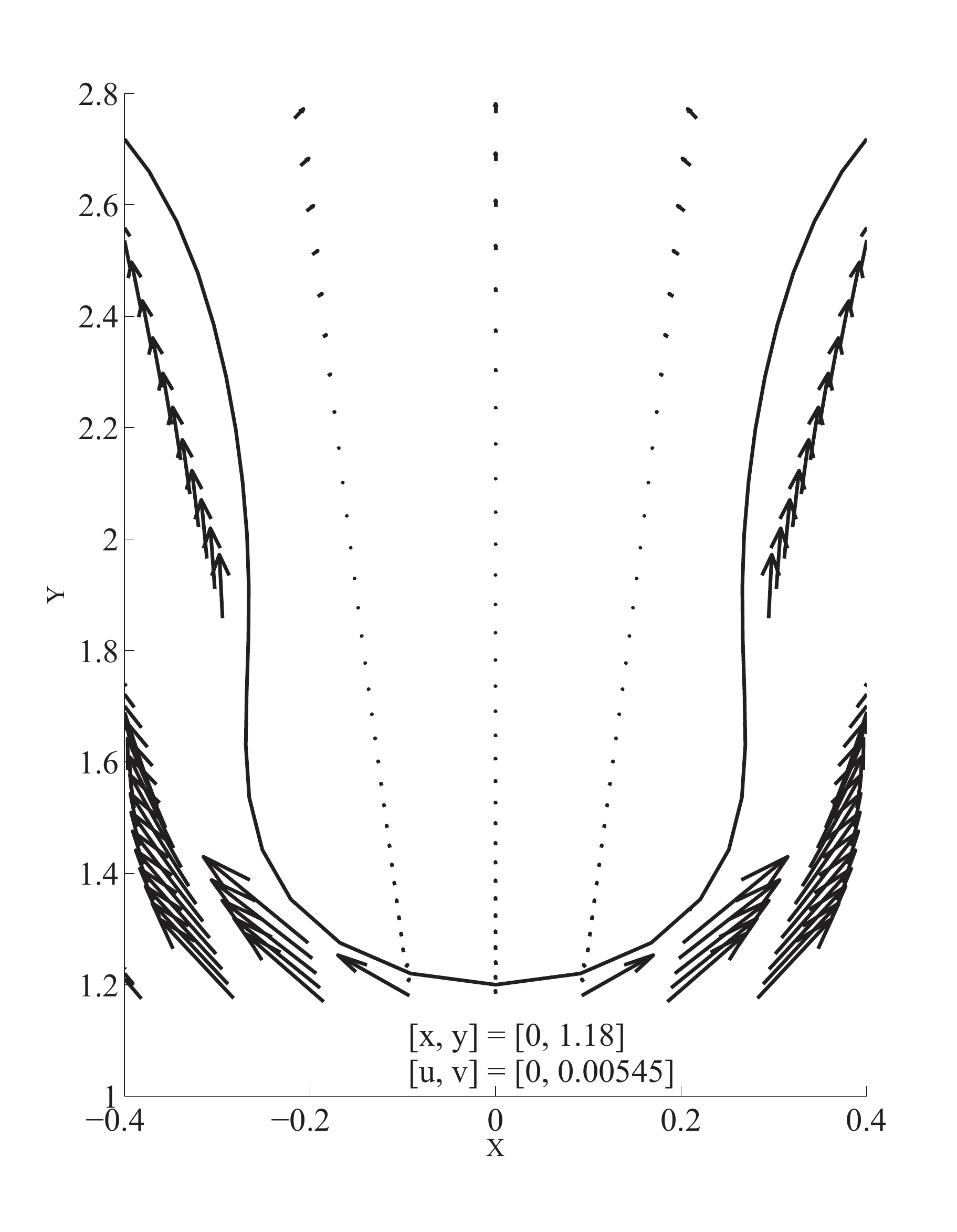}
                \caption{$\beta = 1000$}
                \label{beta1000}
        \end{subfigure}%
        \caption{Fluid velocity vector plots at the base of a finger in the $\beta = 10$ and $\beta = 1000$ cases in figure (\ref{mobility}) at t = 20.}
        \label{velocity_vectors}
\end{figure}

\section{Capillary Number Effects}

As a consequence of the $N^2$ scaling of the convergent series numerical method, high capillary number cases can be accurately resolved using many elements. High injection rates typically found in $CO_2$ sequestration processes give rise to large capillary numbers that promote finger instability. In section \ref{numerical_Stab_section}, large capillary numbers were found to create highly unstable interfaces between the fluids, requiring many elements to accurately solve. 

Several high capillary number flows are shown in figure (\ref{capillary}) with a mobility ratio of 10. The number of fingers created on the first bifurcation increases from two to six through the range of capillary numbers due to a decrease in critical length at which bifurcation occurs. The critical length scale of bifurcation of a finger growing into a parallel flow is given by \cite{couder2000}:

\begin{align}
 \label{crit_len} L_c = \frac{\pi b}{a} \sqrt{\frac{\gamma}{\mu V}}
\end{align}
Where, $V$ represents the front velocity. Observations confirm that generally the number of fingers at the first bifurcation increase with capillary number, although there are several different modes of bifurcation that can occur. This means that bifurcations such as side branching can occur in preference to generating more primary fingers at several values of capillary number, as can be seen in figure (\ref{capillary}).

\captionsetup[subfigure]{labelformat=empty}
\begin{figure}
        \centering
               \begin{subfigure}{0.5\textwidth}
                \includegraphics[width=\textwidth]{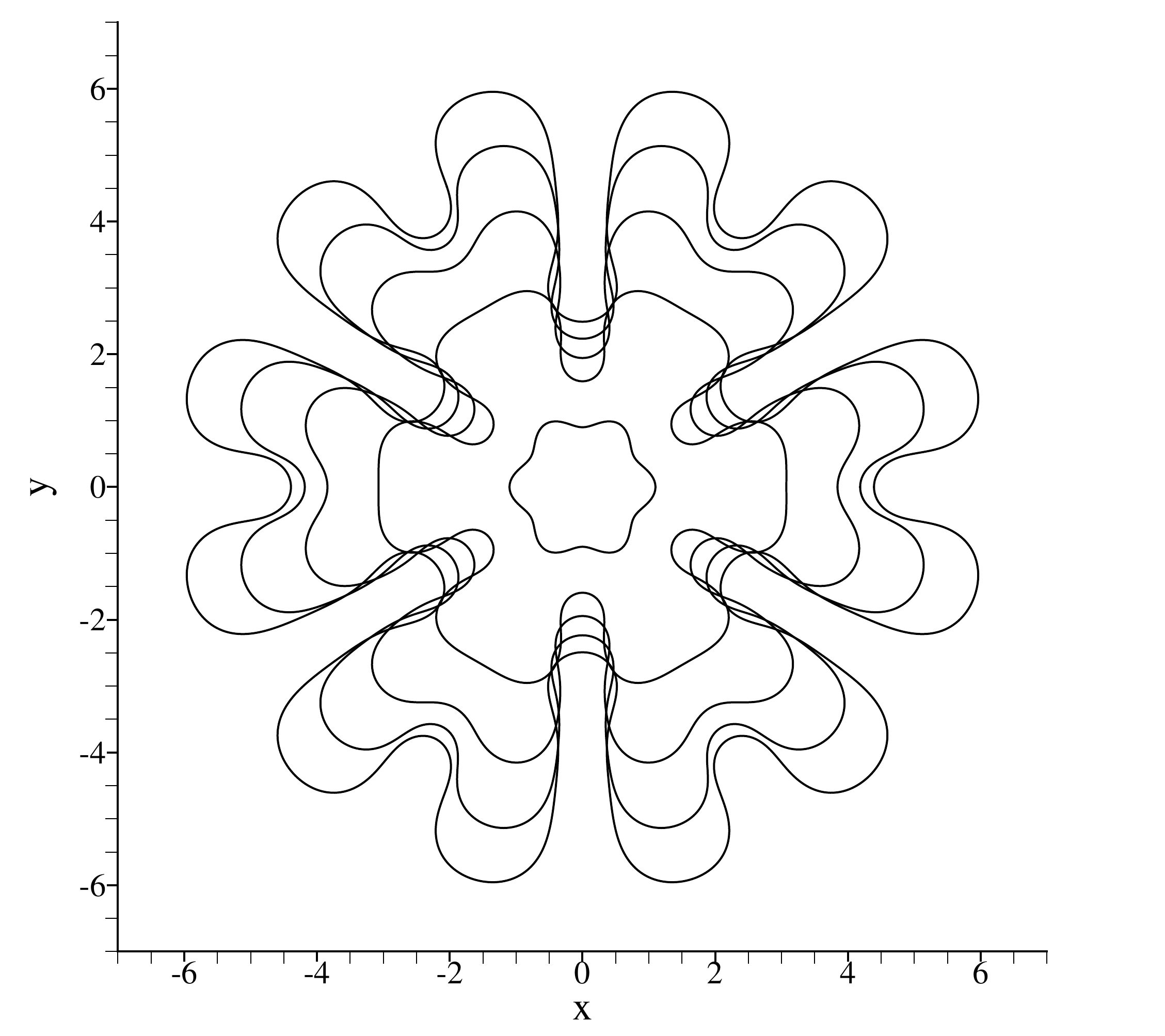}
                \caption{(a) $Ca = 2000$}
                \label{ca=2}
        \end{subfigure}%
        ~ 
        \begin{subfigure}{0.5\textwidth}
                \includegraphics[width=\textwidth]{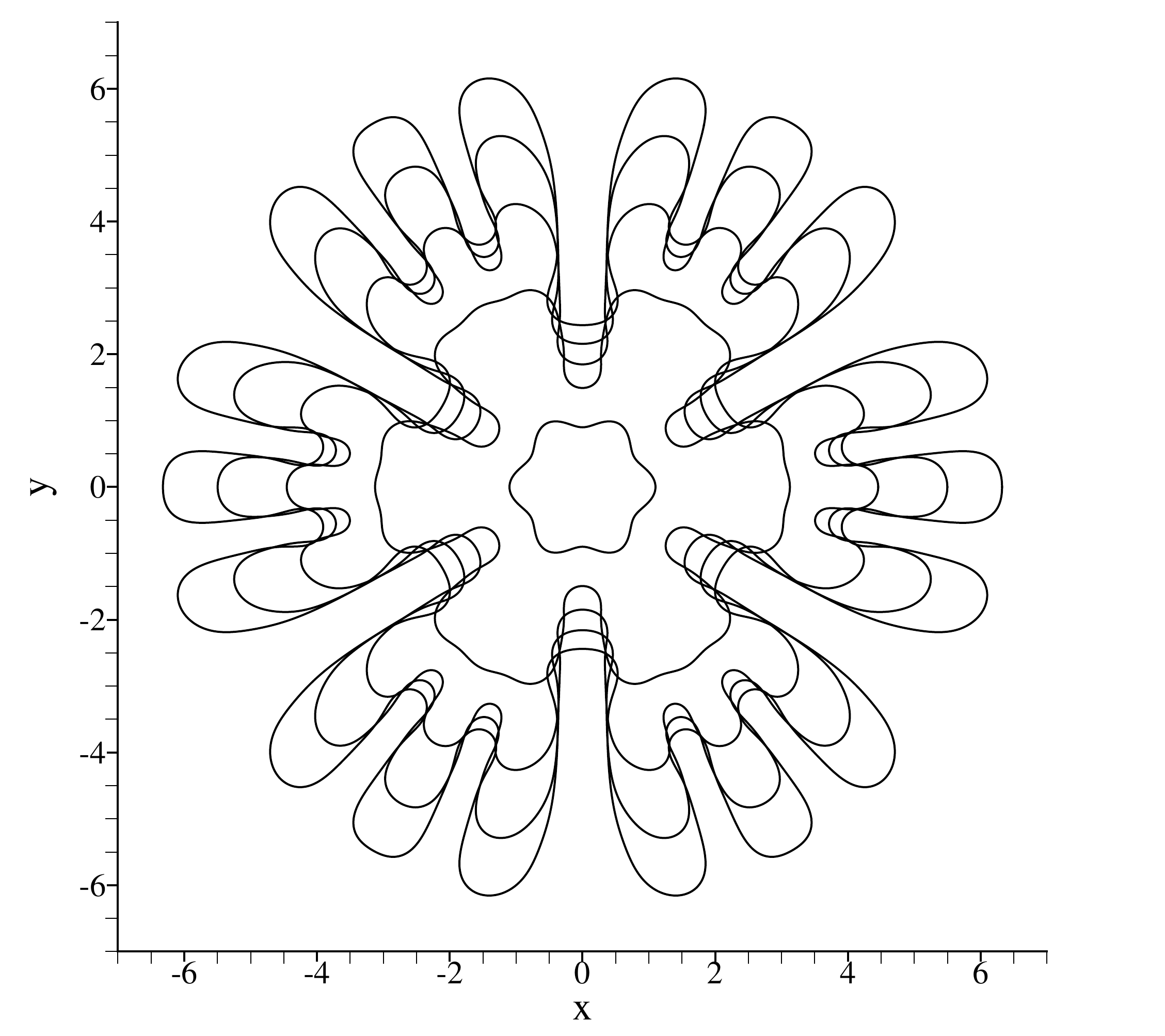}
                \caption{(d) $Ca = 8000$}
                \label{ca=8}
        \end{subfigure}%
        
        \begin{subfigure}{0.5\textwidth}
                \includegraphics[width=\textwidth]{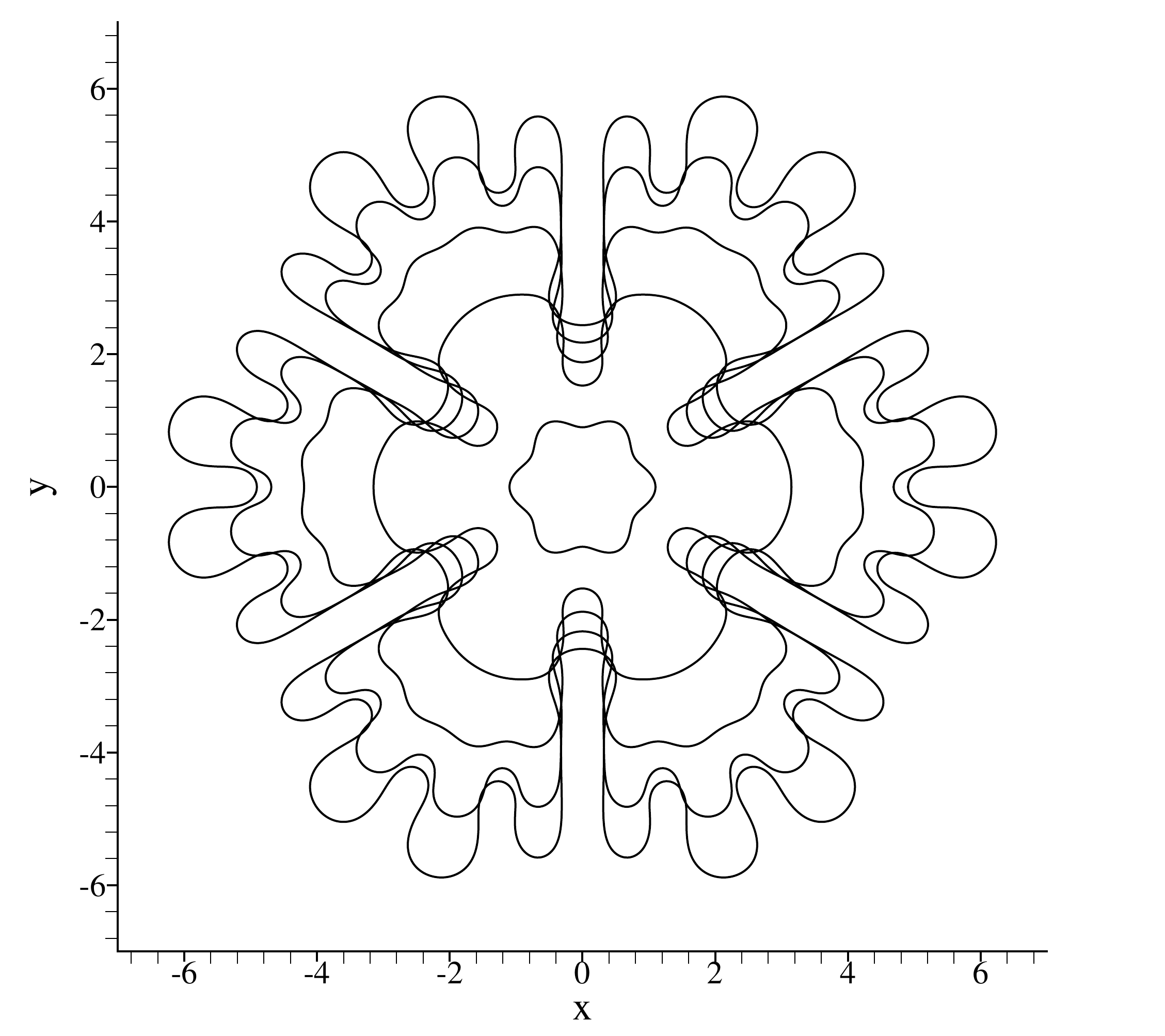}
                \caption{(b) $Ca = 4000$}
                \label{ca=4}
        \end{subfigure}%
        ~ 
                \begin{subfigure}{0.5\textwidth}
                \includegraphics[width=\textwidth]{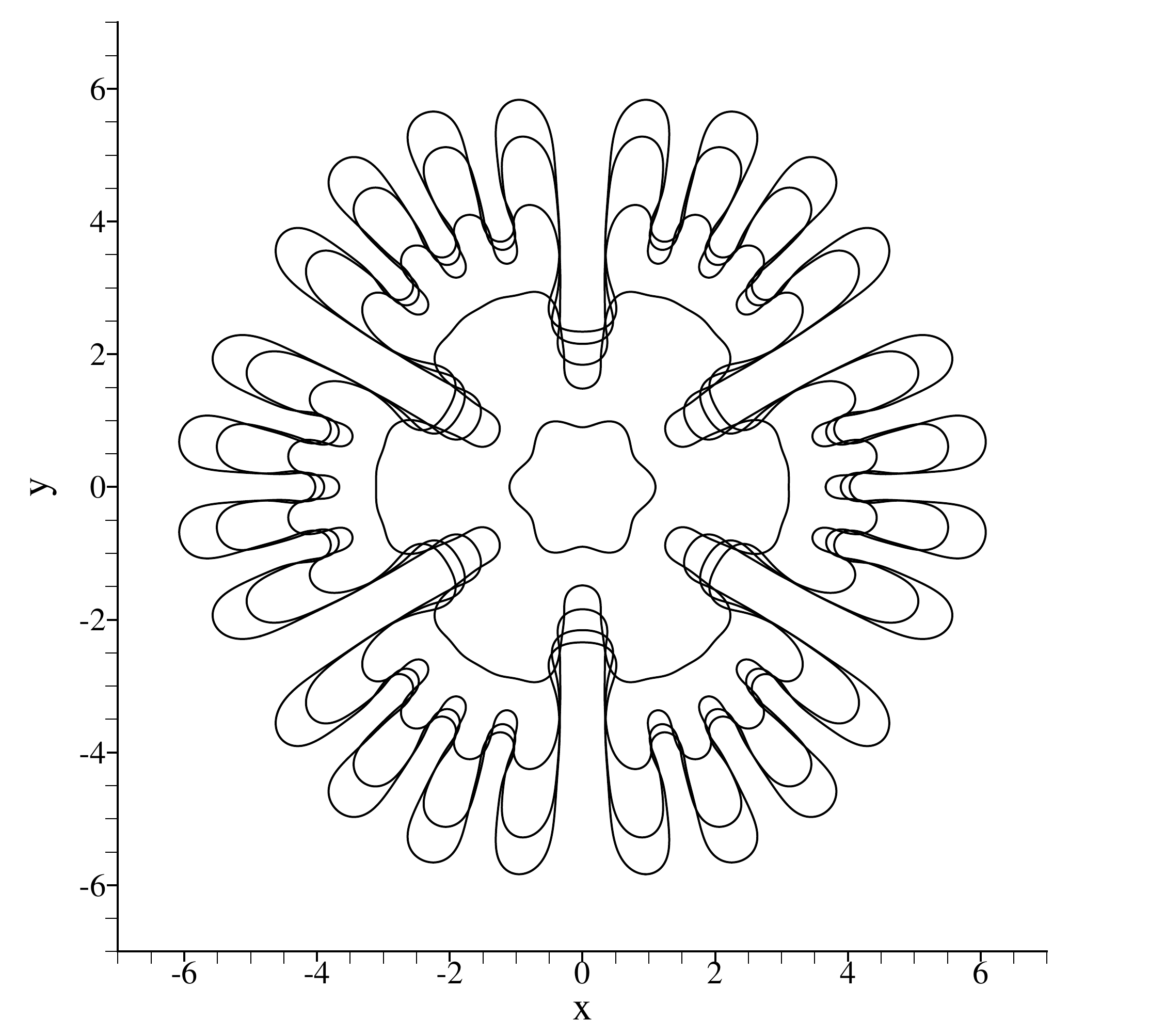}
                \caption{(e) $Ca = 10000$}
                \label{ca=10}
        \end{subfigure}%
        
        \begin{subfigure}{0.5\textwidth}
                \includegraphics[width=\textwidth]{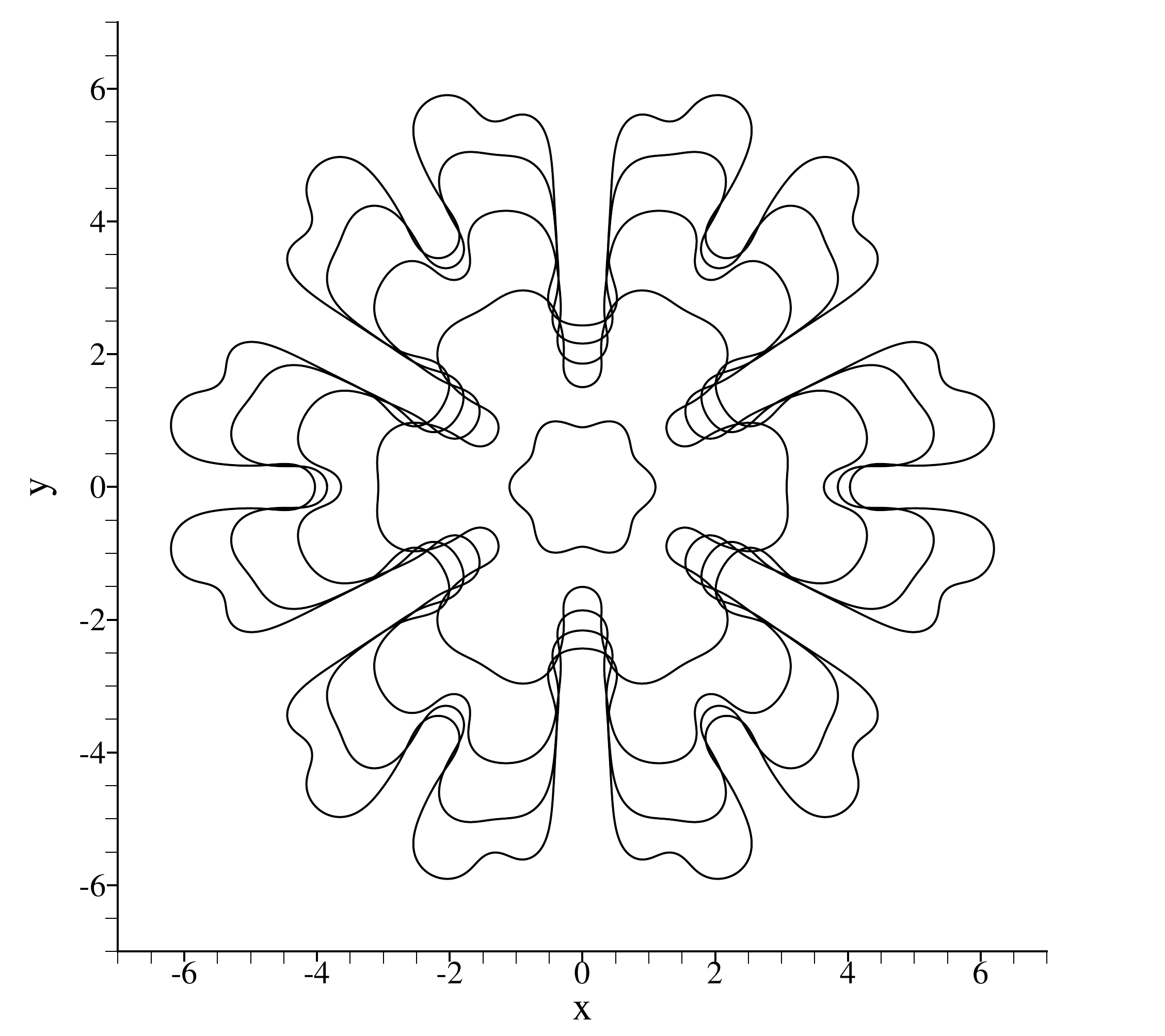}
                \caption{(c) $Ca = 6000$}
                \label{ca=6}
        \end{subfigure}%
        ~ 
                \begin{subfigure}{0.5\textwidth}
                \includegraphics[width=\textwidth]{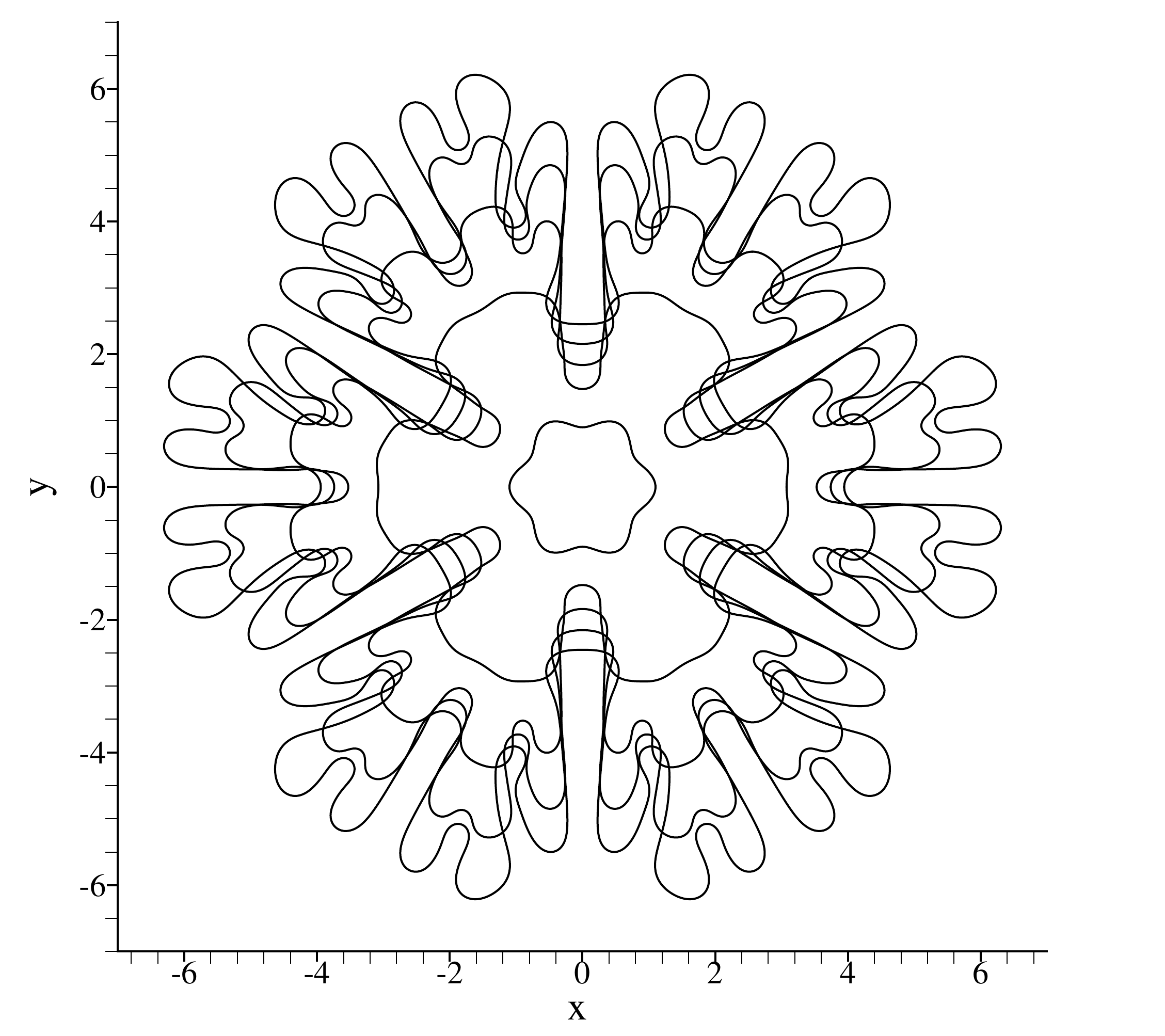}
                \caption{(f) $Ca = 12000$}
                \label{ca=12}
        \end{subfigure}%
        \caption{Bubble evolution plots showing the effect of varying capillary number. Each sub-plot shows the interface of the bubble every $\Delta t=20$ from $t = 0-80$.}
        \label{capillary}
\end{figure}

An important point about the position of the profiles shown in figures (\ref{ca=2}) to (\ref{ca=12}) is that the base positions show almost exactly the same radial evolution with time, but their fronts show vastly differing profiles. In figure (\ref{cap_stab}), the evolution of the base of the fingers is shown for varying capillary number where it can be seen that as the capillary number is increased, the base position starts to converge to the same value. 

As the capillary number is increased, preferential movement of the inner fluid causes the finger fronts to exhibit vastly different profiles, while the base positions remain fairly constant. The convergence of the base position is due to the relatively large, smooth curvature of the finger base profile acting to stabilise the flow. As the driving force of the fluid increases with  capillary number, the lower curvature finger front becomes more likely to destabilise and the flow is forced to this region in preference to the finger base, leaving it unaffected. 

\begin{figure}
\centering
\adjincludegraphics[scale=0.6]{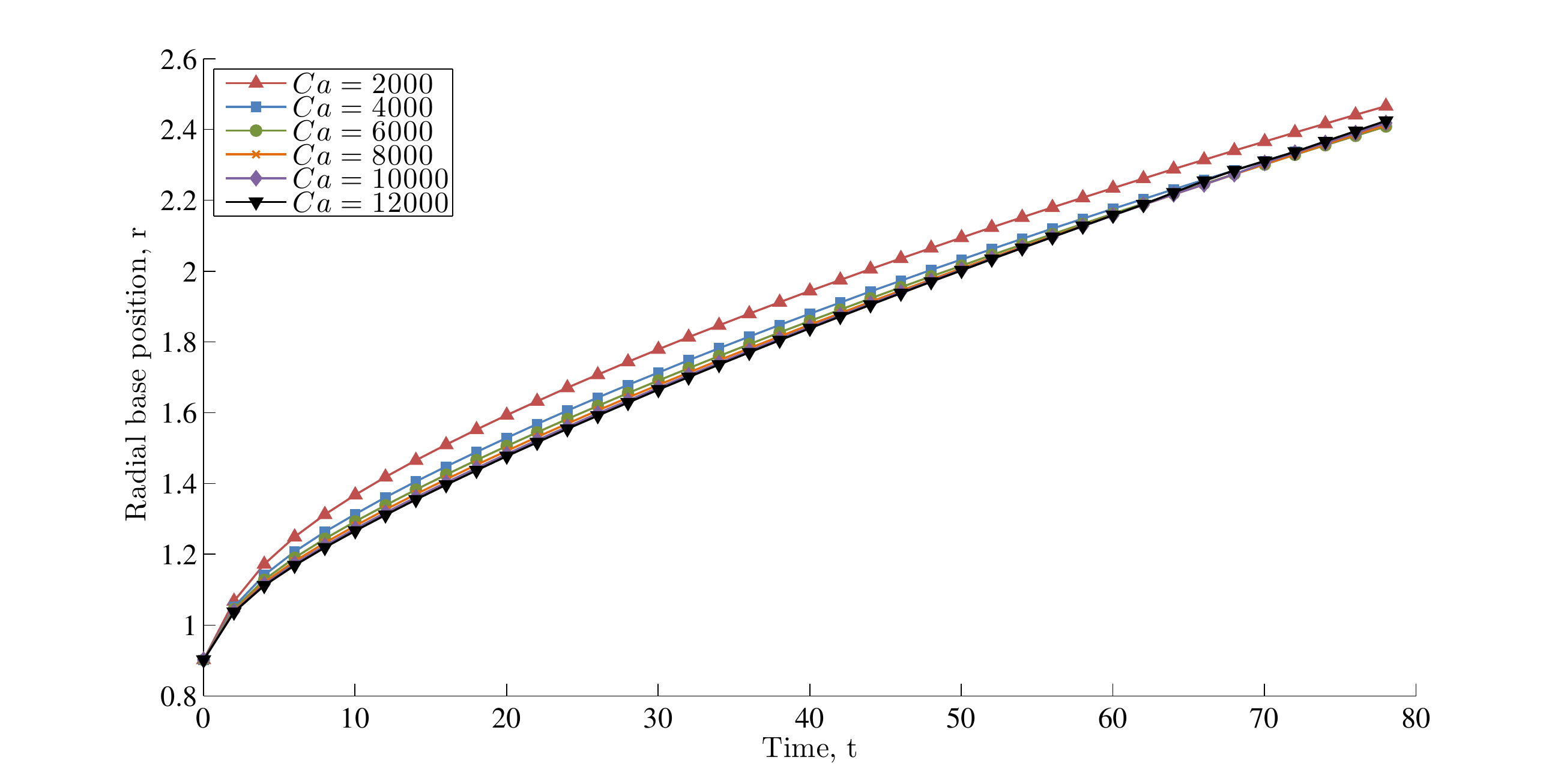}
\centering
\vspace{-20pt}
\caption{Position of the finger base with time and varying capillary number}
\label{cap_stab}
\end{figure}
The different finger front profiles in the two cases presented in figure (\ref{capillary}) are caused by the critical length of bifurcation being smaller for the higher capillary number case. At the point of the first bifurcation, there are more 'flat' sections of the finger larger than the critical length and hence more fingers are able to form. 

\begin{figure}
\centering
\adjincludegraphics[scale=0.5]{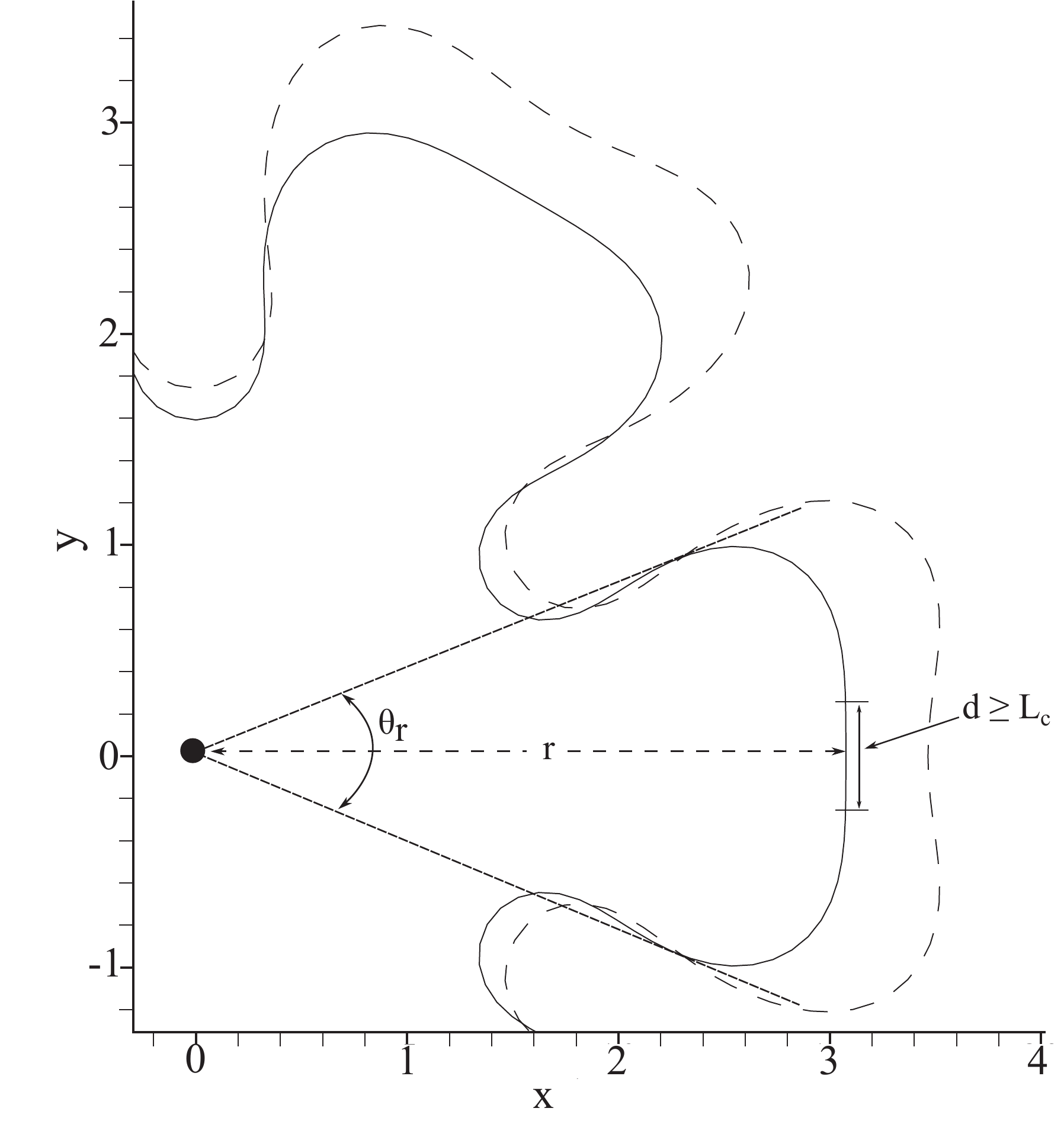}
\centering
\vspace{-20pt}
\caption{Initial bifurcation of two viscous fingers, showing critical length scales. Dashed line shows the interface position $\Delta t = 20$ after the solid line interface.}
\label{bif}
\end{figure}
Once the front velocity, $V$ reaches a low enough speed and there is a flat section in the curve that exceeds the critical length scale, the front starts to destabilise and bifurcate. The exact point of bifurcation is difficult to identify, due to the small length scale at which bifurcation initiates. Therefore, a dimensionless parameter, $W(\theta)$ can be used to provide a robust measure of the point at which the first bifurcation occurs in radial Hele-Shaw flow \cite{couder2000}. This parameter defines when bifurcation occurs based on the interior angle occupied by the finger, $\theta_r$, the radial extend of the finger, $r$ and the critical length of bifurcation, $L_c$. These parameters are shown in figure (\ref{bif}), at the first bifurcation of a radial viscous finger. 

\begin{align}
 \label{w_thet} W(\theta) = \frac{\theta_r r}{L_c}
\end{align}
The value of $W(\theta)$ at which bifurcation occurs shows much less variance than the critical length scale, hence, if the time at which bifurcation first occurs is misjudged slightly, the value of $W(\theta)$ changes by only a fraction of a percent. 

By performing similar tests to \cite{couder2000} at varying capillary numbers, the first point of bifurcation could be measured. In the results of \cite{couder2000}, the front velocity was kept constant unlike the simulations run here that have a constant volume flux injection and hence decreasing front speed with increasing radial distance from the source. Also, \cite{couder2000} used a wedge shaped cell with only one finger, compared to the fully circular cell with multiple fingers used here. 

In figure (\ref{w_graph}) the value of $W(\theta)$ increases with increasing capillary number, as expected due to the critical length scale decreasing. However the mobility ratio does not alter the value of $W(\theta)$ at which bifurcation occurs. In \cite{couder2000} the value of $W(\theta)$ is constant for all capillary numbers, however by having a non-constant front velocity, this value is able to change with capillary number. In figure (\ref{w_graph}) the bifurcation value of $W(\theta)$ follows an explicit trend given by:

\begin{align}
 \label{w_stab} W_{crit}(\theta) = \sqrt{\frac{Ca}{96 \pi}}
\end{align}
At any point in the simulation, if the front velocity and internal angle of the finger is great enough to cause $W(\theta)$ to be above the critical value defined in equation (\ref{w_stab}), bifurcation will occur. This expression works for all capillary numbers up until the first bifurcation, after which the fingers can grow non-linearly and interact significantly with each other. $W(\theta)$ does not predict the type of bifurcation that will occur, only the point at which a bifurcation will occur. Whether the front splits into 2 fingers or 5 fingers is determined by the critical length scale at that point in time and the number of flat sections running perpendicular to the flow that are longer than this length scale.

\begin{figure}
\centering
\adjincludegraphics[scale=0.7]{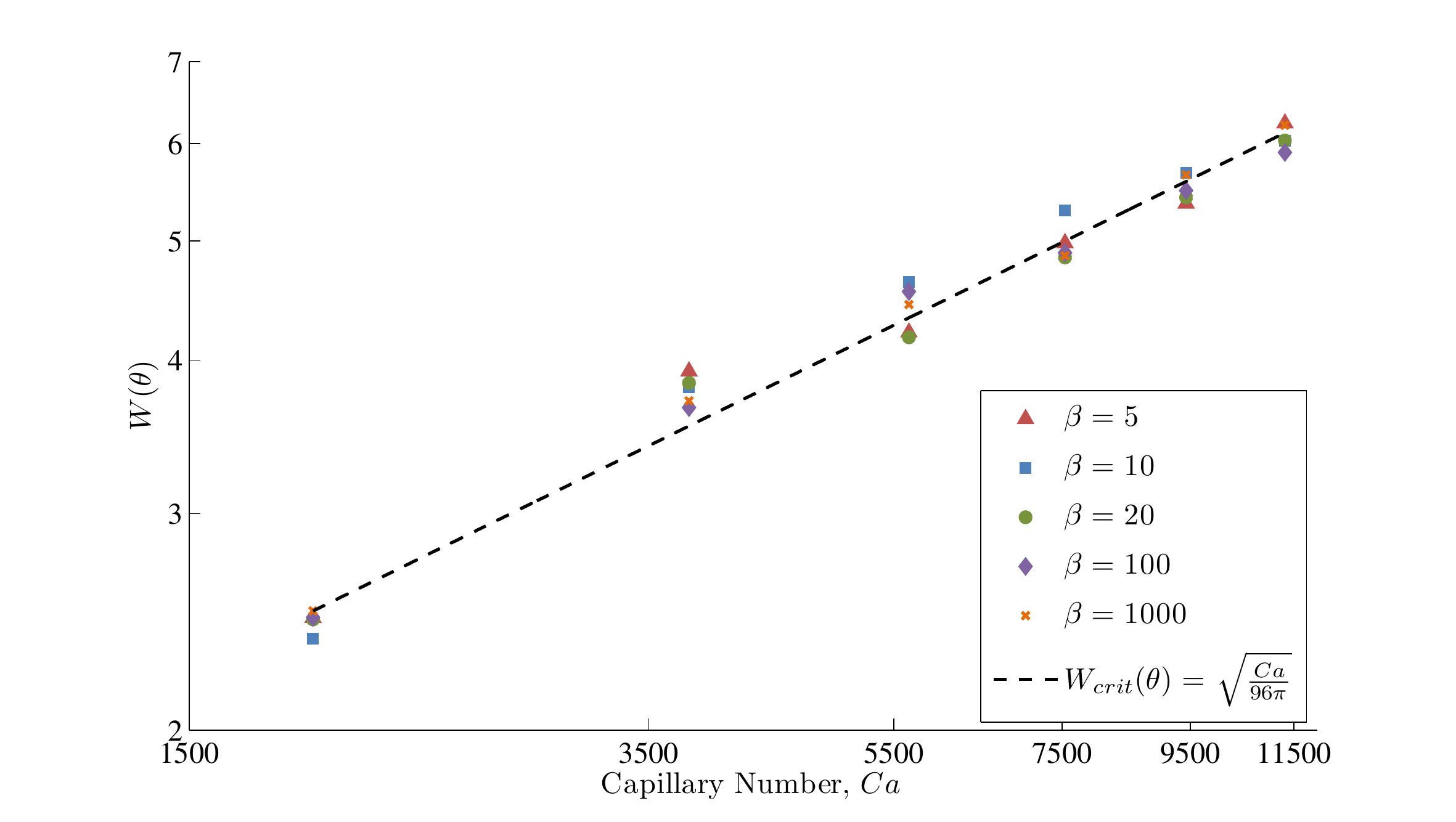}
\vspace{-10pt}
\caption{Variance of W$(\theta)$ with capillary number and mobility ratio.}
\label{w_graph}
\end{figure}

\section{Long time scale evolution}
\label{long_time}

The work presented in this paper has been motivated by the need to investigate low mobility ratio, high capillary number flows that occur in supercritical $CO_2$ sequestration in deep subsurface aquifers. While the current model cannot predict the full complexity of the $CO_2$ plume evolution in the injected porous media aquifer, it can provide qualitative understanding of the mechanisms and plume growth that can occur due to the low mobility ratio, high capillary number flow regime, using the assumption of a perfectly sharp interface between the $CO_2$ and brine. 

To understand the long term mechanisms that occur during low mobility ratio flows, long time evolutions of the interface are presented in this section, in order to see the effect of finger interaction when shielding is inhibited. \cite{moore2003}  and \cite{alvarex2004} show that mechanisms such as finger coalescing and finger break-off could be observed under certain flow regimes. In reaction models and diffuse-interface models, similar mechanisms have also been observed \cite{sun2008}\cite{power2013}\cite{wit1999}, however with truly immiscible models with sharp fronts, these mechanisms have not been explored, due to most former models concentrating on cases of high mobility ratio (generally using very viscous displaced fluids) where shielding inhibits break-off and coalescing \cite{degregoria1986}\cite{hadavinia1995}}. These mechanisms will be explored by studying long term interface growth using realistic injection parameters. 

\begin{table}
\begin{tabular}{|l|l|}
\hline
\textbf{Property}  & \textbf{Value (SI Units)}  \\ \hline
Gaseous injection depth    & 100m                      \\ \hline
Supercritical injection depth    & 1000m                      \\ \hline
T$_{CO_2 (sc)}$         & 50 $^\circ$C               \\ \hline
T$_{CO_2 (g)}$         & 25 $^\circ$C               \\ \hline
P$_{CO_2 (sc)}$         & 20 MPa                     \\ \hline
P$_{CO_2 (g)}$         & 1 MPa                     \\ \hline
NaCl concentration in brine & 0.5 mol/kg                 \\ \hline
$\mu_{{Brine}}$    & 7.60 x $10^{-4}$ Pa.s       \\ \hline
$\mu_{{CO_2 (sc)}}$     & 7.00 x $10^{-5}$ Pa.s       \\ \hline
$\mu_{{CO_2 (g)}}$     & 1.52 x $10^{-5}$ Pa.s       \\ \hline
$M_{{Brine}}$    & 0.1827 x $10^{-6}$ m$^3$.s/kg       \\ \hline
$M_{{CO_2 (sc)}}$     &1.9834 x $10^{-6}$ m$^3$.s/kg       \\ \hline
$M_{{CO_2 (g)}}$     & 9.1342 x $10^{-6}$ m$^3$.s/kg       \\ \hline
$ \beta $  $CO_2$(sc) - brine             & 10.86				        \\ \hline
$ \beta $  $CO_2$(g) - brine             & 50				        \\ \hline
$ \beta $  Infinite mobility ratio model            & $\infty$				        \\ \hline
Intrinsic permeability, k & 1.4 x 10$^{-10}$m$^2$                 \\ \hline
$\gamma$     & 0.03 $kg/s^2$       \\ \hline
$Ca$               & 4561	\\ \hline

\end{tabular}
\caption{Injection Properties of gaseous and supercritical $CO_2$}
\label{props}
\end{table}

\begin{figure}
\centering
\adjincludegraphics[scale=0.51]{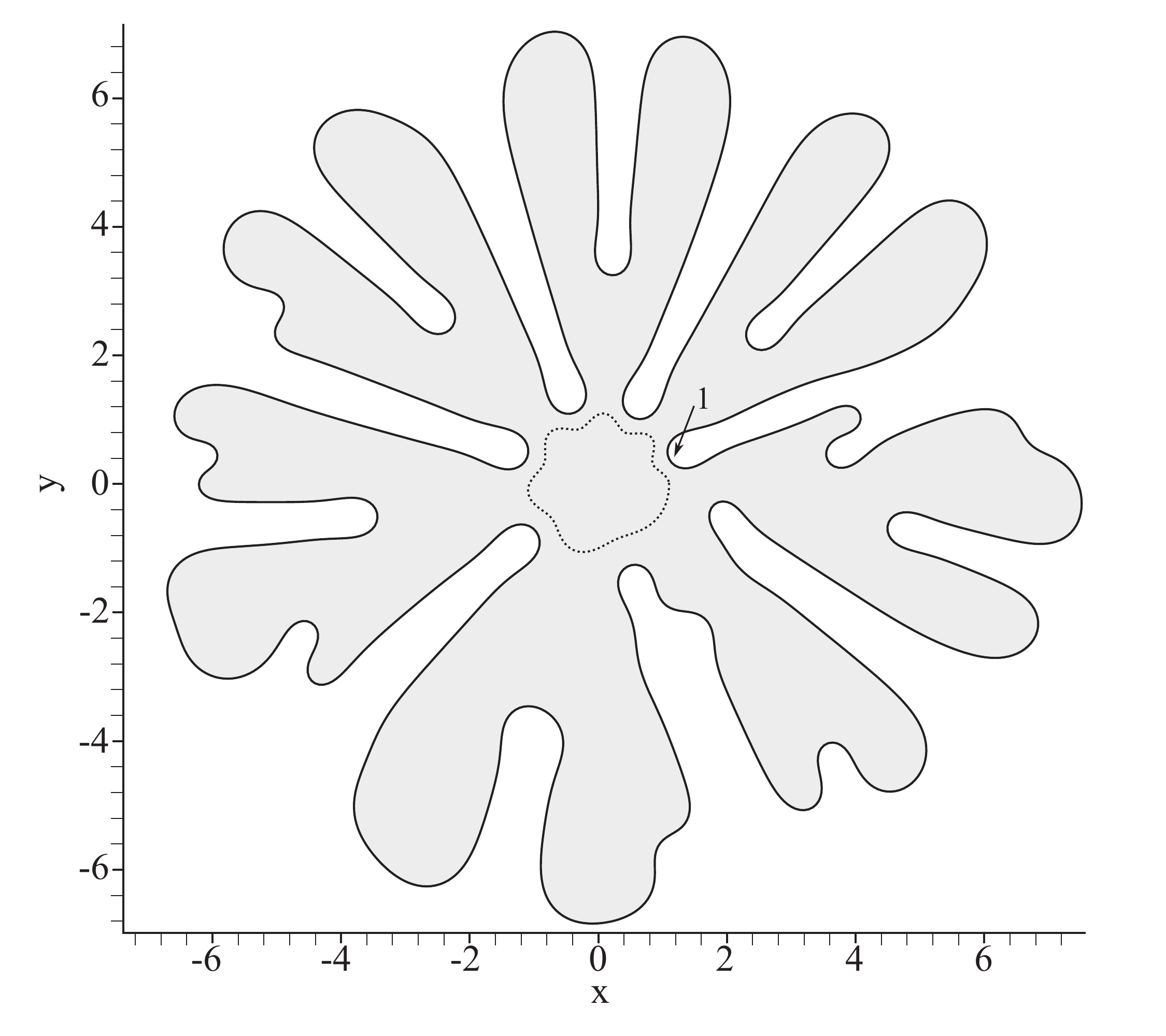}
\centering
\vspace{-20pt}
\caption{Bubble Interface of asymmetric, infinite mobility ratio injection at t = 90, $\beta = \infty$. 1 shows base tracking location in figure (\ref{base_track}).}
\label{co2_single}
\end{figure}

\begin{figure}
\centering
\adjincludegraphics[scale=0.5]{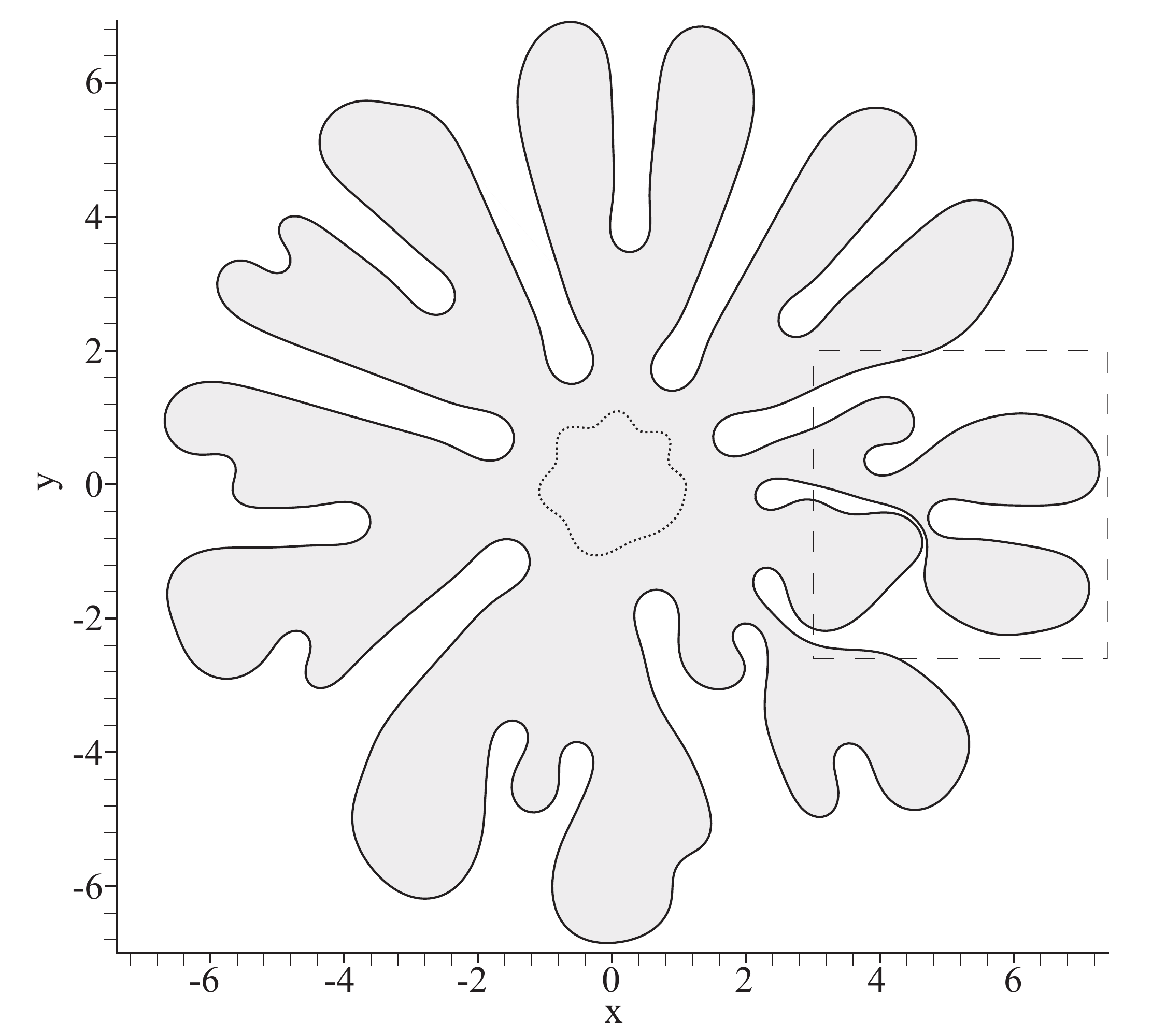}
\centering
\vspace{-20pt}
\caption{Bubble Interface of asymmetric gaseous $CO_2$ injection at t = 90, $\beta = 50$. Dashed box shows zoomed area for subfigures (a) - (c) in figure (\ref{asym_breaking}).}
\label{co2_gas}
\end{figure}

\begin{figure}
\centering
\adjincludegraphics[scale=0.5]{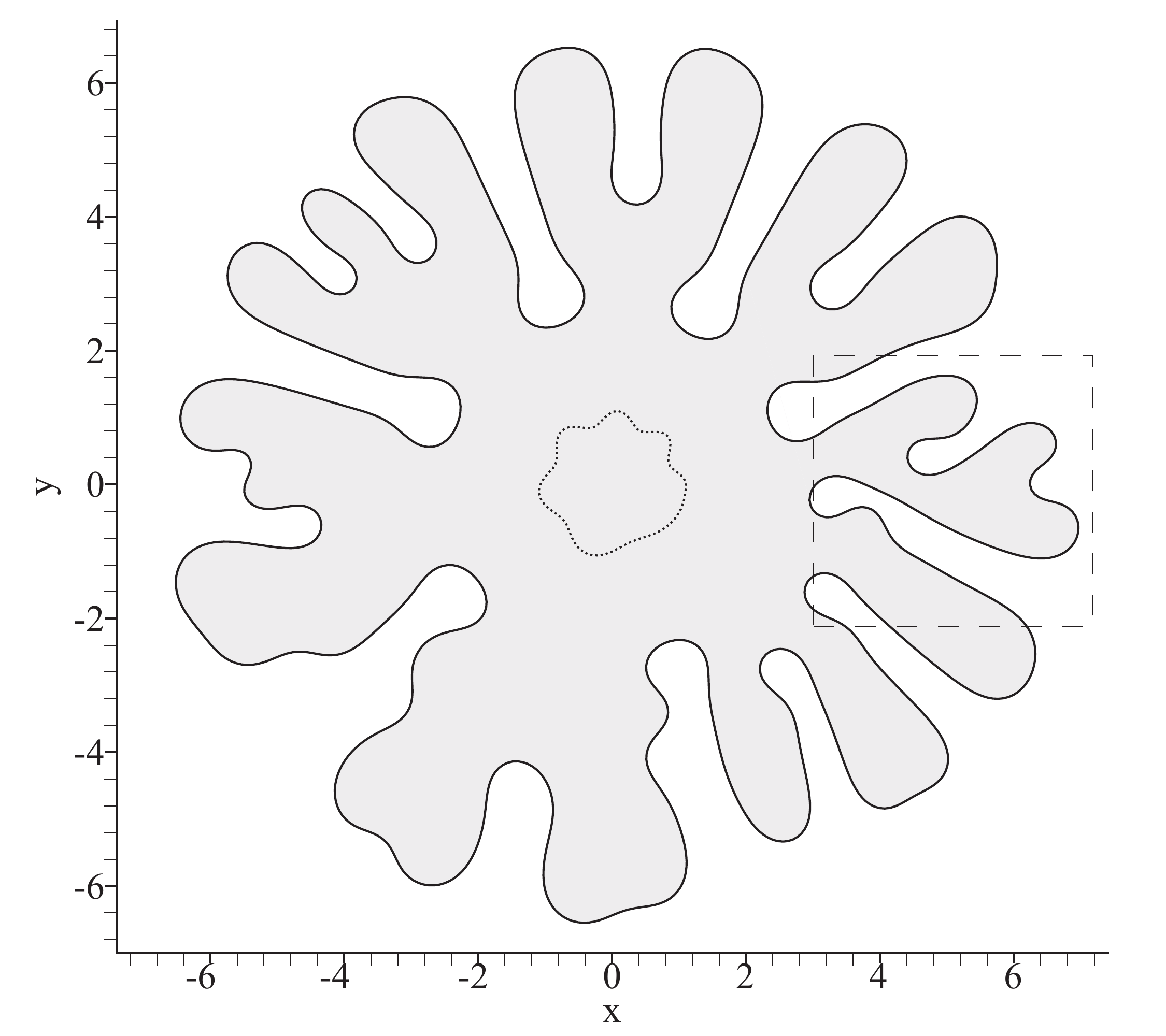}
\centering
\vspace{-20pt}
\caption{Bubble Interface of asymmetric supercritical $CO_2$ injection at t = 90, $\beta = 10.86$. Dashed box shows zoomed area for subfigures (d) - (f) in figure (\ref{asym_breaking}).}
\label{co2_super}
\end{figure}

\begin{figure}
\centering
\adjincludegraphics[scale=0.6]{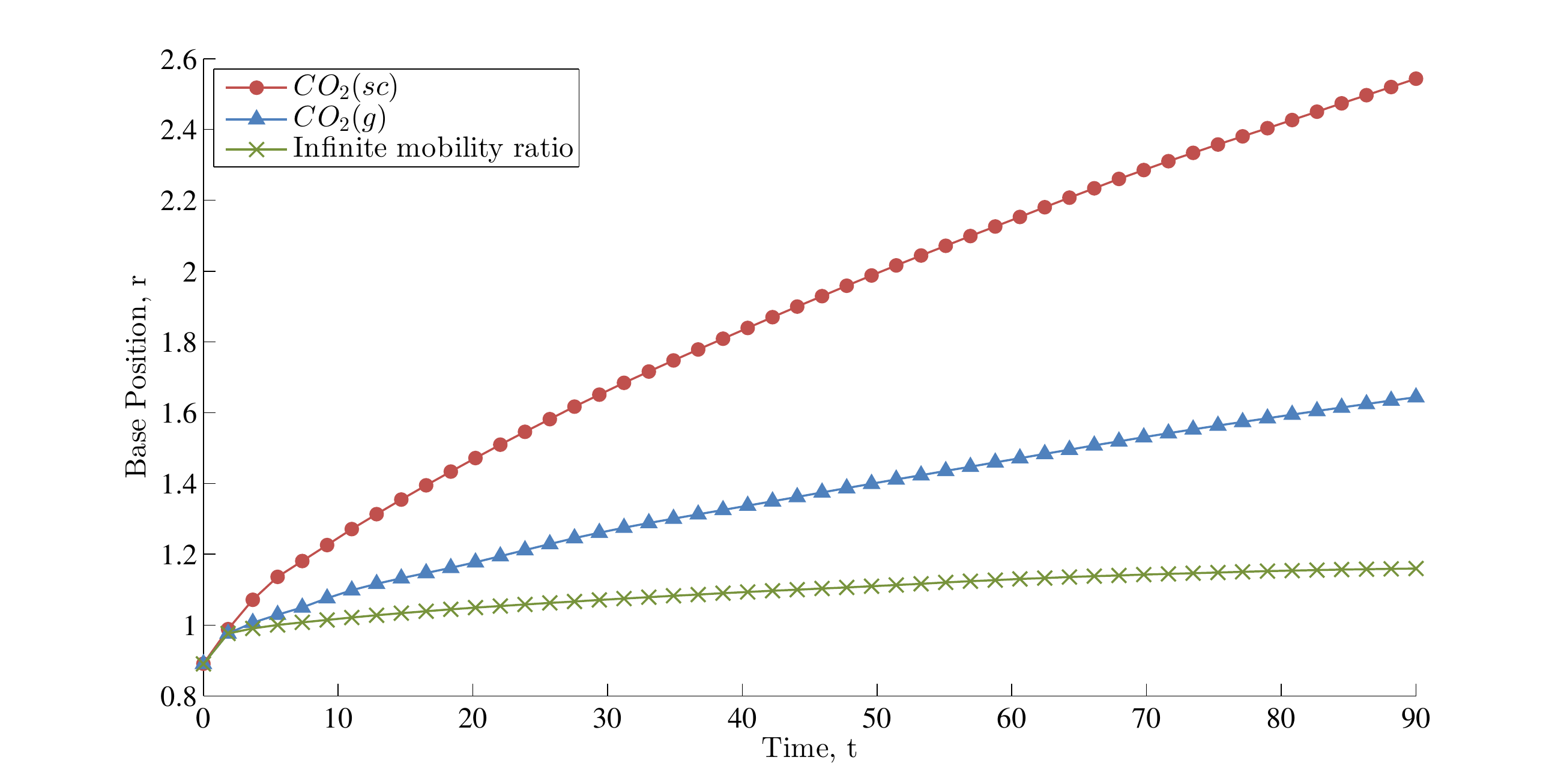}
\centering
\vspace{-20pt}
\caption{Evolution of finger bases (at location 1 in figure (\ref{co2_single})) for infinite mobility ratio ($\beta = \infty$), gaseous ($\beta = 50$) and supercritical ($\beta = 10.86$) $CO_2$ injection. }
\label{base_track}
\end{figure}

\begin{figure}
        \centering
        \begin{tabular}{l c|c r }
           \begin{subfigure}{0.4\textwidth}
                \includegraphics[width=\textwidth]{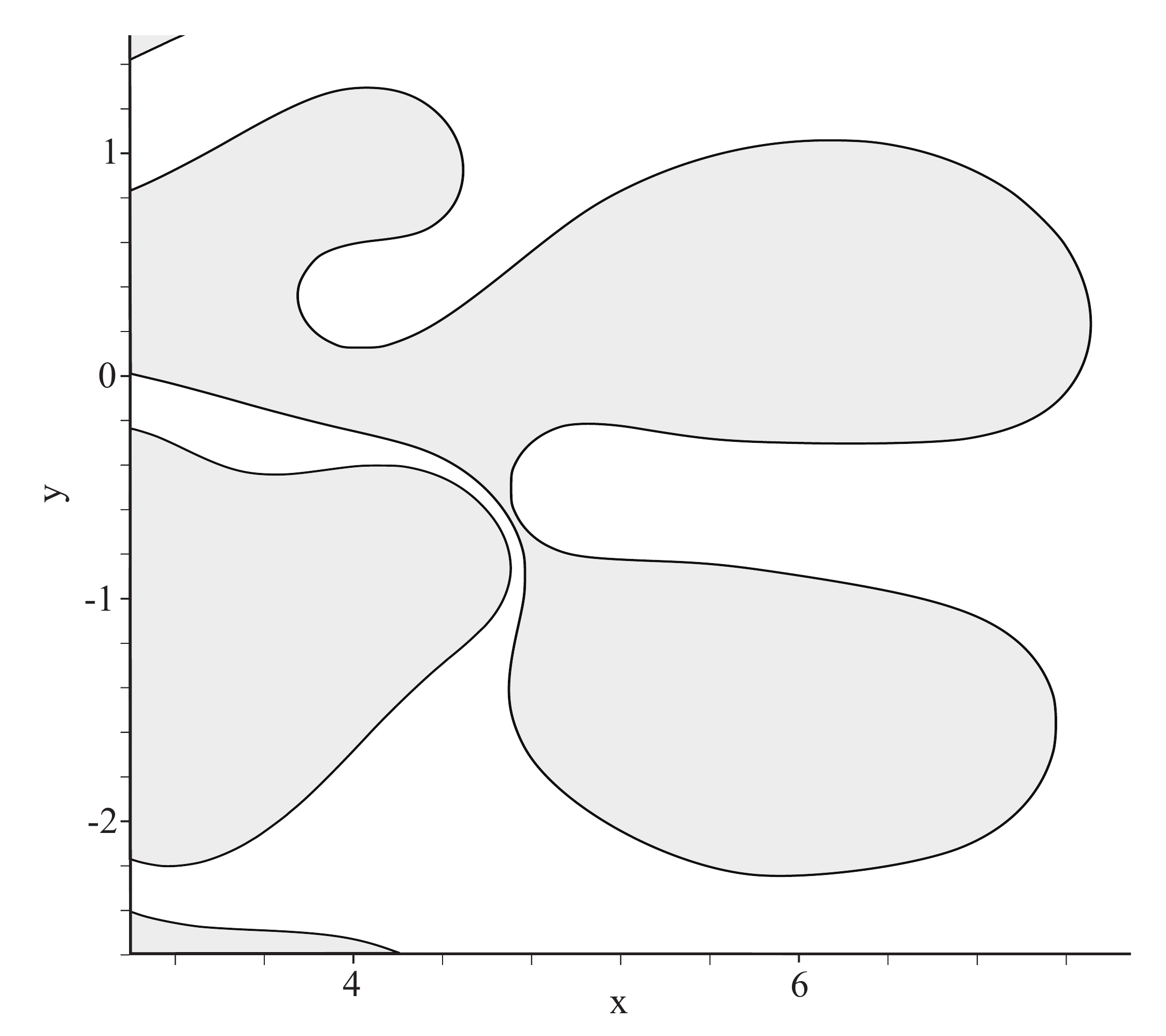}
                \caption{(a) Gaseous $CO_2$, t = 91.6}
                \label{g_break_1}
        \end{subfigure}%
        
        & 
        & 
        & 
         \begin{subfigure}{0.4\textwidth}
                \includegraphics[width=\textwidth]{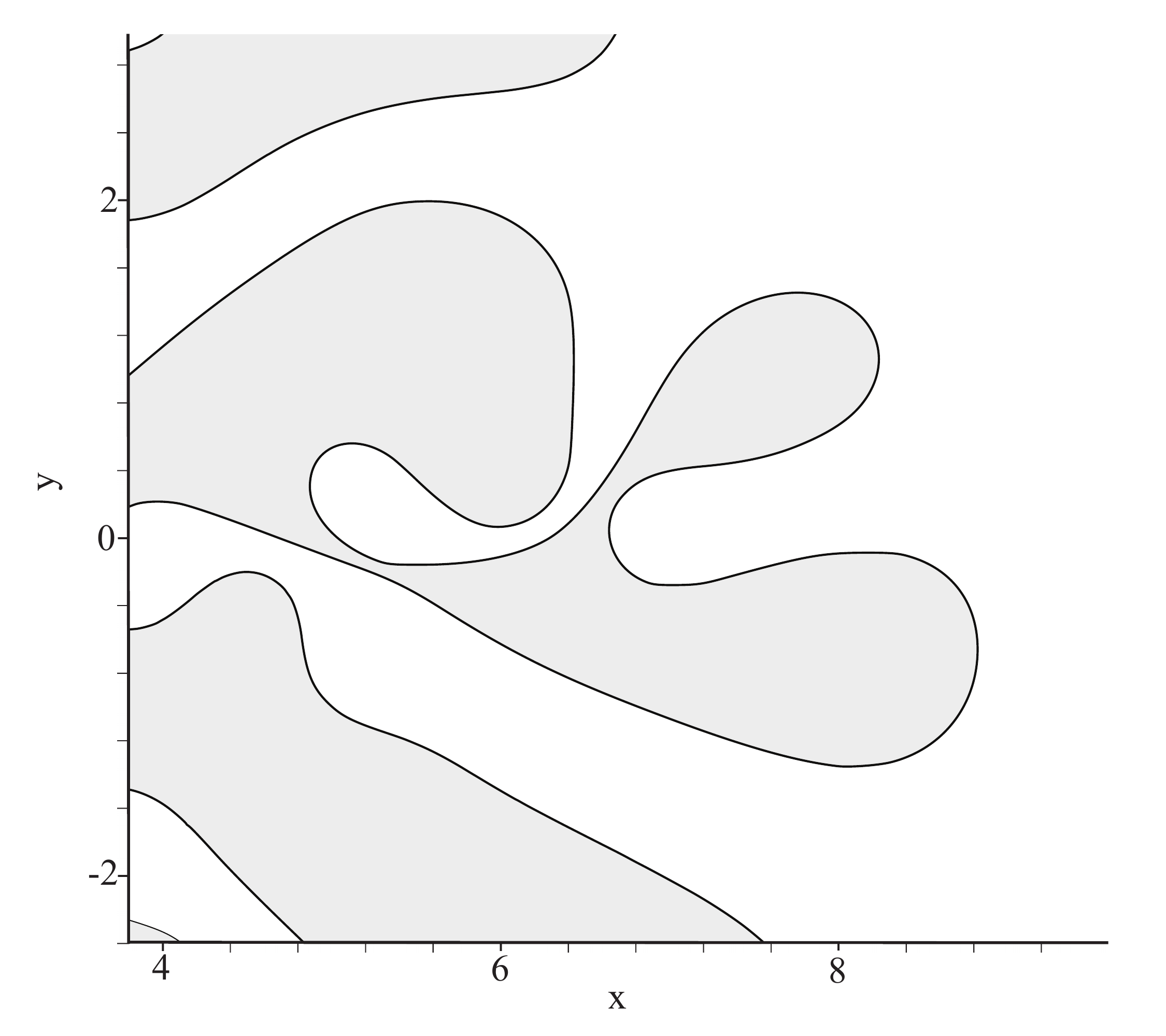}
                \caption{(d) Supercritical $CO_2$, t = 144}
                \label{s_break_1}
                
        \end{subfigure}%
        
         \\
        & 
        & 
        & 
         \\
       \begin{subfigure}{0.4\textwidth}
                \includegraphics[width=\textwidth]{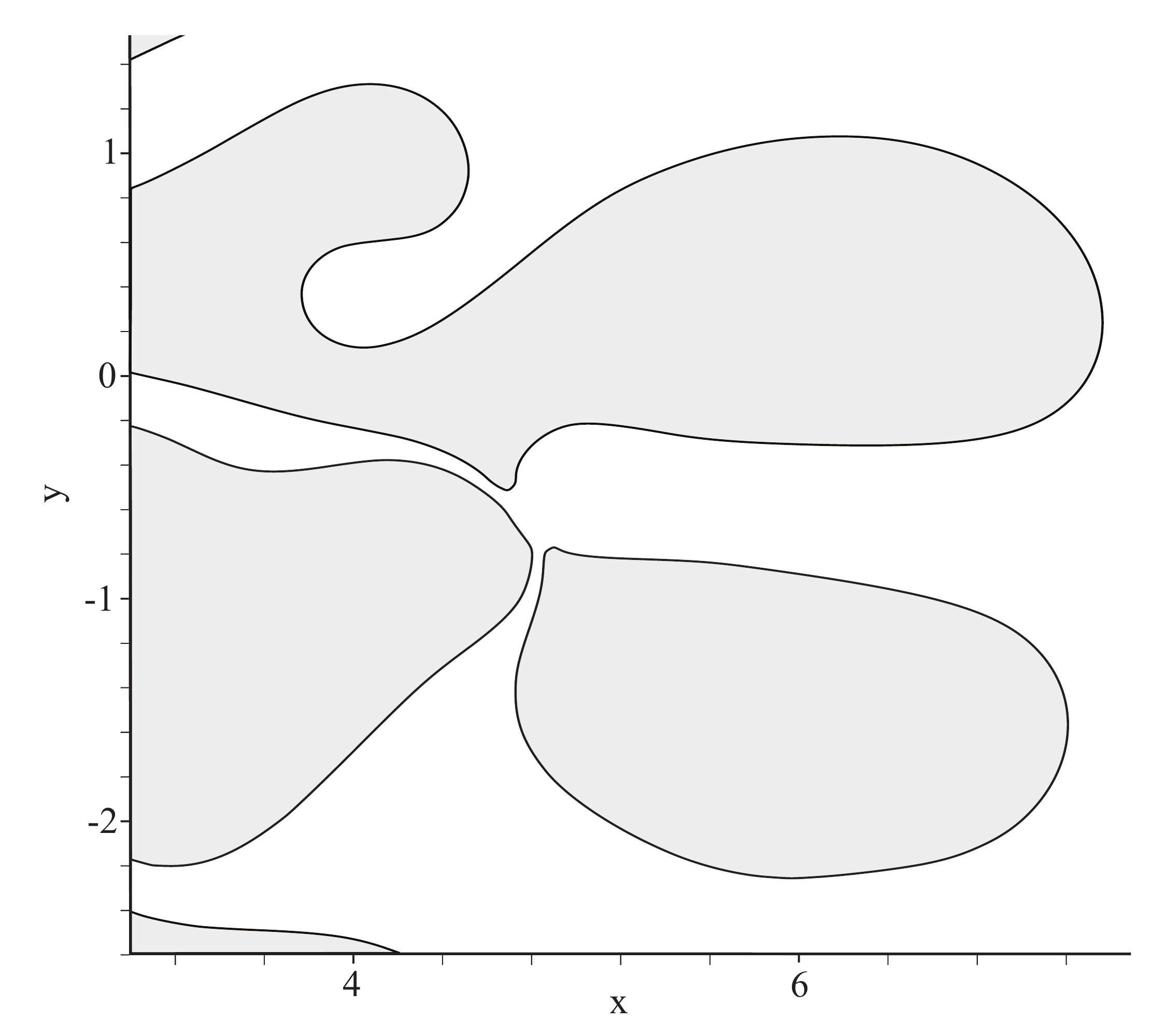}
                \caption{(b) Gaseous $CO_2$, t = 92.6}
                \label{g_break_2}
           \end{subfigure}%
        &
        & 
        &
           \begin{subfigure}{0.4\textwidth}
                \includegraphics[width=\textwidth]{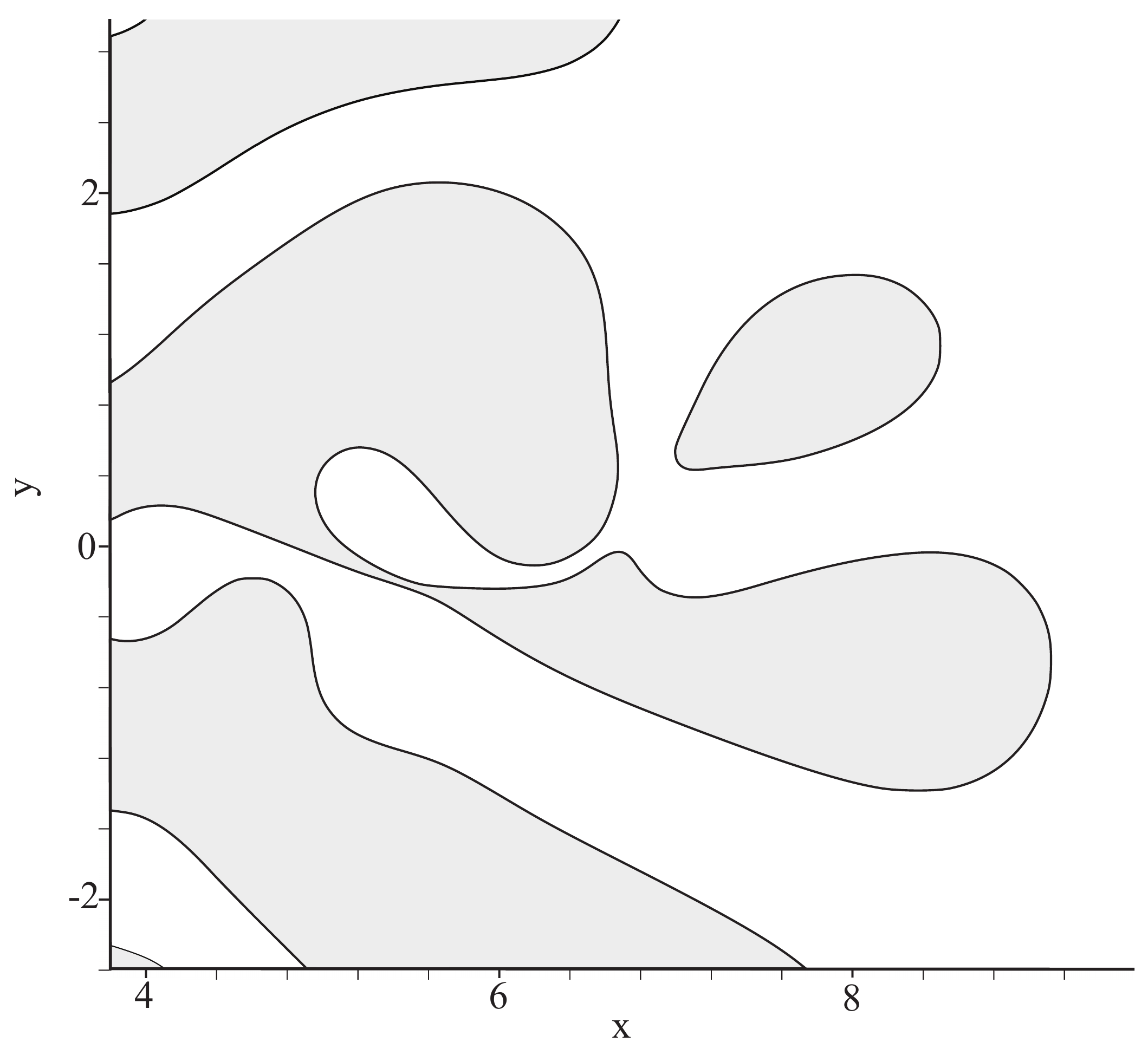}
                \caption{(e) Supercritical $CO_2$, t = 154}
                \label{s_break_2}
        \end{subfigure}%
        
\\
        & 
        & 
        &
\\
        \begin{subfigure}{0.4\textwidth}
                \includegraphics[width=\textwidth]{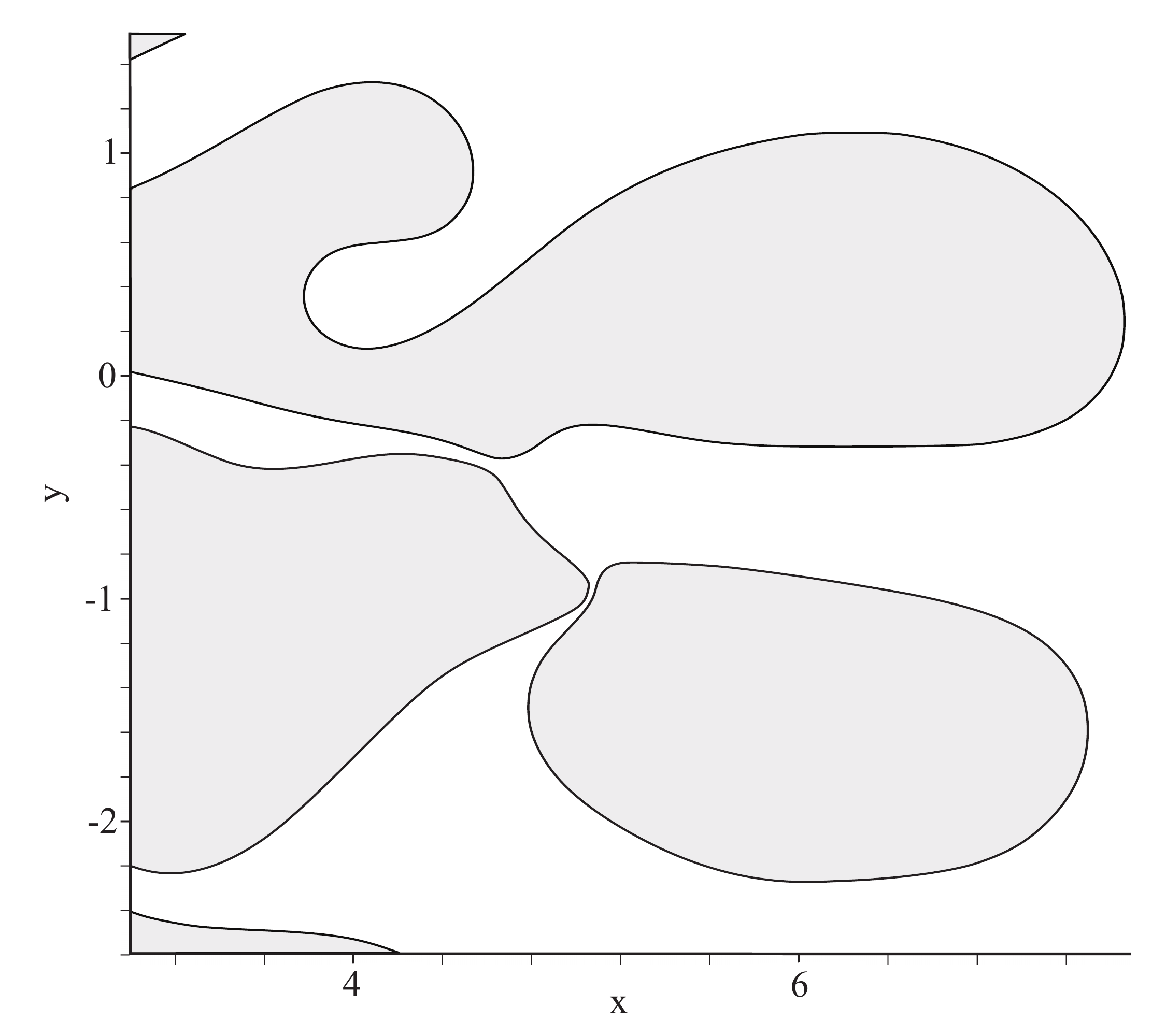}
                \caption{(c) Gaseous $CO_2$, t = 95}
                \label{g_break_3}
        \end{subfigure}%
        &
        & 
        &
          \begin{subfigure}{0.4\textwidth}
                \includegraphics[width=\textwidth]{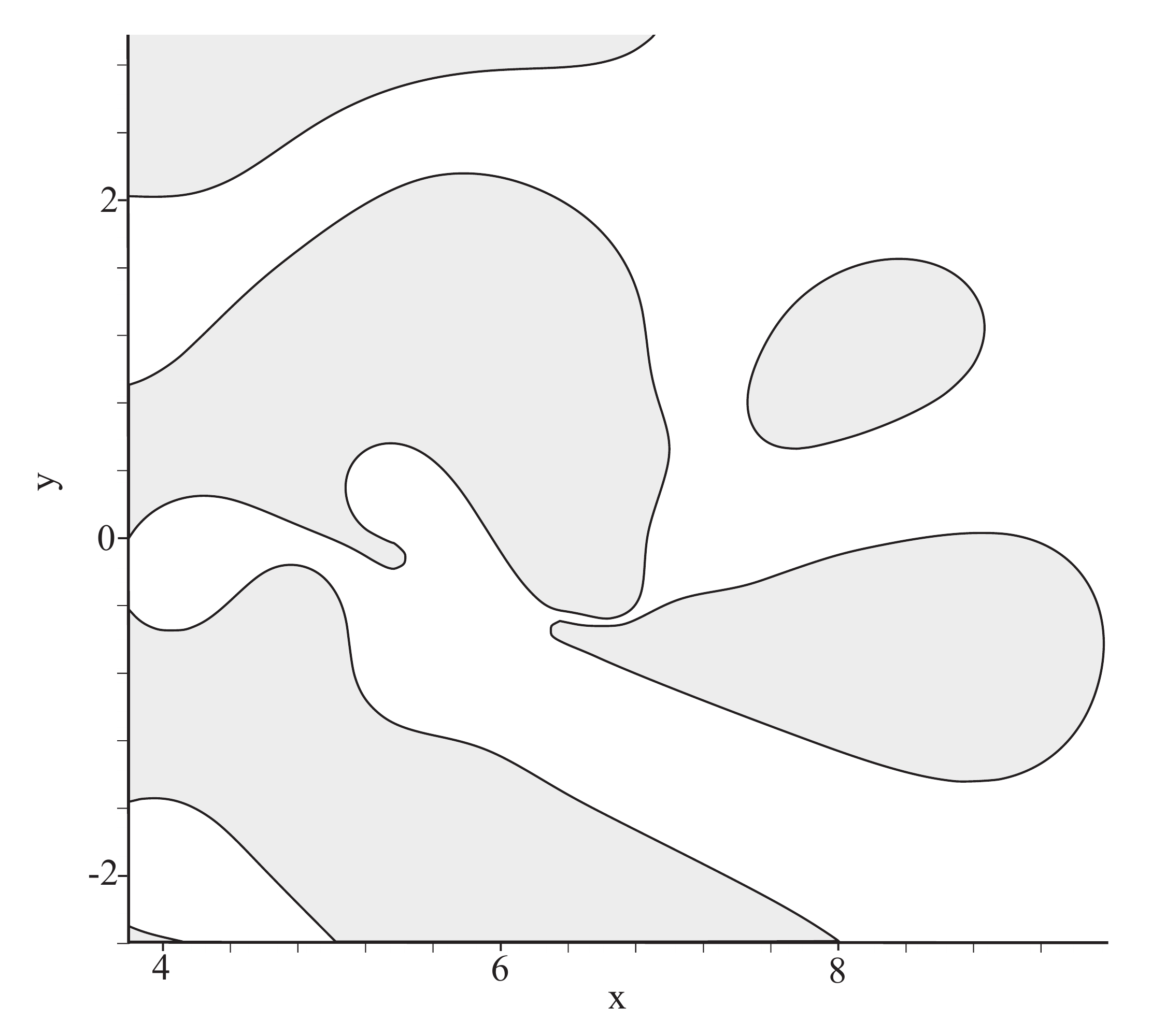}
                \caption{(f) Supercritical $CO_2$, t = 170}
                \label{s_break_3}
        \end{subfigure}%
        
         \end{tabular}
        \caption{Zoomed in plots of finger break-off for the gaseous $CO_2$ injection (left), $\beta$ = 50 and supercritical $CO_2$ injection (right), $\beta$ = 10.86}
        \label{asym_breaking}
        
\end{figure}

Presented in figures (\ref{co2_single}) - (\ref{asym_breaking}) are long time evolutions of three different injection scenarios using fluid parameters defined for supercritical $CO_2$, gaseous $CO_2$ injection and an infinite mobility ratio brine displacement. All simulations use brine as the displaced fluid, with a suitable deep aquifer salinity, and only differ by the mobility of the injecting fluid. Table (\ref{props}) shows the fluid properties for the test cases, with supercritical $CO_2$ properties calculated using \cite{ooyang2011}. The brine salinity, viscosity and surface tension were calculated using standard chemical data tables, under deep subsurface aquifer ambient conditions \cite{kestin1981}\cite{bachu2009}. 


To investigate the complex interfacial dynamics and non-linear growth that could occur in an injection, asymmetry can be introduced into the starting perturbed bubble interface. By including different wavelengths of perturbation along the interface, multiple length scales are produced, mimicking that which would be found in reality due to random noise and disturbance. The initial displacement of the bubble is given by the asymmetric condition below.

\begin{equation}
 \label{asym_bound} r = a + 0.1a \cos \left( 6 \sqrt{\frac{\theta^3}{2 \pi}} \right)
\end{equation}

Equation (\ref{asym_bound}) allows different wavelengths of perturbation to be produced along the interface of the bubble in a controlled manner, in contrast to using a randomly generated noise function. By utilising the asymmetric condition, the effect of different wavelengths of perturbation can be analysed in a reproducible manner, allowing the specific effects of the low mobility ratio environment to be studied accurately. 

Similar to porous media, the flow in a Hele-Shaw cell has an intrinsic permeability, given by the ratio $b^2/12$. In table (\ref{props}), the intrinsic permeability of the Hele-Shaw cell configuration being used corresponds to that of oil reservoir/fractured rock in porous media flow \cite{bear1972}. The three simulations utilising the injection parameters outlined in table (\ref{props}) can be seen in figures (\ref{co2_single}), (\ref{co2_gas}) and (\ref{co2_super}), with dotted lines showing the initial bubble perturbation. 


In figures (\ref{co2_single}), (\ref{co2_gas}) and (\ref{co2_super}), differences in interface patterns can be seen, due to the different mobility ratios in each simulation. The infinite mobility ratio and gaseous injection cases share several similarities, most prominently the near stagnant finger bases that have not moved significantly from their starting positions. This is a common feature of infinite mobility ratio models, and due to the relatively high mobility ratio of the gaseous injection, the finger bases show considerable likeness. This is further emphasised by figure (\ref{base_track}), where the base position of the finger at location 1 in figure (\ref{co2_single}) has been tracked with time for the three different injection scenarios. For the infinite mobility ratio case, the base can be seen to move very little once the initial profile has been set up, however, both the $CO_2(g)$ and $CO_2(sc)$ injection cases show considerable base movement. 

At the moderate mobility ratio of 50, the gaseous injection can be seen to exhibit significant finger interaction in figure (\ref{co2_gas}). The finger shielding effect present in the infinite mobility ratio case has not occurred as significantly in the gaseous injection case and the fingers shown in the dashed box in figure (\ref{co2_gas}) are moving into each other. At the moderate mobility ratio of gaseous $CO_2$ injection,  the inner fluid still possesses some velocity, and hence shielding is inhibited. It is therefore inadequate to use a infinite mobility ratio model for gaseous $CO_2$ injection into brine, with infinite mobility ratio models only being applicable for cases of mobility ratio of 100 or more, common in gas-oil displacements. 

The supercritical $CO_2$ injection case shown in figure (\ref{co2_super}) shows less shielding than the gaseous injection case. The smallest wavelength fingers on the right of the domain have a large interaction with each other, with severe base thinning occurring at two different locations. Due to the relatively large capillary number, the critical length scale is small, allowing side branching to form on some of the larger fingers. More fingers have been allowed to develop, unhindered by the growth of larger fingers, resulting in a larger number of fingers present in the domain. 

In both the gaseous and supercritical $CO_2$ injection cases, competing fingers can grow very close to each other, creating a small immiscible lubrication layer between them. This kind of interaction needs to be monitored closely to study what happens when the fingers are separated by a very small distance. There are two possible outcomes when the fingers grow very close to each other, either the two fingers merge together creating one finger and the inclusion of a brine bubble, or one finger causes the base of the other to thin to such an extent that the finger breaks off.


A rudimentary breaking algorithm was developed that detects the separation between adjacent sections in the viscous fingers. These sections can be external (i.e two fingers are travelling into each other) or internal (i.e the finger's base is thinning). By constructing local B-spline curves, the separation can be analysed efficiently to see whether fingers are likely to collide and merge, or if a finger's base is thinning and the finger will break off. The algorithm then pinches off fingers that have a sufficiently thin base, or merges fingers that are sufficiently close to each other, forming new B-spline curves around each of the disconnected domains. The integral equation formulation allows easy extension of surface integrals over the new disconnected bubbles.





The specific time that the interfaces are captured in figures (\ref{co2_single}), (\ref{co2_gas}) and (\ref{co2_super}) allows the interaction between fingers to be seen and the difference in mobility ratio assessed. After this time, the interaction between fingers in the finite mobility ratio cases becomes highly complex, in particular from fingers that have evolved from the small wavelength perturbations, shown in the dashed boxes. To analyse this interaction and the effect of the lubrication layer between the competing fingers, zoomed in plots of several events have been shown in figure (\ref{asym_breaking}).

\captionsetup[subfigure]{labelformat=empty}

Figure (\ref{asym_breaking}) shows the breaking mechanism of two events in the gaseous and supercritical $CO_2$ injection cases. These occur at different stages in the overall interface evolution due to the difference in finger shielding between the cases. In the gaseous injection, there is still significant finger shielding present, which has caused the finger growing into the base of the primary finger in figure (\ref{asym_breaking}a) to be hindered in its early growth, meaning it did not develop into a primary finger advancing at the forefront of the evolution. However, as it still possesses some velocity, the finger has continued to grow, and eventually thins the base of the primary finger causing it to break off, seen in figure (\ref{asym_breaking}b). This breaking is a combined effect of shielding and the velocity of the inner fluid. If the mobility ratio were pushed much higher, the shielding would have been more significant and the finger would have been almost completely hindered in its early growth and would never have grown to any scale to affect the primary finger.

After the primary finger has broken off in figure (\ref{asym_breaking}c), the secondary finger grows towards it. The breaking of the fingers occurs in a 'snapping' reaction, whereby the fingers break back from each other very rapidly. As the secondary finger is still being fed by the inner fluid, it continues to grow with a significant rate, which is faster than the detached finger. However, a small immiscible lubrication layer of the resident brine separates the two, which is maintained throughout the subsequent evolution and prevents the two fingers from coalescing. The two fingers are immiscible with the brine, and as such cannot transfer mass across the interface and coalesce if this layer is maintained. The secondary finger continues to distort the detached bubble, which will eventually cause it to split. This mechanism is very difficult to capture, due to the very large amount of elements needed to maintain the lubrication layer.

A similar process occurs in the supercritical injection case shown in figures (\ref{asym_breaking}d) - (\ref{asym_breaking}f), in which two fingers detach from the main injection plume. The first finger to detach is caused by the primary finger growing behind it, however the second finger break-off is due to a combination of thinning caused by fingers growing into the left and right side of it. After the second break-off, the finger that is still attached to the main plume acts to push the trailing edge of the detached bubble, which along with surface tension quickly forms it to a droplet shape. The attached finger left from the second break-off quickly recedes due to the very high curvature and re-stabilising effect of surface tension.

This kind of breaking has been reported before in \cite{sun2008} and \cite{guan2003}, however using a diffuse interface model or volume tracking technique does not as accurately resolve the interface between the fluids, introducing a level of uncertainty about the exact position. As the interface is captured implicitly, the lubrication layer between advancing fingers cannot be as accurately defined, meaning that the preference of coalescing or breaking of fingers is unclear. The lubrication layer is vital in defining the movement of fingers in the sharp interface model and as such needs to be accurately resolved. With the current method, the lubrication layer between fingers can be maintained efficiently, allowing the breaking to be seen in much greater detail, with events being explicitly tracked.

In all simulations run, if a sufficiently high enough element density was used, finger break-off would always occur in preference to coalescence. Both mechanisms occur due to the same process of finger interaction and the inhibition of shielding, but as long as the immiscible lubrication layer between the two fluids is maintained, breaking will always occur in preference. The breaking algorithm will detach bubbles that have a neck width under a certain prescribed 'breaking distance'. If this breaking distance is decreased to a very small value much smaller than the element size, the point of separation occurs just before the two surfaces of the interface actually overlap in the next time step, indicting clearly that a break should have occurred. 



Experimentally in \cite{moore2003}, coalescing and breaking of fingers are reported for immiscible flows. Due to the large aspect ratio of the channel being used, there is considerable interaction between fingers, and breaking is found with air injection into resident silicone oil. The finger interaction is promoted by the aspect ratio rather than a low mobility ratio of the fluids, as is the case in the results presented here. Also, in \cite{moore2003} coalescing  occurs between competing fingers under certain conditions after they initially broke. This is due to the fluids having some small amount of miscibility with each other, allowing the rapidly growing finger to coalesce with the bubble that had just detached.

Due to the microscopic scale of fluid interaction that occurs experimentally, and the fact that the fluids under consideration in \cite{moore2003} have a small level of miscibility with each other, there will be a very small miscible region between the fluids. If the fingers are moving with sufficient speed the fingers could overcome the miscible region separating the fingers, and coalesce. However as there is no microscopic miscible layer in the sharp interface model presented here, coalescing cannot occur at the macroscopic scale. The sharp interface model assumes the displacement of the resident fluid occurs much faster than the miscible mixing of the fluids and that there is a discontinuity of properties over the fluid interface. The fluids are considered to be completely immiscible and therefore no coalescence should occur, as a lubrication layer should always separate the two fluids. 

\section{Conclusion}

A BEM formulation for solving finite mobility ratio flows has been developed and used to investigate radial viscous fingering mechanisms in a Hele-Shaw cell. The Hele-Shaw model was used to qualitatively investigate the mechanisms and plume evolution associated with viscous fingering that could occur during $CO_2$ injection and storage in deep porous media aquifers, during carbon sequestration. 

The finite mobility ratio model allowed investigation into the effects that a low mobility ratio and high capillary number have on the $CO_2$ plume evolution.  When the mobility ratio of the two fluids is of order 10 - 50, the fingering characteristics are vastly different to those predicted by infinite mobility ratio models. Finger base movement was found to be independent of capillary number, but strongly dependant on mobility ratio, with the bases moving significantly away from their starting positions. The near stagnation points on the bases of the fingers found in infinite mobility ratio flows were not found when using the finite mobility ratio model for low mobility ratio flows. 

The developed numerical method has been validated and its performance characteristics verified, showing its applicability to finite mobility ratio flows. The quadratic scaling of the solution time shows large improvements over traditional matrix solvers and lifts the restriction of previous models on short time scale solutions. The numerical stability of the solution has been shown to rely heavily on the temporal and spacial discretisation, providing an upper limit for stability. Similarly, the physical instability of bifurcating fingers was analysed with an expression found to predict the point at which the first bifurcation of a finger will occur.

Long time interface evolutions were run to showcase the numerical method for predicting the large time scale dynamics of viscous fingering. Finger interaction was found to be much more significant than in infinite mobility ratio models, and on small wavelength perturbations could lead to base thinning and eventual finger breaking. After breaking, the detached bubbles would continue with the velocity of the surrounding fluid. The numerical method allowed the resolution of the immiscible lubrication layer between fingers meaning the finger breaking and coalescing mechanisms could be explored more explicitly than in previous models. 

Future work will focus on 3D effects that can occur during immiscible displacement in a Hele-Shaw cell, which have recently received signification attention. The wettability of the fluids and the thin film of trailing fluid left behind by the displaced fluid require explicit inclusion in the pressure jump condition at the interface of the fluids, and can significantly alter the resulting displacement \cite{dong2010}\cite{anjos2013}. Viscous effects caused by tangential shear stresses at the interface can also be considered using a Brinkman flow model, as opposed to the Potential flow model used here. Second order velocity components can have significant effects on the displacement when shear stresses are non-negligible \cite{brinkman1947}\cite{nagel2013}.


\section*{Acknowledgements}

The authors would like to thank the University of Nottingham HPC team for the use of the Minerva supercomputer cluster and the IJNMF reviewer, whose comments greatly improved the paper. The present work has been partially supported by an EPSRC post-graduate research scholarship and the European Commission project PANACEA (Project Reference 282900), seventh framework program.


\begin{thebibliography}{10}

\bibitem{saffman1958}
P.G. Saffman and G.I. Taylor.
\newblock {The penetration of a fluid into a porous medium or Hele-Shaw cell
  containing a more viscous liquid.}
\newblock {\em Proceedings of the Royal Society of London. Series A:
  Mathematical and Physical Sciences}, 245(1242):312 -- 329, 1958.

\bibitem{patterson1981}
L.~Patterson.
\newblock {Radial fingering in a Hele Shaw cell}.
\newblock {\em Journal of Fluid Mechanics}, 113:513 -- 529, 1981.

\bibitem{homsy1987}
G.M. Homsy.
\newblock Viscous fingering in porous media.
\newblock {\em Annual Review of Fluid Mechanics}, 19:271--311, 1987.

\bibitem{howison1986}
S.D. Howison.
\newblock {Fingering in Hele-Shaw cells}.
\newblock {\em Journal of Fluid Mechanics}, 167(3):439--453, 1986.

\bibitem{tanveer2000}
S.~Tanveer.
\newblock Surprises in viscous fingering.
\newblock {\em Journal of Fluid Mechanics}, 409:273 -- 308, 2000.

\bibitem{maxworthy1989}
T.~Maxworthy.
\newblock Experimental study of interface instability in a hele-shaw cell.
\newblock {\em Physical Review A}, 39(11):5863 -- 5866, 1989.

\bibitem{miranda1998}
J.A. Miranda and M.~Widom.
\newblock {Radial fingering in a Hele-Shaw cell: A weakly nonlinear analysis}.
\newblock {\em Physica D}, 120:315 -- 328, 1998.

\bibitem{chouke1959}
R.L Chouke, P.~Meurs, and C.~Van der Poel.
\newblock The instability of slow, immisicible, viscous liquid-liquid
  displacements in permeable media.
\newblock {\em Transactions of the American Institute of Mining, Metallurgical
  and Petroleum Engineers}, 216:188 -- 194, 1959.

\bibitem{weitz1987}
D.A. Weitz, J.P. Stokes, R.C. Ball, and A.P. Kushnick.
\newblock Dynamic capillary pressure in porous media: Origin of the viscous
  fingering length scale.
\newblock {\em Physics Review Letters}, 59:2967 -- 2970, 1987.

\bibitem{garcia2003}
J.E. Garcia and K.~Pruess.
\newblock {Flow instabiltiies during injection of $CO_2$ into saline aquifers}.
\newblock In {\em TOUGH symposium 2003}, Lawrence Berkeley National Laboratory,
  Berkeley, California, May 2003.

\bibitem{riaz2006}
A.~Riaz and H.A. Tchelepi.
\newblock Numerical simulation of immiscible two-phase flow in porous media.
\newblock {\em Physics of Fluids}, 18(1), 2006.

\bibitem{gorell1983}
S.B. Gorell and G.M. Homsy.
\newblock A theory of optimal policy of oil recovery by secondary displacement
  process.
\newblock {\em SIAM Journal of Applied Mathematics}, 43:79 -- 98, 1983.

\bibitem{hickernell1986}
F.J. Hickernell and Y.C. Yortsos.
\newblock Linear stability of miscible displacement processes in porous media
  in the absence of dispersion.
\newblock {\em Studies in Applied Mathematics}, 74:93 -- 115, 1986.

\bibitem{zhao1995}
K.X.H Zhao, L.C. Wrobel, and H.~Power.
\newblock {Numerical simulation of viscous fingering using B-spline boundary
  elements}.
\newblock {\em Transactions on Modelling and Simulation}, 11:1--10, 1995.

\bibitem{degregoria1986}
A.J. DeGregoria and L.W. Shwartz.
\newblock A boundary-integral method for two-phase displacement in hele-shaw
  cells.
\newblock {\em Journal of Fluid Mechanics}, 164:383--400, 1986.

\bibitem{li2007}
S.~Li, J.S.; Lowengrub, and P.H. Leo.
\newblock {A rescaling scheme with application to the long-time simulation of
  viscous fingering in a Hele-Shaw cell}.
\newblock {\em Journal of Computational Physics}, 25(1):554--567, 2007.

\bibitem{hadavinia1995}
H.~Hadavinia, S.G. Advani, and R.T. Fenner.
\newblock {The evolution of radial fingering in a Hele-Shaw cell using $C^1$
  continuous Overhauser boundary element method}.
\newblock {\em Engineering Analysis with Boundary Elements}, 16:183--195, 1995.

\bibitem{hansen1999}
E.B. Hansen.
\newblock A numerical study of unstable {Hele-Shaw } flow.
\newblock {\em Computers and Mathematics with Applications}, 38:217--230, 1999.

\bibitem{jawson1977}
M.A. Jawson and G.T. Symm.
\newblock {\em Integral Equations Methods in Potential Theory and
  Elastostatics}.
\newblock Academic Press, New York, 1977.

\bibitem{power1994}
H.~Power.
\newblock The evolution of radial fingers at the interface between two viscous
  liquids.
\newblock {\em Engineering Analysis with Boundary Elements}, 14(4):297 -- 304,
  1994.

\bibitem{guan2003}
X.~Guan and R.~Pitchumani.
\newblock Viscous fingering in a hele-shaw cell with finite viscosity ratio and
  interfacial tension.
\newblock {\em Journal of Fluids Engineering}, 125:354--363, 2003.

\bibitem{sun2008}
Y.~Sun and C.~Beckermann.
\newblock {A two-phase diffuse-interface model for Hele-Shaw flows with large
  property contrasts}.
\newblock {\em Physica D}, 237:3089--3098, 2008.

\bibitem{hou1997}
T.Y. Hou, Z.~Li, S.~Osher, and H.~Zhao.
\newblock {A hybrid method for moving interface problems with application to
  the Hele-Shaw flow}.
\newblock {\em Journal of Computational Physics}, 134:236--252, 1997.

\bibitem{moore2003}
M.G. Moore, A.~Juel, J.M. Burgess, W.D. McCormick, and H.L. Swinney.
\newblock Fluctuations and pinch-offs observed in viscous fingering.
\newblock In {\em Proceedings of the Seventh Experimental Chaos Conference},
  pages 189--194. Springer-Verlag, 2003.

\bibitem{power1995}
H.~Power and L.C. Wrobel.
\newblock {\em Boundary Integral Methods in Fluid Mechanics}.
\newblock Computational Mechanics Publications, Southampton, 1995.

\bibitem{batchelor1967}
G.K. Batchelor.
\newblock {\em An Introuduction to Fluid Dynamics}.
\newblock Cambridge University Press, 1967.

\bibitem{power2013}
H.~Power, D.~Stevens, K.A. Cliffe, and A.~Golin.
\newblock {A boundary element study of the effect of surface dissolution on the
  evolution of immiscible viscous fingering within a Hele-shaw cell}.
\newblock {\em Engineering Analysis with Boundary Elements}, 37:1318--1330,
  2013.

\bibitem{cabral1990}
J.J.S.P Cabral, L.C. Wrobel, and C.A. Brebbia.
\newblock {A BEM formulation using B-splines: I - Uniform blending functions}.
\newblock {\em Engineering Analysis with Boundary Elements}, 7(3):136--144,
  1990.

\bibitem{press1996}
W.H. Press, B.P. Flannery, S.A. Teukolsky, and W.T. Vetterling.
\newblock {\em Numerical Recipes in Fortran 90: The Art of Parallel Scientific
  Computing}.
\newblock Cambridge University Press, 2nd edition edition, 1996.

\bibitem{chorin1985}
A.J. Chorin.
\newblock Curvature and solidification.
\newblock {\em Journal of Computational Physics}, 57:472 -- 490, 1985.

\bibitem{gui1998}
M.~Guigianni.
\newblock Formulation and numerical treatment of boundary integral equations
  with hypersingular kernals.
\newblock In V.~Sladek and J.~Sladek, editors, {\em Singular Intgerals in
  Boundary Element Methods}, pages 85--125. Computational Mechanics
  Publications, 1998.

\bibitem{mikhlin1957}
S.G. Mikhlin.
\newblock {\em Multidimensional singular integral and integral equations}.
\newblock Pergamon Press, New York, 1957.

\bibitem{hadamard1952}
J.~Hadamard.
\newblock {\em Lectures on Cauchy's problem in linear partial differential
  equations}.
\newblock Dover, New York, 1952.

\bibitem{ooyang2011}
L.~Ouyang.
\newblock New correlations for predicting the density and viscosity of
  supercritical carbon dioxide under conditions expected in carbon capture and
  sequestration operations.
\newblock {\em The Open Petroleum Engineering Journal}, 4:13--21, 2011.

\bibitem{couder2000}
E.~Lajeunesse and Y.~Couder.
\newblock On the tip-splitting instability of viscous fingers.
\newblock {\em Journal of Fluid Mechanics}, 419:125--149, 2000.

\bibitem{alvarex2004}
E.~Alvarez-Lacelle, J.~Ortin, and J.~Casademunt.
\newblock Low viscosity contrast fingering in a rotating hele-shaw cell.
\newblock {\em Physics of Fluids}, 16(4):908--924, 2004.

\bibitem{wit1999}
A.~De Wit and G.M. Homsy.
\newblock Viscous fingering in reaction-diffusion systems.
\newblock {\em Journal of Chemical Physics}, 110(17):8663--8675, 1999.

\bibitem{kestin1981}
J.~Kestin, H.E. Khalifa, and R.J. Correia.
\newblock Tables of the dynamic and kinematic viscosity of aqueos {NaCL}
  solutions in the temperature range {20-150 $^\circ C $} and the pressure
  range {0.1 - 35 MPa}.
\newblock {\em Journal of Physical and Chemical Reference Data}, 10:71 -- 87,
  1981.

\bibitem{bachu2009}
S.~Bachu and D.B. Bennion.
\newblock Interfacial tension between $co_2$, freshwater, and brine in the
  range of pressure from 2 to 27mpa, temperature from 20 to 125$^ \circ c$, and
  water salinity from 0 to 334 000 mg/l.
\newblock {\em Journal of Chemical Engineering Data}, 54:765--775, 2009.

\bibitem{bear1972}
Jacob Bear.
\newblock {\em Dynamics of Fluids in Porous Media}.
\newblock Dover, 1972.

\bibitem{dong2010}
B.~Dong, Y.Y. Yang, W.~Li, and Y.~Song.
\newblock Lattice boltzmann simulation of viscous fingering phenomenon of
  immiscible fluids displacement in a channel.
\newblock {\em Computers \& Fluids}, 39:768 -- 779, 2010.

\bibitem{anjos2013}
P.H. Anjos and J.A. Miranda.
\newblock Radial viscous fingering: Wetting effects on pattern-forming
  mechanisms.
\newblock {\em Physical Review E}, 88:053003--1 --7, 2013.

\bibitem{brinkman1947}
H.C. Brinkman.
\newblock A calculation of the viscous force exerted by a floflow fluid on a
  dense swarm of particles.
\newblock {\em Applied Scientific Research}, A1:27 -- 34, 1947.

\bibitem{nagel2013}
M.~Nagel and F.~Gallaire.
\newblock A new prediciton of wavelength selection in radial viscous fingering
  involving normal and tangential stresses.
\newblock {\em Physics of Fluids}, 25, 2013.

\end{thebibliography}
\end{document}